\documentclass[times]{aastex631}
\usepackage{CJK}
\usepackage{amsmath}
\hypersetup{linkcolor=blue,citecolor=blue,filecolor=blue,urlcolor=blue}

\begin{document}
\begin{CJK*}{UTF8}{gbsn}
\shorttitle{B in IRDC G28.34}
\shortauthors{Liu et al.}

%\title{Dark Dragon Breaks Magnetic Chain: Dynamical Early-stage Clumps/Cores Formed in Equilibrium Environments in IRDC G28.34. supercritical clumps/cores in IRDC G28.34 formed in subcritical environments}
%\title{Dark Dragon Unleashed: Dynamical Substructures of IRDC G28.34 Form in Magnetically-supported Environments}
%\title{Dark Dragon Breaks Magnetic Chain: Dynamical Substructures of IRDC G28.34 Form in Quasi-Equilibrium Environments}
\title{Dark Dragon Breaks Magnetic Chain: Dynamical Substructures of IRDC G28.34 Form in Supported Environments}
%\title{Dark Dragon Breaking Magnetic Chain but Resisted by Gas Pressure in High-Density: IRDC G28.34 in Quasi-Equilibrium}
%\title{Dark Dragon Transits from Magnetically-Supported to Gas Pressure-Supported: IRDC G28.34 in Quasi-Equilibrium}

\correspondingauthor{Junhao Liu}
\email{liujunhao42@outlook.com; junhao.liu@nao.ac.jp}

\author[0000-0002-4774-2998]{Junhao Liu (刘峻豪)}
\affiliation{National Astronomical Observatory of Japan, 2-21-1 Osawa, Mitaka, Tokyo 181-8588, Japan}
\affiliation{East Asian Observatory, 660 N. A`oh\={o}k\={u} Place, University Park, Hilo, HI 96720, USA}

\author[0000-0003-2384-6589]{Qizhou Zhang}
\affiliation{Center for Astrophysics $\vert$ Harvard \& Smithsonian, 60 Garden Street, Cambridge, MA 02138, USA}

\author[0000-0001-9299-5479]{Yuxin Lin}
\affiliation{Max-Planck-Institut f{\"u}r Extraterrestrische Physik, Giessenbachstr 1, D-85748 Garching bei M{\"u}nchen, Germany}

\author[0000-0002-5093-5088]{Keping Qiu}
\affiliation{School of Astronomy and Space Science, Nanjing University, 163 Xianlin Avenue, Nanjing 210023, Jiangsu, People's Republic of China}
\affiliation{Key Laboratory of Modern Astronomy and Astrophysics (Nanjing University), Ministry of Education, Nanjing 210023, Jiangsu, People's Republic of China}

\author[0000-0003-2777-5861]{Patrick M. Koch}
\affiliation{Institute of Astronomy and Astrophysics, Academia Sinica, 11F of Astronomy-Mathematics Building, AS/NTU No.1, Sec. 4, Roosevelt Rd, Taipei 10617, Taiwan, Republic of China}

\author[0000-0003-2300-2626]{Hauyu Baobab Liu}
\affiliation{Department of Physics, National Sun Yat-Sen University, No. 70, Lien-Hai Road, Kaohsiung City 80424, Taiwan, Republic of China} 
\affiliation{Center of Astronomy and Gravitation, National Taiwan Normal University, Taipei 116, Taiwan, Republic of China}

\author[0000-0002-7402-6487]{Zhi-Yun Li}
\affiliation{Astronomy Department, University of Virginia, Charlottesville, VA 22904-4325, USA}

\author[0000-0002-3829-5591]{Josep Miquel Girart}
\affil{Institut de Ci\`{e}ncies de l'Espai (ICE, CSIC), Can Magrans s/n, E-08193 Cerdanyola del Vall\`{e}s, Catalonia, Spain}
\affil{Institut d'Estudis Espacials de de Catalunya (IEEC), E-08034 Barcelona, Catalonia, Spain}

\author[0000-0003-2133-4862]{Thushara G.S. Pillai}
\affiliation{MIT Haystack Observatory, 99 Millstone Road, Westford, MA, 01826, USA}

\author[0000-0003-1275-5251]{Shanghuo Li}
\affiliation{Max Planck Institute for Astronomy, Konigstuhl 17, D-69117 Heidelberg, Germany}

\author[0000-0002-9774-1846]{Huei-Ru Vivien Chen}
\affiliation{Institute of Astronomy and Astrophysics, Academia Sinica, 11F of Astronomy-Mathematics Building, AS/NTU No.1, Sec. 4, Roosevelt Rd, Taipei 10617, Taiwan, Republic of China}

\author[0000-0001-8516-2532]{Tao-Chung Ching}
\affiliation{National Radio Astronomy Observatory, P.O. Box O, Socorro, NM 87801, USA}

\author[0000-0002-3412-4306]{Paul T. P. Ho}
\affiliation{East Asian Observatory, 660 N. A`oh\={o}k\={u} Place, University Park, Hilo, HI 96720, USA}
\affiliation{Institute of Astronomy and Astrophysics, Academia Sinica, 11F of Astronomy-Mathematics Building, AS/NTU No.1, Sec. 4, Roosevelt Rd, Taipei 10617, Taiwan, Republic of China}

\author[0000-0001-5522-486X]{Shih-Ping Lai}
\affiliation{Institute of Astronomy and Department of Physics, National Tsing Hua University, Hsinchu 30013, Taiwan, Republic of China}
\affiliation{Institute of Astronomy and Astrophysics, Academia Sinica, 11F of Astronomy-Mathematics Building, AS/NTU No.1, Sec. 4, Roosevelt Rd, Taipei 10617, Taiwan, Republic of China}

\author[0000-0002-1407-7944]{Ramprasad Rao}
\affiliation{Center for Astrophysics $\vert$ Harvard \& Smithsonian, 60 Garden Street, Cambridge, MA 02138, USA}
\affiliation{Institute of Astronomy and Astrophysics, Academia Sinica, 11F of Astronomy-Mathematics Building, AS/NTU No.1, Sec. 4, Roosevelt Rd, Taipei 10617, Taiwan, Republic of China}

\author[0000-0002-0675-276X]{Ya-Wen Tang}
\affiliation{Institute of Astronomy and Astrophysics, Academia Sinica, 11F of Astronomy-Mathematics Building, AS/NTU No.1, Sec. 4, Roosevelt Rd, Taipei 10617, Taiwan, Republic of China}

\author[0000-0002-7237-3856]{Ke Wang}
\affiliation{Kavli Institute for Astronomy and Astrophysics, Peking University, 5 Yiheyuan Road, Haidian District, Beijing 100871, People's Republic of China}

\begin{abstract}

We have comprehensively studied the multi-scale physical properties of the infrared dark cloud (IRDC) G28.34 (the Dragon cloud) with dust polarization and molecular line data from Planck, FCRAO-14m, JCMT, and ALMA. 
We find that the averaged magnetic fields of clumps tend to be either parallel with or perpendicular to the cloud-scale magnetic fields, while the cores in clump MM4 tend to have magnetic fields aligned with the clump fields.
Implementing the relative orientation analysis (for magnetic fields, column density gradients, and local gravity), Velocity Gradient Technique (VGT), and modified Davis-Chandrasekhar-Fermi (DCF) analysis, we find that: G28.34 is located in a trans-to-sub-Alfv\'{e}nic environment ($\mathcal{M}_{A}=0.74$ within $r=15$ pc); the magnetic field is effectively resisting gravitational collapse in large-scale diffuse gas, but is distorted by gravity within the cloud and affected by star formation activities in high-density regions; and the normalized mass-to-flux ratio tends to increase with increasing density and decreasing radius.  
Considering the thermal, turbulent, and magnetic supports, we find that the environmental gas of G28.34 is in a super-virial (supported) state, the infrared dark clumps may be in a near-equilibrium state, and core MM4-core4 is in a sub-virial (gravity-dominant) state.
In summary, we suggest that magnetic fields dominate gravity and turbulence in the cloud environment at large scales, resulting in relatively slow cloud formation and evolution processes. Within the cloud, gravity could overwhelm magnetic fields and turbulence, allowing local dynamical star formation to happen. 
%We find some correlations between the large-scale and small-scale magnetic fields, suggesting that the magnetic field in molecular clouds are linked instead of  formation and 

\end{abstract}



\section{Introduction} \label{sec:intro}

Magnetic fields and turbulence are two major forces resisting the gravitational collapse of molecular clouds in star formation regions \citep{2007ARA&A..45..565M}. Observational studies of the magnetic field are important to understand how it regulates star formation and how it is affected by star formation. Observations of polarized dust emission \citep[produced by dust grain alignment,][]{2007JQSRT.106..225L, 2007MNRAS.378..910L, 2015ARA&A..53..501A} have been the most widely used technique to trace the plane-of-sky (POS) magnetic field orientation in star-forming molecular clouds \citep[e.g.,][]{1984ApJ...284L..51H}. There has been an increasing number of both single-dish and interferometric dust polarization observations in molecular clouds\footnote{Cloud, clump, core, and condensation scales corresponds to $\sim$10 pc, $\sim$1 pc, $\sim$0.1 pc, and $\sim$0.01 pc, respectively.} \citep{2019FrASS...6...15P, 2019FrASS...6....3H}. From the observational magnetic field studies, the recent review papers summarised that magnetically trans-to-super-critical clumps/cores form in sub-critical and trans-to-sub-Alfv\'{e}nic clouds \citep{2022ApJ...925...30L, 2022FrASS...9.3556L}. The substructures of clouds may transit to an averagely trans-to-super-Alfv\'{e}nic state as density increases \citep{2021Galax...9...41L, 2022FrASS...9.3556L}, but this result is less clear due to the uncertainties of the analysis methods. Despite the progress, most of the previous observational studies of magnetic fields were in more evolved star formation regions where significant star-forming activities have taken place \citep[e.g.,][]{2021ApJ...915L..10S, 2023ApJ...945..160L}. The general magnetic field properties of clouds at early star formation stages remain under explored. 
% and averagely trans-to-super-Alfv\'{e}nic individual clumps/cores could still be sub-Alfv\'{e}nic
% The Alfv\'{e}nic state of cloud substructures is less certain, may transit to an averagely trans-to-super-Alfv\'{e}nic state \citep{2021Galax...9...41L, 2022ApJ...925...30L, 2022FrASS...9.3556L}. , 2023ASPC..534..193P

The massive star formation process is relatively less understood than low-mass star formation partly due to a lack of observations at early evolutionary stages. Massive infrared dark clouds (IRDCs) are believed to harbor the early phase of massive star formation \citep{2006A&A...450..569P, 2006ApJ...641..389R, 2012ApJ...756...60S, 2019ApJ...886..102S}, but their weak polarized dust emission makes it more challenging to detect than more evolved regions. So far, there have been only a handful of single-dish \citep{2015ApJ...799...74P, 2018A&A...620A..26J,  2018ApJ...859..151L, 2019ApJ...878...10T, 2019ApJ...883...95S, 2020A&A...644A..52A, 2023ApJ...953...66N} and interferometric \citep{2018A&A...614A..64B, 2019ApJ...884...48C, 2020ApJ...895..142L} studies of magnetic fields in IRDCs, and there is a lack of multi-scale studies of magnetic fields in the same IRDC. Since the dynamic role of the magnetic field may vary from large to small scales \citep{2022ApJ...925...30L, 2022FrASS...9.3556L}, it is essential to comprehensively investigate the multi-scale magnetic field in a single IRDC to advance our understanding of the magnetic field properties in the early stages of massive star formation. 

G28.34+0.06 (hereafter G28.34, also known as the Dragon cloud) is a well-studied massive filamentary IRDC located at a distance of 4.8 kpc \citep[e.g.,][]{1998ApJ...508..721C, 2006A&A...450..569P, 2006ApJ...641..389R, 2008ApJ...672L..33W, 2009ApJ...696..268Z, 2017ApJ...840...22L, 2018RNAAS...2...52W}. The majority of G28.34 is 8 $\mu$m-dark except for the northern end \citep{2009ApJ...696..268Z}. \citet{2006ApJ...641..389R} has identified 18 millimeter dust continuum clumps (MM1-18) within G28.34 and in its vicinity. The clumps in G28.34 are found to further fragment into smaller substructures with higher resolution observations \citep[e.g., ][]{2009ApJ...696..268Z, 2015ApJ...804..141Z, 2019ApJ...873...31K}. All the clumps in G28.34 show signs of star formation activities \citep[e.g.,][]{2006ApJ...651L.125W, 2019ApJ...874..104K}. The large mass reservoir and infrared dark behavior make G28.34 a perfect place to study the extremely early evolutionary stage of massive star formation. 

\citet{2020ApJ...895..142L} presented a study of the small-scale magnetic field in three clumps (MM1, MM4, and MM9) in G28.34 with Atacama Large Millimeter/submillimeter Array (ALMA) polarization observations. They found that even considering both the turbulent and magnetic support, core MM4-core4 is still in a non-equilibrium state dominated by gravity. As a follow-up work of \citet{2020ApJ...895..142L}, in this paper, we utilize the James Clerk Maxwell Telescope (JCMT) dust polarization observations and the Planck dust polarization data to study the multi-scale magnetic field in G28.34 and to determine the multi-scale energy balance in this IRDC. 

%The density structure of G28.34 has been extensively studied from cloud to condensation scales \citep[e.g., ][]{2006ApJ...641..389R, 2009ApJ...696..268Z, 2015ApJ...804..141Z, 2016ApJ...821L...3T, 2017ApJ...840...22L, 2018ApJ...855L..25K, 2019ApJ...873...31K, 2020ApJ...895..142L}. %, 2015ApJ...804..141Z, 2016ApJ...821L...3T, 2018ApJ...855L..25K, 2019ApJ...873...31K, 

\section{Observations} \label{sec:observation}
\subsection{ALMA dust polarization and molecular line observations}
Clumps MM4 and MM9 in G28.34 were observed by ALMA between 2017 April 18 and 2018 September 11 under projects 2016.1.00248.S (PI: Qizhou
Zhang) and 2017.1.00793.S (PI: Qizhou Zhang) in configurations C-1, C-3, and C-4. Three spectral windows were configured to observe the dust continuum at $\sim$215.5–219.5 GHz and $\sim$232.5–234.5 GHz (band 6) in the full polarization mode. Four spectral windows were configured to cover the $^{12}$CO (2-1), OCS (19-18), $^{13}$CS (5-4), and N$_2$D$^+$ (3-2) lines with a channel width of 122 kHz (0.16 km s$^{-1}$) over a bandwidth of 58.6 MHz ($\sim$76 km s$^{-1}$). The data in configurations C-1 and C-3 have been previously reported in \citet{2020ApJ...895..142L}. 

The data were calibrated with Common Astronomy Software Applications \citep[CASA, ][]{2007ASPC..376..127M}. We performed two rounds of phase-only self-calibrations
for the dust continuum. We imaged the molecular line cubes and Stokes $I$, $Q$, and $U$ maps of dust continuum using the CASA task \textit{TCLEAN} with a Briggs weighting parameter of robust = 0.5. We adopted a pixel size of $0.1 \arcsec$ for the imaging. The synthesized beam of the three configurations-combined images is $\sim0.7 \arcsec \times 0.5\arcsec$ ($\sim$0.016-0.012 pc at a distance of 4.8 kpc), which improves the resolution of our previous two configurations-combined images \citep[$\sim0.9 \arcsec \times 0.7\arcsec$, ][]{2020ApJ...895..142L}. The maximum recoverable scale\footnote{https://almascience.eso.org/observing/observing-configuration-schedule/prior-cycle-observing-and-configuration-schedule} is $\sim13 \arcsec$ ($\sim$0.3 pc at 4.8 kpc). Before primary beam correction, the 1$\sigma$ root-mean-square (RMS) noise is $\sigma_{I}\sim$0.03 and 0.03 mJy beam$^{-1}$ for the Stokes $I$ dust continuum maps and $\sigma_{QU}\sim$0.01 and 0.012 mJy beam$^{-1}$ for the Stokes $Q$ or $U$ dust continuum maps of MM4 and MM9, respectively. The debiased polarized intensity $PI$ and its corresponding uncertainty $\sigma_{PI}$ are calculated as 
$PI = \sqrt{Q^2 + U^2 - \sigma_{QU}^2} $ \citep{2006PASP..118.1340V} and $\sigma_{PI} \sim \sqrt{2}\sigma_{QU}$, where $\sigma_{QU}$ is the 1$\sigma$ rms noise on the background region of the $Q$ or $U$ maps. The polarization position angle $\theta_{\mathrm{p}}$ is estimated with $\theta_{\mathrm{p}} = 0.5 \arctan(U/Q)$. The uncertainty on the polarization position angle \citep{1993A&A...274..968N} is given by  $\delta \theta = 0.5 \sqrt{\sigma_{QU}^2/(Q^2 + U^2}) \sim 20\degr.26 (\sigma_{PI}/PI) \sim 28\degr.65 (\sigma_{QU}/PI)$.
We only adopt the N$_2$D$^+$ (3-2) line data in this study. The RMS noises of the N$_2$D$^+$ line cubes (before primary beam correction) are $\sim$1.6 and 1.9 mJy beam$^{-1}$ at a 0.16 km s$^{-1}$ channel for MM4 and MM9, respectively. All the ALMA images shown in this paper are before primary beam correction. All the continuum fluxes used for calculations of physical parameters are after primary beam correction. 

\subsection{JCMT dust polarization observations and molecular line data}

The 850 $\mu$m polarized emission of G28.34 was observed by SCUBA-2/POL-2 \citep{2013MNRAS.430.2513H, 2016SPIE.9914E..03F} on the JCMT between 2022 February 24 and 2022 June 25 under the project M22AP018 (PI: Junhao Liu). The observations were made with the POL-2 DAISY mode with low noise levels in a central region of a 3$\arcmin$ radius and with increased noises toward the edge. The spatial resolution of JCMT is $\sim14\arcsec$ ($\sim$0.33 pc) at 850 $\mu$m. The center of our DAISY field is in the southern part of G28.34 near MM4. A similar POL-2 observation centered at MM1 has been conducted by the JCMT large program B-Fields in STar-Forming Region Observations (BISTRO). A detailed analysis of the multi-scale magnetic field in the infrared bright clump MM1 will be presented by Hwang, J. et al. in prep. as part of the BISTRO survey.

The POL-2 data were reduced using the SMURF \citep{2013ascl.soft10007J} package of Starlink \citep{2014ASPC..485..391C} in a process similar to \citet{2019ApJ...877...43L}. The final maps were gridded to 7$\arcsec$ pixels (Nyquist sampling). The flux conversion factor (FCF) is 495 Jy beam$^{-1}$ pW$^{-1}$ for SCUBA-2 after 2018 June 30 \citep{2021AJ....162..191M}. Accounting for the additional 
flux losses of the POL-2 with a factor of 1.35, we adopt an FCF of 668 Jy beam$^{-1}$ pW$^{-1}$ to convert the unit of our POL-2 data from pW to Jy beam$^{-1}$. Within the central 3$\arcmin$-radius region, the mean values for the observational uncertainty of $I$, $Q$, and $U$ (i.e., $\delta I$, $\delta Q$, and $\delta U$) are $\sim$1.8, 1.9, and 1.9 mJy beam$^{-1}$, respectively. The debiased polarized intensity $PI$ and its corresponding error $\delta PI$ are calculated as $PI = \sqrt{Q^2 + U^2 - 0.5(\delta Q^2 + \delta U^2)}$ and $ \delta PI = (Q \delta Q +  U \delta U)/\sqrt{Q^2 + U^2}$, respectively. The polarization position angle $\theta_{\mathrm{p}}$ and its uncertainty $\delta \theta$ \citep{1993A&A...274..968N} are estimated with $\theta_{\mathrm{p}} = 0.5 \arctan(U/Q) $ and $\delta \theta = 0.5 \sqrt{(Q^2 \delta U^2 +  U^2 \delta Q^2)/(Q^2 + U^2)^2} \sim 28\degr.65 (\delta PI/PI)$, respectively, where $\delta PI \sim \delta Q \sim \delta U$. The polarization percentage is given by $P=PI/I$ and $\delta P=\sqrt{\delta PI^2/I^2 + \delta I^2(Q^2+U^2)/I^4}$ (see Appendix \ref{sec:PM}).

Additionally, we collect the archival $^{13}$CO (3-2) (program: M10AC06) and HCO$^{+}$ (4-3) (programs: M16BP081 and M17BP087) line data observed with the Heterodyne Array Receiver Program (HARP) and Auto-Correlation Spectrometer
and Imaging System \citep[ACSIS, ][]{2009MNRAS.399.1026B} on the JCMT. The spatial and spectral resolutions are $\sim14 \arcsec$ and 0.055 km s$^{-1}$, respectively, for both lines. To increase the signal-to-noise ratio (S/N), we smooth the two lines to a spectral resolution of 0.2 km s$^{-1}$. We estimate the main beam temperature ($T_{\mathrm{mb}}$) from the corrected antenna temperature ($T^\ast _{\mathrm{A}}$) adopting a main beam efficiency of 0.64 \citep{2009MNRAS.399.1026B}. The typical observation uncertainties of $^{13}$CO (3-2) and HCO$^{+}$ (4-3) are 0.5-1 K and 0.2-0.4 K (in $T_{\mathrm{mb}}$), respectively, per 0.2 km s$^{-1}$ channel within our studied region. 

\subsection{Planck 353 GHz dust polarization data}\label{sec:obs_planck}
We adopt the Planck High Frequency Instrument \citep[HFI, ][]{2010A&A...520A...9L} 353 GHz Stokes $I$, $Q$, and $U$ maps of the thermal dust emission \citep[version R3.00, ][]{2020A&A...641A...4P} toward G28.34 and its surrounding area constructed with the Generalized Needlet Internal Linear Combination method \citep[GNILC, ][]{2011MNRAS.418..467R}. We also adopt the earlier released dust optical depth ($\tau_{353}$) and temperature maps \citep[version R1.02, ][]{2014A&A...571A..11P}. The resolution of the Planck maps is 5$\arcmin$ ($\sim$7 pc) at 353 GHz. The pixel size of the Planck maps is 1.71$\arcmin$. Within our considered map area ($1\degr \times 1\degr$), the mean values for the uncertainties of $Q$ and $U$ (i.e., $\delta Q$ and $\delta U$) are $\sim$3 and 4 $\mu$K$_{\mathrm{CMB}}$, respectively. The debiased polarized intensity $PI$ and its corresponding uncertainty $\delta PI$ are calculated as $PI = \sqrt{Q^2 + U^2 - 0.5(\delta Q^2 + \delta U^2)}$ and $ \delta PI \sim \sqrt{(Q^2 \delta Q^2 +  U^2 \delta U^2)/(Q^2 + U^2)}$, respectively. We compute polarization position angles in the equatorial coordinates with $\theta_{\mathrm{p}} = 0.5 \arctan(U/Q) - \Delta\theta_{\mathrm{p}}^{\mathrm{g-e}}$, where
\begin{equation}
    \Delta\theta_{\mathrm{p}}^{\mathrm{g-e}} =  \arctan( \frac{ \cos(l-32.9\degr) }{ \cos b \cot 62.9\degr - \sin b  \sin(l-32.9\degr)} )
\end{equation}
is the angle between the galactic and equatorial reference directions \citep{1998MNRAS.297..617C}. For G28.34 at $l=28.34\degr$ and $b=0.06\degr$, we adopt $\Delta\theta_{\mathrm{p}}^{\mathrm{g-e}} \approx 62.82\degr$. The uncertainty on the Planck polarization position angle is given by $\delta \theta \sim 28\degr.65 (\delta PI/PI)$.

The Planck and JCMT observations are at the same frequency, so we could compare their intensities to investigate the consistency of the two datasets. To compare the Planck and JCMT data, we convolve the JCMT $I$ and $PI$ maps to the same resolution as the Planck maps. At 5$\arcmin$ resolution, the JCMT peak intensity is 98 and 3.1 mJy beam$^{-1}$ for $I$ and $PI$, respectively. At the same position, the Planck intensity is 240 and 3.7 mJy beam$^{-1}$ for $I$ and $PI$, respectively. The JCMT peak $I$ and $PI$ values are $\sim$41\% and $\sim$84\% of the Planck values, respectively. This comparison suggests that the JCMT POL-2 data filters out a significant amount of the total intensity\footnote{POL-2 data reduction filters out the atmospheric emission as well as the extended
emission for $I$, $Q$, and $U$ maps.} but recovers the majority of the polarized intensity. 

\subsection{FCRAO-14m $^{13}$CO (1-0) data}
The FCRAO-14m $^{13}$CO (1-0) data are adopted from the Galactic Ring Survey \citep[GRS;][]{2006ApJS..163..145J}. The spatial and spectral resolutions of the $^{13}$CO (1-0) data are $\sim$46$\arcsec$ ($\sim$1 pc) and 0.21 km s$^{-1}$, respectively. The pixel size is 22.14$\arcsec$. The typical sensitivity of the GRS survey is 0.13 K (in $T^\ast _{\mathrm{A}}$). The main beam temperature is estimated from $T^\ast _{\mathrm{A}}$ adopting a main beam efficiency of 0.48. The $^{13}$CO (1-0) data were previously reported in \citet{2020A&A...638A..44B}. 

%\subsection{CfA-1.2m $^{12}$CO (1-0) data}
%The $^{12}$CO (1-0) data is adopted from the CfA-1.2m CO survey \citep{2001ApJ...547..792D}. The typical spatial and spectral resolutions of the $^{12}$CO (1-0) data are $\sim$8.8$\arcmin$ ($\sim$12.2 pc) and 1.3 km s$^{-1}$, respectively. 

\section{Results and analyses} \label{sec:results}

\subsection{Dust polarization and magnetic fields}

Here we briefly overview the multi-scale magnetic field structures in G28.34 revealed by Planck, JCMT, and ALMA. Assuming that the observed linear polarization of the dust continuum is due to dust grain alignment, we rotate the dust polarization position angle by 90$\degr$ to infer the magnetic field orientation. 

\begin{figure}[!htbp]
 \gridline{\fig{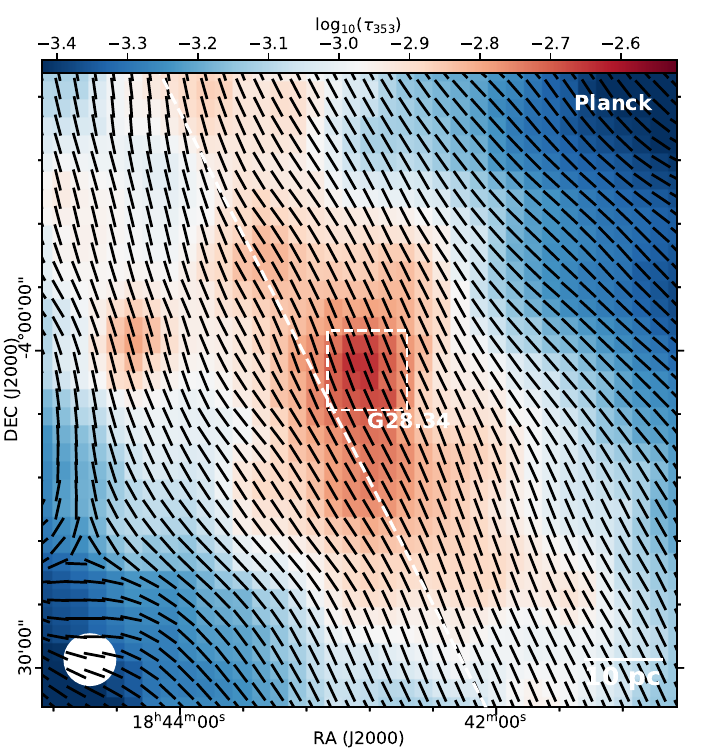}{0.48\textwidth}{(a)}
 \fig{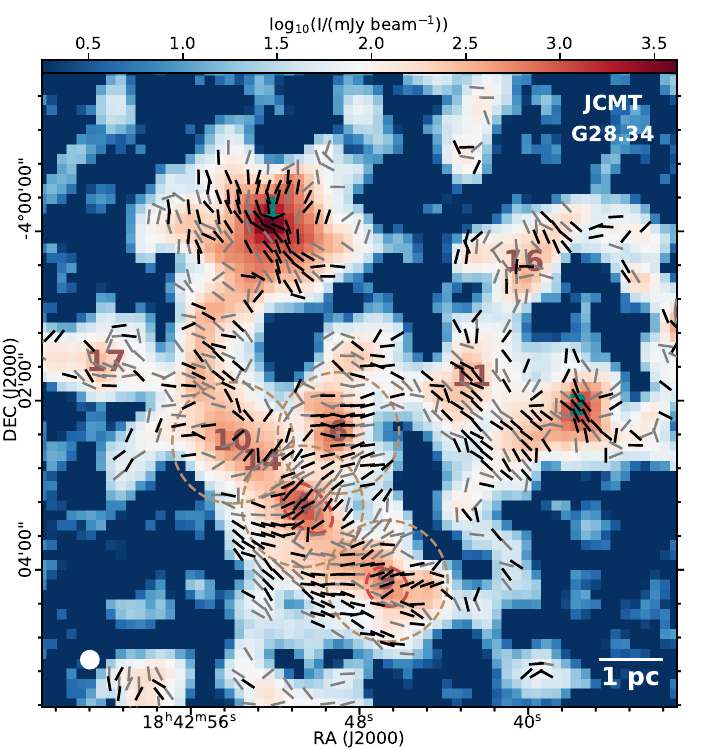}{0.48\textwidth}{(b)}
 }
 
\caption{(a). Magnetic field orientations revealed by Planck (black line segments) 0.85 mm dust polarization data overlaid on the Planck dust optical depth map (color scales) toward the IRDC G28.34 and its surrounding materials. Line segments are of arbitrary length. The 5$\arcmin$  ($\sim$7 pc) beam (white circle) and a scale bar of 10 pc are indicated in the lower left and right corners, respectively. A white dashed line indicates the galactic plane ($b=0\degr$). The dashed rectangle indicates the JCMT map area in (b). (b). Magnetic field orientations revealed by JCMT POL-2 0.85 mm dust polarization observations overlaid on the JCMT 0.85 mm total intensity map (color scales. Only area with S/N($I$)$>$25 is shown) toward the IRDC G28.34. Black and grey line segments indicate S/N($PI$)$>$3 and 2$<$S/N($PI$)$<$3, respectively. Line segments are of arbitrary length. The 14$\arcsec$ ($\sim$0.33 pc) beam (black circle) and a scale bar of 1 pc are indicated in the lower left and right corner, respectively. The infrared-bright and infrared-dark molecular clumps identified by \citet{2006ApJ...641..389R} are labeled with green and brown numbers, respectively. Red dashed contours indicate the ALMA fields of MM4 and MM9 corresponding to primary-beam responses of 0.5. Orange dashed circles mark the regions for MM4, MM6, MM9, and MM10 within which we use the polarization position angles to calculate the magnetic field strength in Section \ref{sec:B_clump}. \label{fig:G28_B_large}}
\end{figure}

Figure \ref{fig:G28_B_large}(a) shows the large-scale magnetic field orientation surrounding G28.34 revealed by Planck. The well-ordered large-scale magnetic field shows a predominant northeast-southwest orientation along the galactic plane.

%Full Width at Half Maximum (FWHM) field of view of
%The cyan contour indicates the region with S/N($I$)=25 for the JCMT data. 

Figure \ref{fig:G28_B_large}(b) shows the magnetic field orientation in G28.34 and several nearby massive clumps revealed by JCMT. The magnetic field morphology in G28.34 is complex. Along the ridge of the main straight dark filament (containing MM10, MM14, MM4, and MM9), the magnetic field seems to be perpendicular to the filament spine, which is in agreement with previous observational studies of magnetic fields in filamentary IRDCs \citep{2018ApJ...859..151L, 2019ApJ...883...95S, 2019ApJ...878...10T, 2020A&A...644A..52A} and might be a result of gravitational accretion flows \citep{2018MNRAS.480.2939G, 2018MNRAS.473.4220L}. In the northwestern clumps (MM1, MM2, MM11, and MM16) and in the diffuse region to the southeast of MM4 and MM9, the magnetic field tends to be parallel to the main straight dark filament, which may suggest that the magnetic field has kept its initial configuration inheriting from the large-scale magnetic field in these regions. 
% and is also consistent with simulations of IRDCs \citep[e.g.,][]{}. The observed perpendicular alignment between magnetic fields and the main filament.  , or alternatively that the magnetic field in these regions has been compressed by global converging gas flows along the northwest-southeast direction \citep{2020A&A...638A..44B}

\begin{figure}[!htbp]
 \gridline{\fig{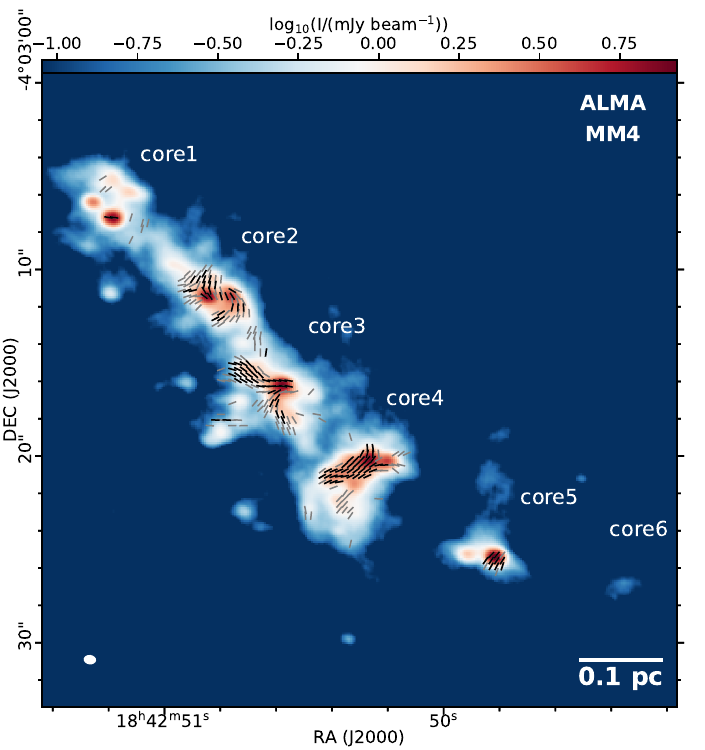}{0.48\textwidth}{}
 \fig{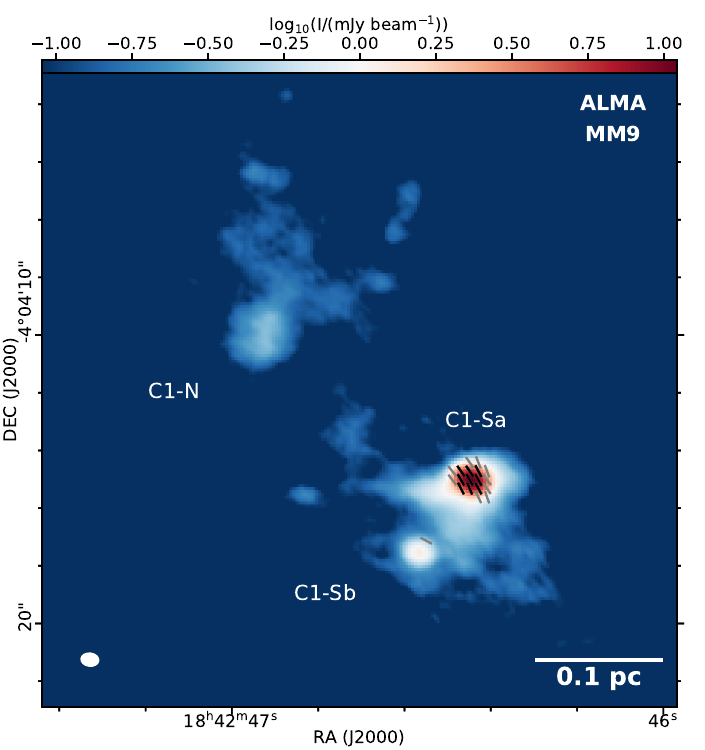}{0.48\textwidth}{}
 }
\caption{Magnetic field orientations revealed by ALMA 1.3 mm dust polarization observations overlaid on the ALMA 1.3 mm total intensity map (color scales) toward MM4 and MM9. Black and grey line segments indicate S/N($PI$)$>$3 and 2$<$S/N($PI$)$<$3, respectively. Line segments are of arbitrary length. The $\sim0.7 \arcsec \times 0.5\arcsec$ ($\sim$0.016-0.012 pc) synthesized beam (black ellipse) and a scale bar of 0.1 pc are indicated in the lower left and right corners, respectively. \label{fig:G28_B_ALMA}}
\end{figure}

Figure \ref{fig:G28_B_ALMA} shows the magnetic field orientation in clumps MM4 and MM9 revealed by ALMA. Overall, our three configurations-combined (C-1, C-3, and C-4) ALMA images exhibit magnetic field morphologies similar to the two configurations-combined (C-1 and C-3) ALMA images reported by \citet{2020ApJ...895..142L}. 
%There is only marginal dust polarization detection toward MM9. Thus, we do not show the magnetic field orientation map of MM9 here.

%Purple contours indicate the Full Width at Half-Maximum (FWHM) field of view of our ALMA observations. 
% The dashed circle indicates the central 3$\arcmin$ radius region of our POL-2 field. 

\subsubsection{Comparing multi-scale magnetic fields}

\begin{figure}[!tbp]
\gridline{\fig{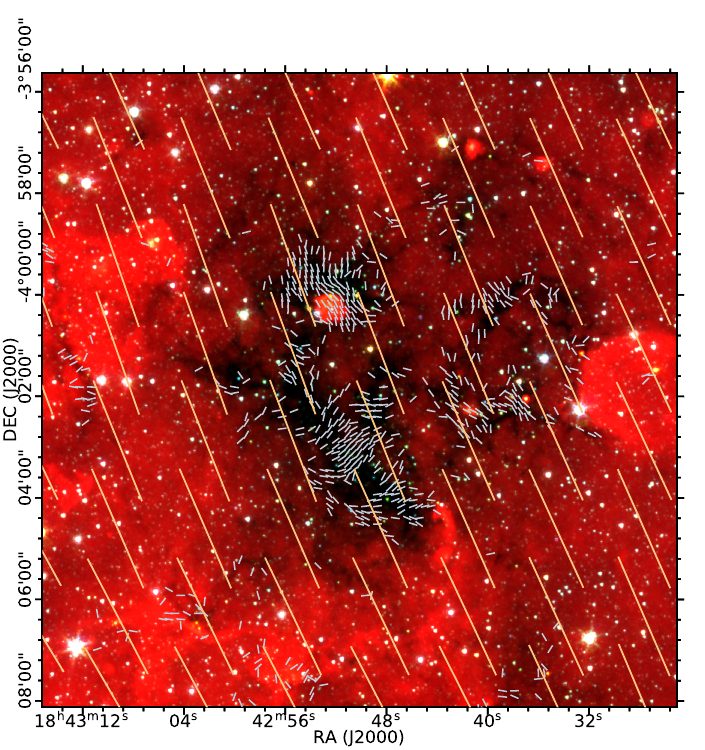}{0.45\textwidth}{}}
\caption{Magnetic field orientations revealed by Planck (orange lines) and JCMT (white lines) overlaid on the Spitzer three-color composite image (red/green/blue = 8.0/4.5/3.6 $\mu$m). The Spitzer data is from the GLIMPSE project \citep{2009PASP..121..213C}. \label{fig:rgb}}
\end{figure}

\begin{figure}[!htbp]
 \gridline{\fig{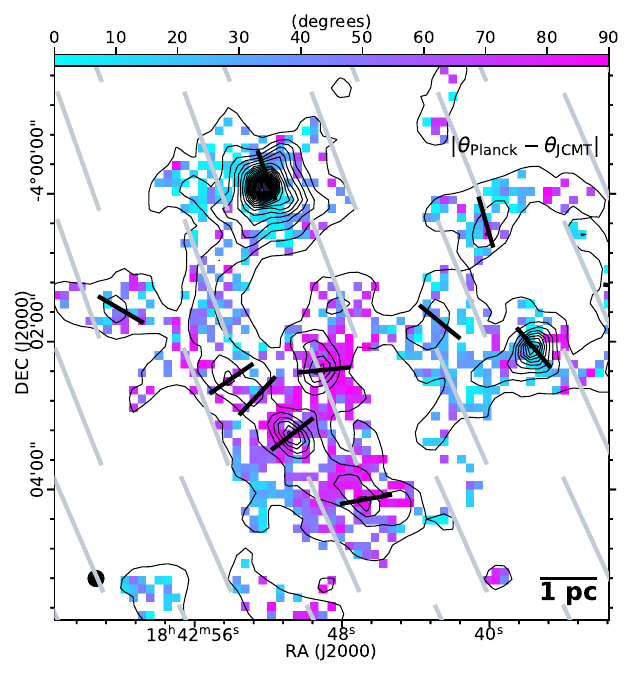}{0.48\textwidth}{(a)}
 \fig{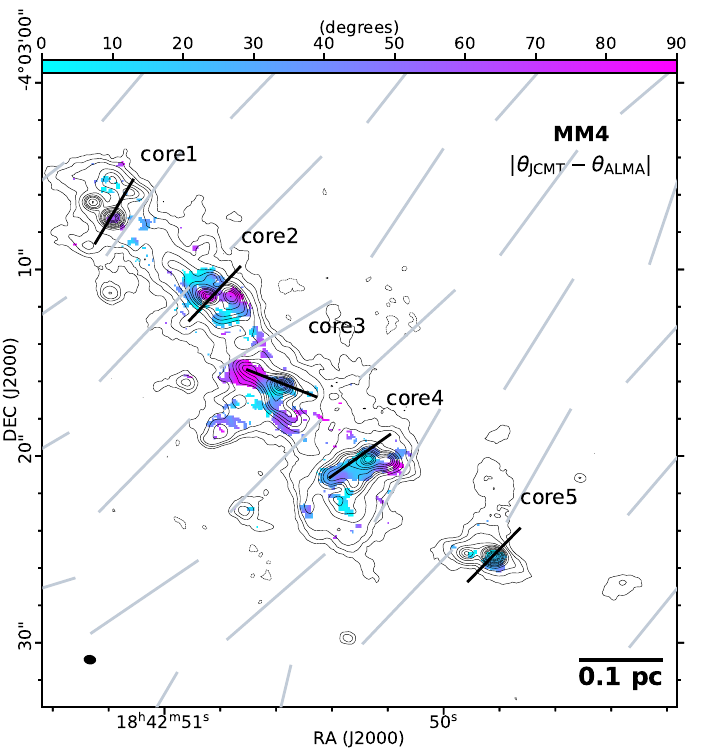}{0.48\textwidth}{(b)}
 }
\caption{(a). Map of absolute angular differences (color scale) between magnetic field orientations revealed by Planck and JCMT in G28.34. The magnetic field orientation revealed by Planck is shown as grey line segments. The average magnetic field orientation within each 1-pc clump is indicated as black line segments. Contour starts at 50 mJy beam$^{-1}$ and continues at 150 mJy beam$^{-1}$. (b). Map of absolute angular differences (color scale) between magnetic field orientations revealed by JCMT and ALMA in MM4. The magnetic field orientation revealed by JCMT is shown as grey line segments. The average magnetic field orientation within each 0.1-pc core is indicated as black line segments. Contour levels are ($\pm$3, 6, 10, 20, 30, 40, 50, 70, 90, 110, 150, 180, 210, 250, 290, 340, 390, 450) $\times \sigma_{I}$. \label{fig:G28_mulB}}
\end{figure}

\begin{figure}[!htbp]
 \gridline{\fig{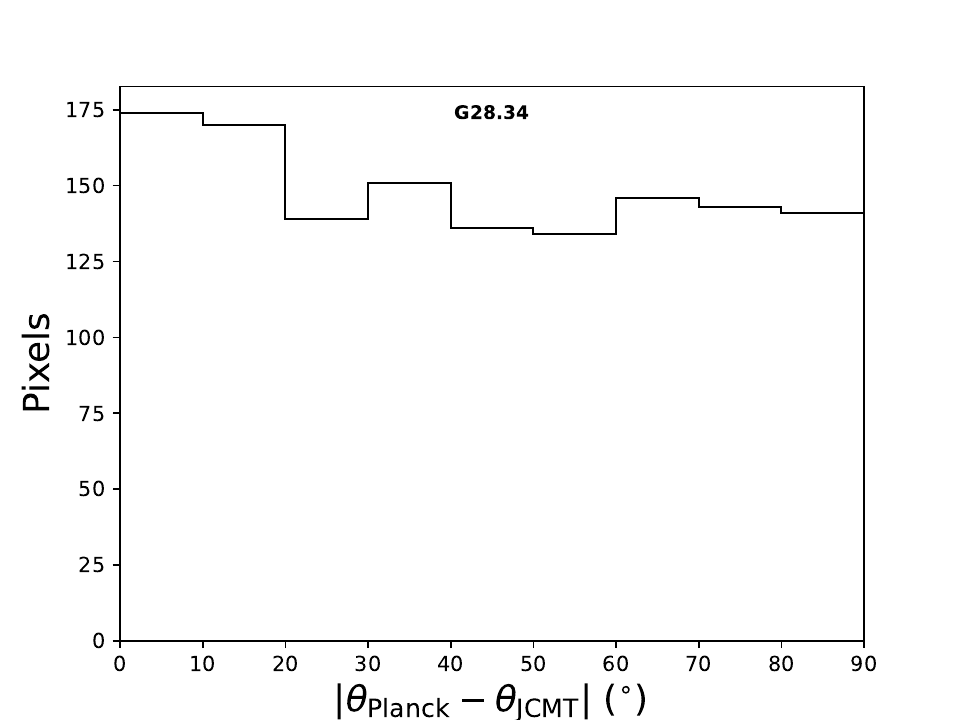}{0.45\textwidth}{(a)}
 \fig{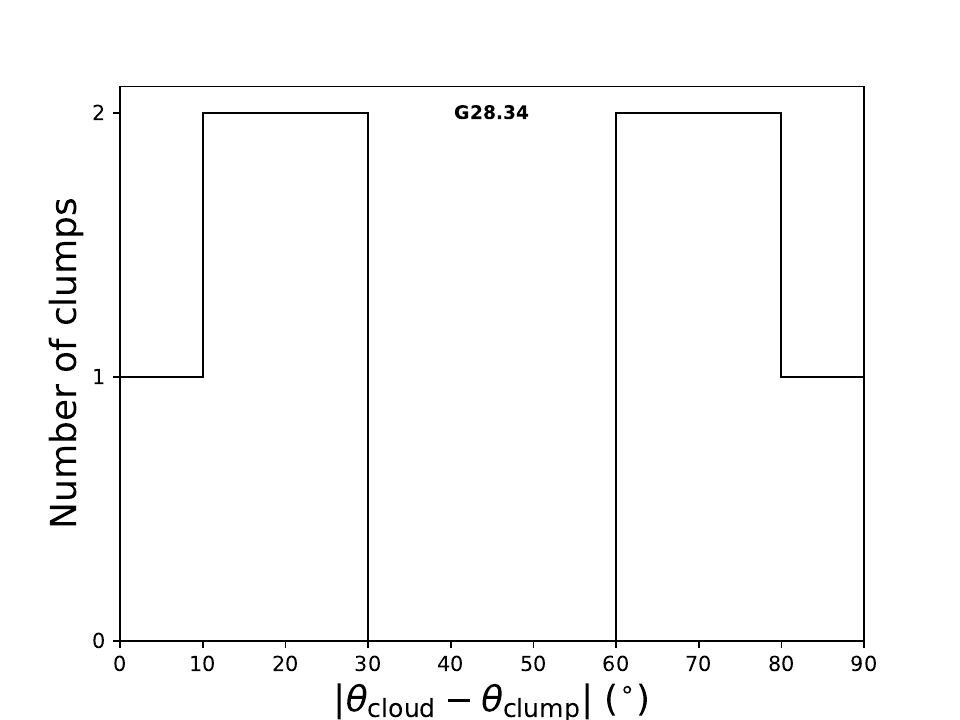}{0.45\textwidth}{(b)}
 }
\gridline{
 \fig{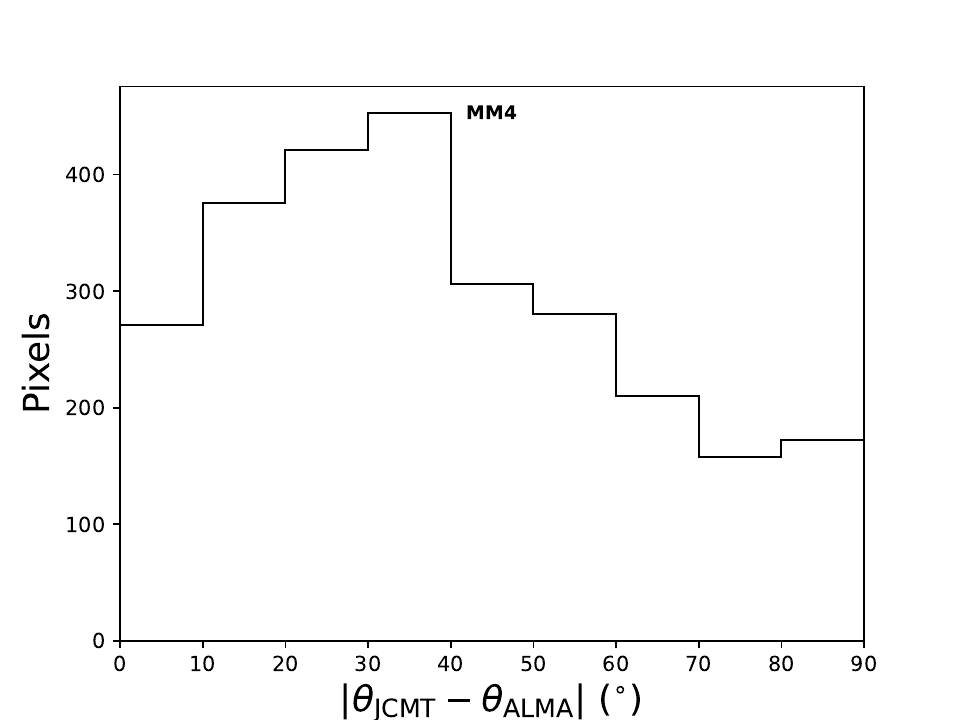}{0.45\textwidth}{(c)}
  \fig{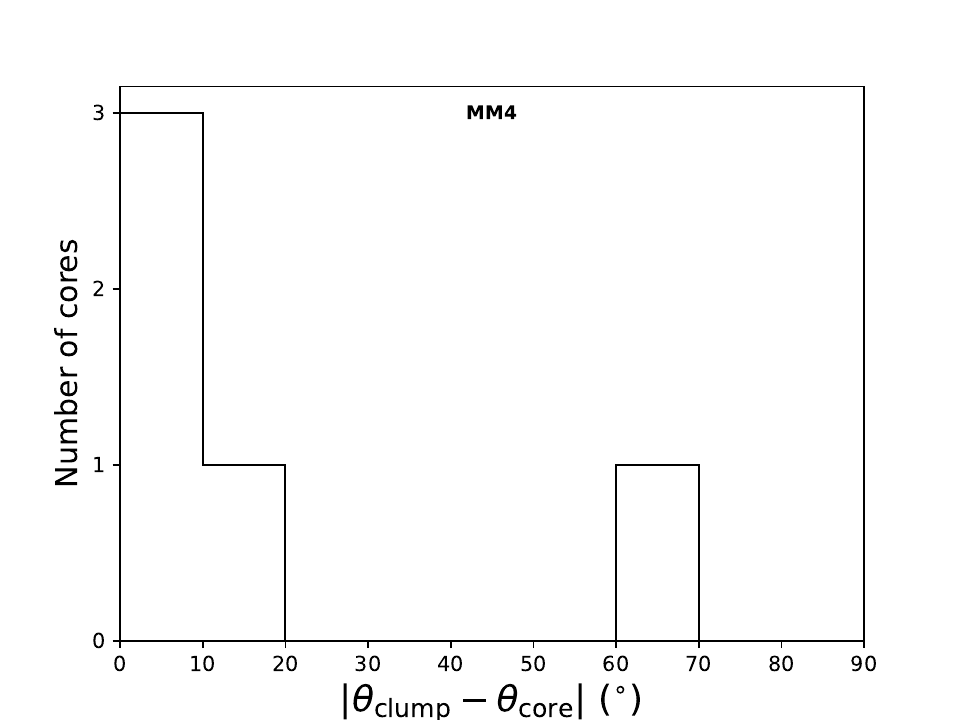}{0.45\textwidth}{(d)}
 }
\caption{(a). Histogram of angular differences between magnetic field orientations revealed by Planck and JCMT in G28.34. (b). Histogram of angular differences between averaged cloud- and clump-scale magnetic fields in G28.34. (c). Histogram of angular differences between magnetic field orientations revealed by JCMT and ALMA in MM4. (d). Histogram of angular differences between averaged clump- and core-scale magnetic fields in MM4. \label{fig:G28_mulBhist}}
\end{figure}

One question to be addressed is how the small-scale magnetic field is correlated with the large-scale magnetic field \citep[e.g.,][]{2009ApJ...704..891L,2014ApJ...792..116Z,2015Natur.520..518L}. To study this, we first compare the JCMT magnetic field orientation ($\theta_{\mathrm{JCMT}}$) in every pixel of the JCMT detection area (S/N($PI$)$>$2) with the Planck magnetic field orientation ($\theta_{\mathrm{Planck}}$) at the nearest pixel of the Planck map and calculate their angular difference with approaches similar to \citet{2014ApJ...792..116Z}. Figure \ref{fig:rgb} overlays the magnetic field orientations revealed by Planck and JCMT. Figure \ref{fig:G28_mulB}(a) shows the spatial distribution of the absolute angular difference between the Planck and JCMT magnetic field orientations. Figure \ref{fig:G28_mulBhist}(a) shows the histogram of the absolute angular difference between the Planck and JCMT magnetic field orientations. Although overall there is no strong relation between the Planck and JCMT magnetic field orientations (Figure \ref{fig:G28_mulBhist}(a)), the spatial distribution for their angular difference is not random (as mentioned above and see Figure \ref{fig:G28_mulB}(a)). The difference between global and local statistics signifies the importance of investigating the local distributions of the angular difference. Thus, we further investigate the relation between the cloud- and clump-scale averaged magnetic fields \citep[e.g.,][]{2009ApJ...704..891L}. To do this, we average the Planck polarization data (i.e., $Q$ and $U$) toward the cloud within a circle of 10 pc and the JCMT polarization detection toward each clump within a circle of 1 pc, and recalculate the polarization position angles. We find that half of the clumps (MM1, MM2, MM11, MM16, and MM17) within our studied region have averaged magnetic fields aligned within 30$\degr$ of the cloud-scale magnetic field, while the other half of the clumps (MM4, MM6, MM9, MM10, and MM14) have averaged magnetic fields misaligned at 60$\degr$-90$\degr$ with respect to the cloud-scale magnetic field (see Figure \ref{fig:G28_mulBhist}(b)). The bimodal distribution implies that the clump-scale magnetic field is organized with respect to the uniform cloud-scale Planck field. Otherwise, a randomly-orientated small-scale magnetic field cannot produce the observed bimodal distribution \citep{2014ApJ...792..116Z}. Future larger-sample observational studies are required to understand whether the bimodal distribution for angular differences between cloud- and clump-scale averaged magnetic fields is ubiquitous, and future theoretical and numerical studies will be needed to understand the physical mechanism behind this bimodal distribution.

Similarly, we investigate the angular difference between the polarization position angles revealed by JCMT and ALMA. Figures \ref{fig:G28_mulB}(b) and \ref{fig:G28_mulBhist}(b) show the spatial distribution and histogram, respectively, for the angular difference between the JCMT and ALMA magnetic field orientations toward MM4. The local spatial distribution of the angular difference (Figure \ref{fig:G28_mulB}(b)) is complex and we refrain from describing them in detail. The histogram of angular difference (Figure \ref{fig:G28_mulBhist}(c)) shows that the magnetic field on scales of cores and condensations in MM4, as revealed by ALMA, is preferentially aligned with the clump-scale magnetic field revealed by JCMT. We further compare the averaged clump-scale and core-scale magnetic fields for core1$-$5 in MM4. To calculate the averaged magnetic fields of cores, we adopt the core positions in \citet{2009ApJ...696..268Z} and average the polarization detections of each core within a radius of 0.1 pc in a way similar to the derivation of the averaged clump-scale magnetic fields. We find that most cores (i.e., core1, 2, 4, and 5) have averaged magnetic fields aligned within 20$\degr$ of the clump-scale magnetic field of MM4 (see Figure \ref{fig:G28_mulBhist}(d)). The preservation of magnetic field orientation could suggest that the magnetic field plays a crucial role in the collapse of this clump and in the formation of dense cores within \citep{2015MNRAS.452.2500L}. The clump-scale magnetic field is perpendicular to the chain of cores in MM4, which could suggest that the fragmentation in this clump is regulated by the magnetic field in the parental clump \citep{2008ApJ...687..354N}. We refrain from comparing the averaged condensation-scale magnetic fields with large-scale magnetic fields in MM4 because the fragmentation in this clump at condensation level is complex \citep{2015ApJ...804..141Z, 2019ApJ...873...31K}. On the other hand, the marginal polarization detection in MM9 does not support any statistics. With a rough comparison, we find that the angle between the clump-scale magnetic field of MM9 and the condensation-scale magnetic field in C1-Sa is $\sim$65$\degr$. 

%outflow and B?

\subsection{Density and mass} \label{sec:N}
Here we briefly describe the estimation of the gas density and mass from dust continuum and molecular line data. Then we investigate the mass-radius relation and density-radius relation.

\subsubsection{Core} \label{sec:N_core}
For the dense cores revealed by ALMA observations, we estimate the dust mass with 
\begin{equation}
M_{\mathrm{dust}} = \frac{F_{\mathrm{\nu}} d^2}{ \kappa_{\nu} B_{\nu} (T)}, \label{eq:Mdust}
\end{equation}
where $F_{\mathrm{\nu}}$ is the flux density at frequency $\nu$, $d$ is the distance, $\kappa_{\nu} = (\nu / 1 \mathrm{THz})^{\beta}$ is the dust opacity \citep{1983QJRAS..24..267H} in m$^2$ kg$^{-1}$, and $B_{\nu} (T)$ is the Planck function at temperature $T$. We adopt a dust emissivity index ($\beta$) of $\sim$1.5 \citep[e.g.,][]{2007A&A...466.1065B, 2007ApJ...654L..87C} and a gas temperature of 15 K \citep{2018RNAAS...2...52W}. Adopting a gas-to-dust ratio of $\Lambda = 100$ \citep{1972ApJ...172..491S}, the gas mass can be estimated with $M_{\mathrm{gas}} = \Lambda M_{\mathrm{dust}}$.  Then the gas column density is estimated with   
\begin{equation}
N_{\mathrm{H_2}} = \frac{M_{\mathrm{gas}}}{\mu_{\mathrm{H_2}} m_{\mathrm{H}} A},
\label{eq:N}
\end{equation}
where $\mu_{\mathrm{H_2}} = 2.8$ is the mean molecular weight per hydrogen molecule \citep{2008A&A...487..993K}, $m_{\mathrm{H}}$ is the atomic mass of hydrogen, and $A$ is the area. 

\subsubsection{Clump and cloud}  \label{sec:N_clump}
We adopt the column density map (hereafter the COMB map) at intermediate scales from \citet{2017ApJ...840...22L}. The COMB map was made by performing an iterative SED fitting procedure for multi-wavelength (70 $\mu$m to 850 $\mu$m) continuum data from Herschel, Caltech Submillimeter Observatory (CSO), JCMT, and Planck, using image combination techniques to recover extended emission for ground-based telescopes while preserving the high angular resolution. Details about basic data combinations and SED fitting procedures can be found in \citet{Lin16}. In particular, we used an updated combination procedure similar to \citet{Jiao22}, where the Planck map is first deconvolved based on an extrapolated model image. The updated combined image benefits from a better dynamical range to recover more extended emissions. The resolution of the resulting column density maps is 10$\arcsec$. The size of the COMB map is $\sim$600$\arcsec$ ($\sim$7 pc). With the column density map, the gas mass is derived using Equation \ref{eq:N}. 
%but with a slightly different approach, we derive the column density for the main filamentary cloud and the dense clumps with combined datasets (hereafter the COMB data) of ... (to be filled in by Yuxin) {The dust opacity at 1.3 mm is adopted following \citet{OH94} for thin gas mantles at gas density of 10$^{3}\,$cm$^{-3}$ (OH5 model, $\kappa_{\mathrm{1.3\,mm}}$ = 0.9 cm$^{2}$g$^{-1}$). }

\subsubsection{Environmental gas}\label{sec:N_envir}
For the environmental gas surrounding IRDC G28.34, we calculate the gas column density from the Planck $\tau_{353}$ map and the FCRAO-14m $^{13}$CO (1-0) data.

For the Planck data, we convert $\tau_{353}$ to the column density of hydrogen atoms ($N_{\mathrm{H}}$) with the relation \citep{2014A&A...571A..11P, 2016AA...586A.138P}
\begin{equation}
\tau_{353}/N_{\mathrm{H}} = 1.2 \times 10^{-26} \mathrm{cm}^{2}.
\end{equation}
Note that $N_{\mathrm{H}}$ accounts for the column density of both the atomic gas and molecular gas \citep{2014A&A...571A..11P, 2016AA...586A.138P} along the line of sight (LOS). The gas mass is given by 
\begin{equation}
M_{\mathrm{gas}} = \mu_{\mathrm{H}} m_{\mathrm{H}} A N_{\mathrm{H}},
\label{eq:M}
\end{equation}
where we assume the mean atomic weight per hydrogen atom is $\mu_{\mathrm{H}} \sim 1.4$.
%$N_{\mathrm{H}} = N_{\mathrm{HI}} + 2X_{\mathrm{CO}}W_{\mathrm{CO}}$ in \citet{2014A&A...571A..11P, 2016AA...586A.138P} accounts for the column density of both the atomic gas and molecular gas, where $N_{\mathrm{HI}}$ is the column density of HI, $X_{\mathrm{CO}}$ is the conventional X factor, and $W_{\mathrm{CO}}$ is the integrated intensity of $^{12}$CO (1-0). 

For the $^{13}$CO (1-0) data, we estimate the integrated line intensity with $W_{\mathrm{13CO}} = \Sigma_i^{N_{\mathrm{ch}}} I_{i} \Delta v_{\mathrm{ch}}$, where $I_i$, $\Delta v_{\mathrm{ch}}$, and $N_{\mathrm{ch}}$ are the 
line intensity, channel width, and the number of integrated channels, respectively. The $^{13}$CO (1-0) line is integrated between 71 and 86 km s$^{-1}$ to cover the main velocity component of the cloud \citep{2020A&A...638A..44B}. Because the $^{13}$CO (1-0) line peak brightness temperature is much smaller than the gas temperature, this line is very likely optically thin \citep{2020A&A...638A..44B}. In the optically thin case, the upper state column density is given by \citep{1999ApJ...517..209G}
\begin{equation}
N_u=\frac{8\pi k_B \nu^2 W_{\mathrm{13CO}}}{hc^3 A_{ul}},
\end{equation}
where $k_B$, $h$, and $A_{ul}$ are the Boltzmann constant, Planck constant, and Einstein coefficient, respectively. Assuming local thermal equilibrium, the total column density of $^{13}$CO is given by
\begin{equation}
N_{^{13}CO} = \frac{N_{\mathrm{u}}Z}{g_\mathrm{u} e^{-E_\mathrm{u}/k_BT}},
\end{equation}
where $g_\mathrm{u}$, $E_\mathrm{u}$, and $Z$ are the statistical weight of the upper state, the upper energy level, and the partition function, respectively. Here we adopt $T=15$ K as well. The values of $A_{ul}$, $g_\mathrm{u}$, $E_\mathrm{u}$, and $Z$ are adopted from the CDMS \citep{2001AA...370L..49M} and LAMDA \citep{2005AA...432..369S} databases. Thus, the column density of H$_2$ can be estimated using the standard abundance \citep{2013tra..book.....W}
\begin{equation}
N_{H_2} = N_{^{13}CO} \times 4.6 \times 10^5. 
\end{equation}
The standard abundance is only valid when $N_{H_2}<5 \times 10^{21}$ cm$^{-2}$ and could present a scatter of factor two to five \citep{2013tra..book.....W}. Because CO isotopes could be depleted in dark clouds \citep{2007ARA&A..45..339B}, $N_{H_2}$ toward IRDC G28.34 estimated with $^{13}$CO may only be a lower limit. With the estimated gas column density, the gas mass is derived with Equation \ref{eq:N}. 

We find that the ratio between $0.5N_{\mathrm{H}}$ estimated with the Planck $\tau_{353}$ map and $N_{H_2}$ estimated with the $^{13}$CO data is $\sim$10 toward the center of G28.34. Their ratio gradually increases to $\sim$10$^3$ toward the edge of the Planck map. As the Planck observations trace all the atomic and molecular gases along the LOS, it is reasonable that the column density traced by Planck should be higher than the molecular gas integrated within a specific velocity range. On the other hand, the systematic lower gas column densities of $^{13}$CO might be due to the $^{13}$CO depletion, the FCRAO-14m observation filtering out the large-scale emission at the extent of the off-position, and the uncertainties in the adopted $^{13}$CO abundance.
%discrepancy between column densities estimated with dust emission and line data has been previously reported by \citep{2015A&A...578A..29S}

\subsubsection{Scaling relations}\label{sec:N_scaling}

The mass-radius ($M-r$) and density-radius ($n-r$) relations are important properties of star formation regions \citep{1981MNRAS.194..809L}. With the estimated column density and gas mass, we investigate the $M-r$ and $n-r$ relations for the cores and clumps in IRDC G28.34 and in its surrounding gas. 
%$N-r$ and $N-r$, 

With the gas mass derived from the Planck $\tau_{353}$ map and from the FCRAO-14m $^{13}$CO (1-0) data, we obtain the mass profile for the environmental gas within circles of different radii (from beam size to 20 pc) centered at G28.34. With the COMB mass map, we obtain the mass profile for each infrared dark clump within circles of different radii (from beam size to 2 pc) centered at the local continuum peak. For simplicity, we only consider clumps with both molecular line detection and sufficient dust polarization detection (i.e., MM4, MM6, MM9, MM10, and MM14) in our analysis. MM10 and MM14 are unseparated at the resolution of the COMB map and are thus considered as one clump in our analysis. We also obtain a mass estimation for the whole COMB map area ($r\sim$7 pc). At the resolution of ALMA, the fragmentation status is complicated in MM4 and MM9. Thus, we only report the total mass and effective radius ($\sqrt{A/\pi}$) for the ALMA area with S/N($I$)$>$5. Similarly, we obtain the profile for the average column density within circles of different radii. Assuming the studied structures are spherical, the average number density is estimated from the average column density with $n = 0.75 N / r$. 

\begin{figure}[!htbp]
 \gridline{\fig{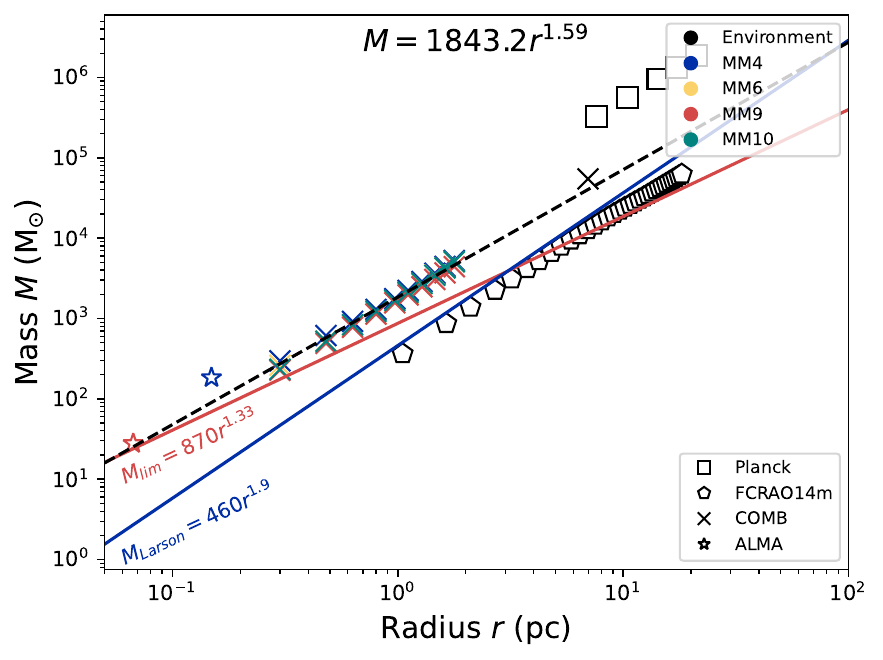}{0.45\textwidth}{(a)}
 }
  \gridline{\fig{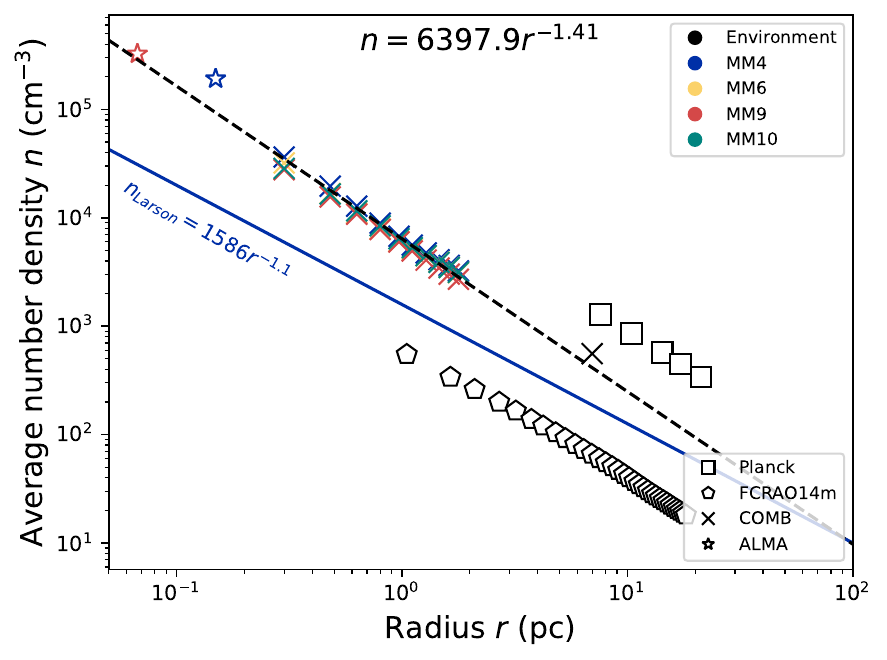}{0.45\textwidth}{(b)}
 }
\caption{(a). Mass-radius relation. The red solid line indicates the empirical threshold for massive star formation \citep[$M_{lim}=870M_{\odot}(r/\mathrm{pc})^{1.33}$,][]{2010ApJ...716..433K}. The blue solid line indicates the Larson's third law \citep[$M_{Larson}=460 M_{\odot}(r/pc)^{1.9}$,][]{1981MNRAS.194..809L}. (b). Average number density-radius relation. The blue solid line indicates the Larson's third law \citep[$n_{Larson}=1586$ cm$^{-3} (r/\mathrm{pc})^{-1.1}$,][]{1981MNRAS.194..809L}. Different colors represent different regions. Different symbols indicate different datasets. The black dashed lines present the results of a simple power-law fit for the ALMA data and the COMB data. \label{fig:G28_mass_n_r}}
\end{figure}

Figure \ref{fig:G28_mass_n_r}(a) shows the obtained mass-radius relation. An obvious trend is that the mass and radius for the ALMA data and the COMB data are positively correlated. With a simple minimum chi-squared power-law fit, we obtain $M\propto r^{1.59\pm0.14}$. Here we assume an uncertainty of a factor of 2 for the mass during the fitting. The ALMA data for MM4 is higher than the fitted power law, which might be due to the deviation from the assumed spherical structure in this clump. The fitted power law is higher than the empirical threshold for massive star formation \citep[$M_{lim}=870M_{\odot}(r/pc)^{1.33}$,][]{2010ApJ...716..433K}, indicating that IRDC G28.34 is potentially forming massive stars. The obtained power-law index of 1.59 is smaller than the value of 1.9 reported by \citet{1981MNRAS.194..809L}. Note that our $M-r$ relation presents the relation in an individual cloud, while Larson's third law presents the relation for an ensemble of cloud samples, so the two relations may not be comparable. The mass derived from the Planck map is higher than the fitted power law, while the mass derived from the FCRAO-14m $^{13}$CO data is lower than the fitted power law. As discussed in Section \ref{sec:N_envir}, this deviation might be because the molecular mass is overestimated by the Planck map and underestimated by the FCRAO-14m $^{13}$CO data.

Figure \ref{fig:G28_mass_n_r}(b) shows the obtained average number density-radius relation. With a power-law fit for the ALMA data and the COMB data, we obtain $n\propto r^{-1.41\pm0.14}$. Here we assume an uncertainty of a factor of 2 for the density during the fitting. The obtained power-law index of -1.41 is steeper than the value of -1.1 reported by \citet{1981MNRAS.194..809L}. Similarly, our $n-r$ relation in an individual cloud may not be comparable to Larson's third law which presents the relation for an ensemble of cloud samples. 

\subsection{Molecular line and velocity fields}\label{sec:line}

Here we briefly overview the multi-scale velocity structures revealed by the FCRAO-14m $^{13}$CO (1-0), JCMT $^{13}$CO (3-2) and HCO$^{+}$ (4-3), and ALMA N$_2$D$^{+}$ (3-2) line data. The information of each line is summarised in Table \ref{tab:line}. The value of line parameters are adopted from the CDMS \citep{2001AA...370L..49M} and LAMDA \citep{2005AA...432..369S} databases. No information on the collision rate coefficient exists for N$_2$D$^{+}$ in LAMDA, so we adopt the values for N$_2$H$^{+}$ instead. A particular line tracer is only sensitive to gas with densities above its critical density ($n_{\mathrm{c}}$) but no more than a factor of 2 orders of magnitude \citep{1998ApJ...504..223G}. We calculate the intensity-weighted velocity centroid $V_c(\boldsymbol{x})$ at position $\boldsymbol{x}$ with 
\begin{equation}
V_c(\boldsymbol{x}) = \frac{\Sigma_i^{N_{\mathrm{ch}}} I_i(\boldsymbol{x}) v_i \Delta v_{\mathrm{ch}}}{\Sigma_i^{N_{\mathrm{ch}}} I_i(\boldsymbol{x}) \Delta v_{\mathrm{ch}}},
\end{equation}
where $I_i(\boldsymbol{x})$, $v_i$, $\Delta v_{\mathrm{ch}}$, and $N_{\mathrm{ch}}$ are the line intensity, line-of-sight velocity, velocity channel width, and number of channels, respectively. The propagated uncertainty of $V_c(\boldsymbol{x})$ is 
\begin{equation}
\delta V_c(\boldsymbol{x}) = \frac{\sigma_{ch} \Delta v_{\mathrm{ch}}\sqrt{\Sigma_i^{N_{\mathrm{ch}}} (v_i - V_c(\boldsymbol{x}))^2}}{\Sigma_i^{N_{\mathrm{ch}}} I_i(\boldsymbol{x}) \Delta v_{\mathrm{ch}}},
\end{equation}
where $\sigma_{ch}$ is the noise of one spectral channel (see Section \ref{sec:observation}). For FCRAO-14m $^{13}$CO (1-0), we only consider line emissions from 71 to 86 km s$^{-1}$ which covers the main velocity component of G28.34 at large scales \citep{2020A&A...638A..44B}. There are no apparent LOS distant gas structures at this velocity superposed at the same POS position for this IRDC \citep{2006ApJ...653.1325S, 2020A&A...638A..44B}. For JCMT observations toward the cloud, we only consider velocities within $\sim$5 km s$^{-1}$ with respect to the local-standard-of-rest (LSR) velocity ($V_{\mathrm{lsr}}\sim 79$ km s$^{-1}$) of G28.34 to avoid potential contamination from high-velocity outflows. For ALMA observations toward MM4 and MM9, we only consider velocities within $\sim$3 km s$^{-1}$ with respect to the $V_{\mathrm{lsr}}$ of each clump \citep[$\sim$79 and 80 km s$^{-1}$ for MM4 and MM9, respectively,][]{2015ApJ...804..141Z} because there is nearly no line emission at $>$3 km s$^{-1}$. All lines are likely to be optically thin as the main-beam temperature of the line peaks is smaller than the gas/dust temperature of $\sim$15 K (see Appendix \ref{sec:appline}).

\begin{deluxetable}{cccccccc}[t!]
\tablecaption{Summary of molecular line data \label{tab:line}}
\tablecolumns{8}
\tablewidth{0pt}
\tablehead{
\colhead{Line} &
\colhead{Frequency} &
\colhead{$E_{\mathrm{u}}/k$ \tablenotemark{a}} & 
\colhead{$n_{\mathrm{c}}$ \tablenotemark{b}} &
\colhead{Instrument} & 
\colhead{$l_{\mathrm{beam}}$  \tablenotemark{c}} & 
\colhead{$l_{\mathrm{MRS}}$  \tablenotemark{d}} & 
\colhead{Targets} \\
\colhead{} &  \colhead{(GHz)} & \colhead{(K)} &  \colhead{(cm$^{-3}$)}  & \colhead{} & \colhead{($\arcsec$)}  & \colhead{($\arcsec$)} & \colhead{}
}
\startdata
N$_2$D$^{+}$ (3-2) & 231.3218 & 22.2 & 1.0 $\times$ 10$^6$ & ALMA & 0.7 & 13 & MM4,MM9\\
HCO$^+$ (4-3) & 356.7342 & 42.8 & 8.4 $\times$ 10$^6$ & JCMT & 14 & ... & G28.34 \\
$^{13}$CO (3-2) & 330.5880 & 31.7 & 2.9 $\times$ 10$^4$ & JCMT & 14 & ... & G28.34 \\ %Y
$^{13}$CO (1-0) & 110.2014 & 5.3 & 1.9 $\times$ 10$^3$ & FCRAO-14m & 46 & ... & ... \\ %Y
\enddata
\tablenotetext{}{Notes. The line information is from the CDMS \citep{2001AA...370L..49M} and LAMDA \citep{2005AA...432..369S} databases.}
\tablenotetext{a}{Upper energy level in units of K.}
\tablenotetext{b}{Critical density at 15 K. For N$_2$D$^{+}$, we adopt the collision rate coefficient of N$_2$H$^{+}$. }
\tablenotetext{c}{Resolution.}
\tablenotetext{d}{Maximum recoverable scale for ALMA.}

\end{deluxetable}

\subsubsection{Velocity centroid maps}
\begin{figure*}[!htbp]
 \gridline{
 \fig{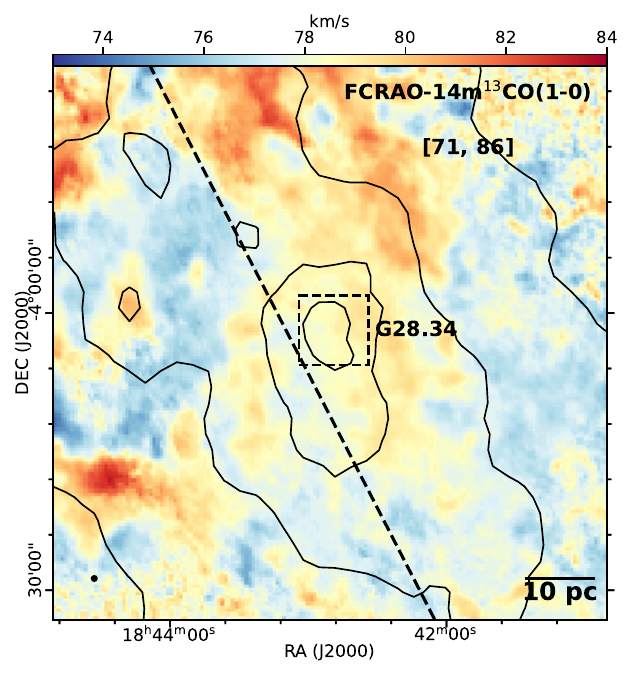}{0.33\textwidth}{(a)}
 \fig{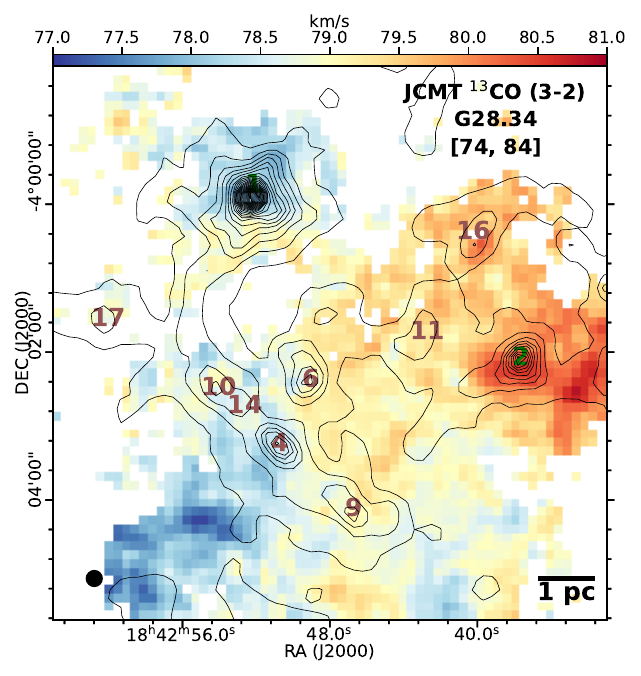}{0.33\textwidth}{(b)}
 \fig{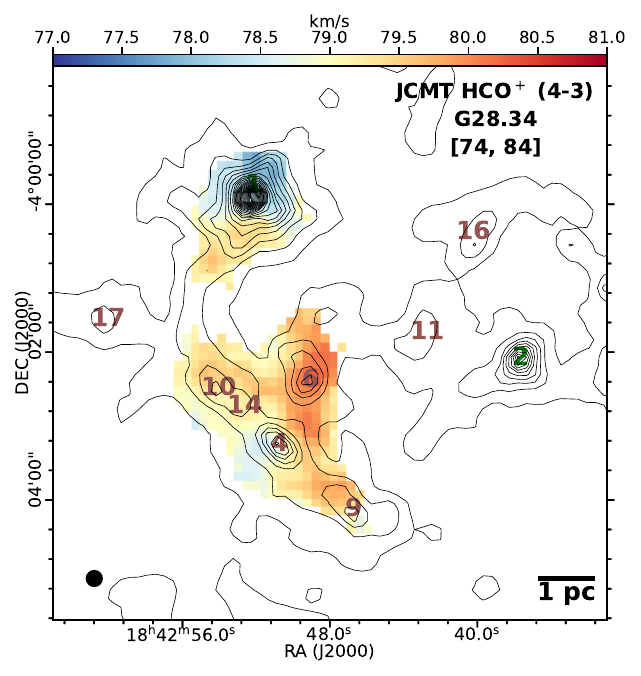}{0.33\textwidth}{(c)}
 }
\caption{(a) Velocity centroid map of the FCRAO-14m $^{13}$CO (1-0) data in the surrounding of G28.34. Black contours correspond to the Planck $\tau_{353}$ map, starting from 0.0005 and continuing with an interval of 0.0005. A dashed line indicates the galactic plane ($b=0\degr$). The dashed rectangle indicates the JCMT map area in (b) and (c). (b)-(c) Velocity centroid maps of the $^{13}$CO (3-2) and HCO$^{+}$ (4-3) emission toward the IRDC G28.34 obtained with JCMT. Black contours correspond to the JCMT 0.85 mm dust continuum map. Contours start at 50 mJy beam$^{-1}$ and continue at a step of 150 mJy beam$^{-1}$. The beam and a scale bar are indicated in the lower left and lower right corner of each panel, respectively.  \label{fig:G28_large_line_m1}}
\end{figure*}

\begin{figure*}[!htbp]
 \gridline{
 \fig{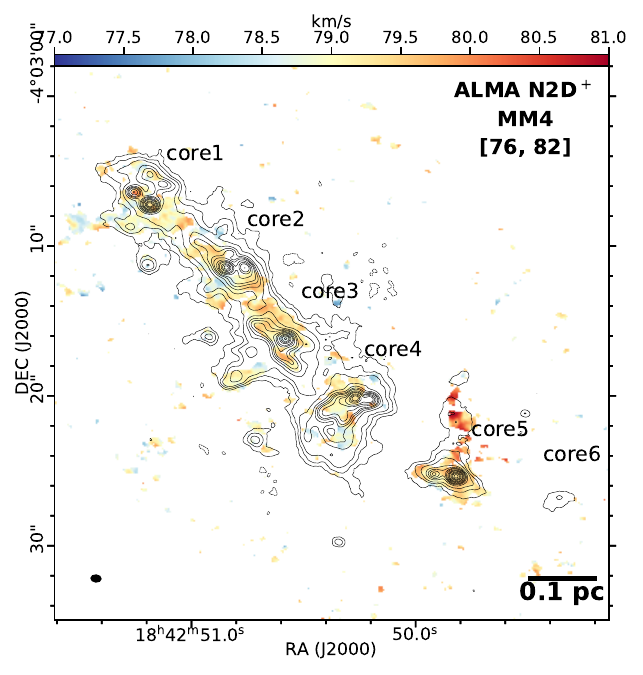}{0.4\textwidth}{(a)}
 \fig{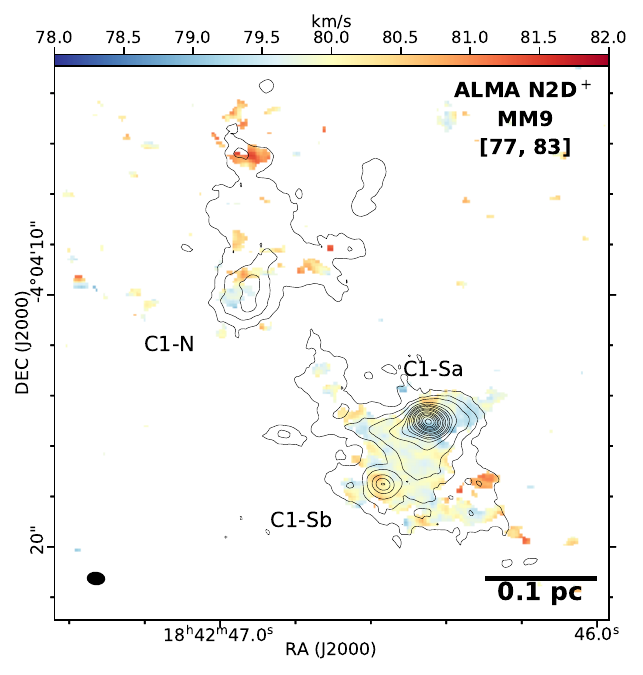}{0.4\textwidth}{(b)}
 }

\caption{Velocity centroid map of the ALMA N$_2$D$^{+}$ (3-2) data toward MM4 and MM9. Contours correspond to the ALMA 1.3 mm dust continuum map. Contour levels are ($\pm$3, 6, 10, 20, 30, 40, 50, 70, 90, 110, 150, 180, 210, 250, 290, 340, 390, 450) $\times \sigma_{I}$. The synthesized beam and a scale bar are indicated in the lower left and lower right corner of each panel, respectively. \label{fig:G28_alma_line_m1}}
\end{figure*}

Figures \ref{fig:G28_large_line_m1} and \ref{fig:G28_alma_line_m1} show the multi-scale velocity structure surrounding and within G28.34. The FCRAO-14m $^{13}$CO (1-0) data (see Figure \ref{fig:G28_large_line_m1}(a)) reveals a velocity gradient across Galactic latitudes near G28.34 \citep{2020A&A...638A..44B}. The origin of this global gradient is still unclear. At higher resolution, the JCMT $^{13}$CO (3-2) data reveals a velocity gradient perpendicular to the main dark filament (see Figure \ref{fig:G28_large_line_m1}(b)) that is consistent with the global velocity gradient. Previous studies have interpreted this gradient as gas flows converging at the position of the filament \citep{2020A&A...638A..44B}. The origin of the converging flow may be related to gravitationally driven collapse or external compression \citep{2020A&A...638A..44B}. The HCO$^{+}$ (4-3) line with a higher critical density than the $^{13}$CO (3-2) transition reveals the velocity structure of most clumps in G28.34 except for MM17 (see Figure \ref{fig:G28_large_line_m1}(c)). Within the dark filament, the higher-density gas traced by the HCO$^{+}$ (4-3) line is more redshifted than the lower-density gas traced by the $^{13}$CO (3-2) transition. With the NH$_3$ observations, \citet{2015ApJ...804..141Z} have identified a longitudinal velocity gradient of 0.6 km s$^{-1}$ along the main dark filament from MM10 to MM14 to MM4, which reverses from MM4 to MM9. This velocity gradient is also seen in the JCMT $^{13}$CO (3-2) and HCO$^{+}$ (4-3) data. Toward MM1, there seems to be a north-south velocity gradient deviating from the trend of the large-scale gradient, which may be an indicator of local gravitational infall in this clump. At smaller scales,  the variation of $V_c$ is overall small in dense cores in MM4 and MM9 as seen by the ALMA N$_2$D$^{+}$ (3-2) observations. There is a redshifted velocity component toward the streamer-like continuum structure in the north of MM4-core5, which may indicate that the core is accreting gas from the clump gas through the streamer. There is a possible sign of a redshifted velocity component in the north of C1-N. Other than these redshifted velocity components, there are no apparent signs of ordered velocity gradients in the major part of cores in MM4 and MM9.

\subsubsection{Multi-scale statistics}\label{sec:vstat}

Here we use the velocity dispersion-radius ($\sigma_v-r$) relation and velocity centroid dispersion-radius ($\sigma_c-r$) relation to explore the multi-scale velocity statistics in G28.34. 

In a similar way to the derivation of the $M-r$ relation (Section \ref{sec:N_scaling}), we derive the $\sigma_v-r$ relation for the average spectra (see Appendix \ref{sec:appline}) of each line within circles of different radii, except for the ALMA data where the spectra are averaged within the whole emission area. Similar to the approaches that most previous studies have adopted, we fit a single-component Gaussian profile for each line to obtain the Full Width at Half Maximum (FWHM. i.e., linewidth). The relation between the FWHM and velocity dispersion is $FWHM=\sqrt{8\ln{2}} \sigma_v$. For the JCMT data, we focus on the infrared dark clumps MM4, MM6, MM9, and MM10 (+MM14). The thermal velocity linewidth ($<$0.06 km s$^{-1}$ for our studied molecules) is much smaller than the non-thermal linewidth in our studied regions due to the low temperature, so we regard the total linewidth as the non-thermal linewidth and do not subtract the thermal contribution. It should be noted that the radius of each considered region does not necessarily reflect the corresponding spatial scale of the measured linewidth. This is because the linewidth is also dependent on the LOS depth, which cannot be easily determined from observations. As we only fit a single-component Gaussian profile for all the lines, the linewidth in some regions could be broadened by the superposition of substructures with different LOS local-standard-of-rest (LSR) velocities at small scales (i.e., indistinguishable multiple velocity components in the line profiles. See Appendix \ref{sec:appline}). Due to these effects, the measured linewidth from the observed spectra likely presents the upper limit of the true non-thermal linewidth at the corresponding scale of the radius. 

\begin{figure}[!htbp]
 \gridline{\fig{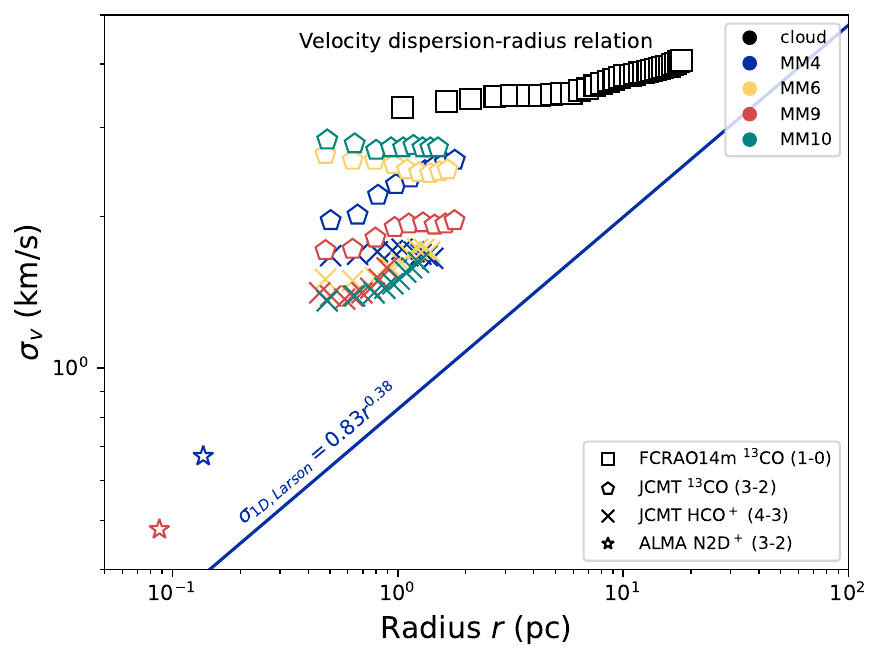}{0.48\textwidth}{(a)}
 \fig{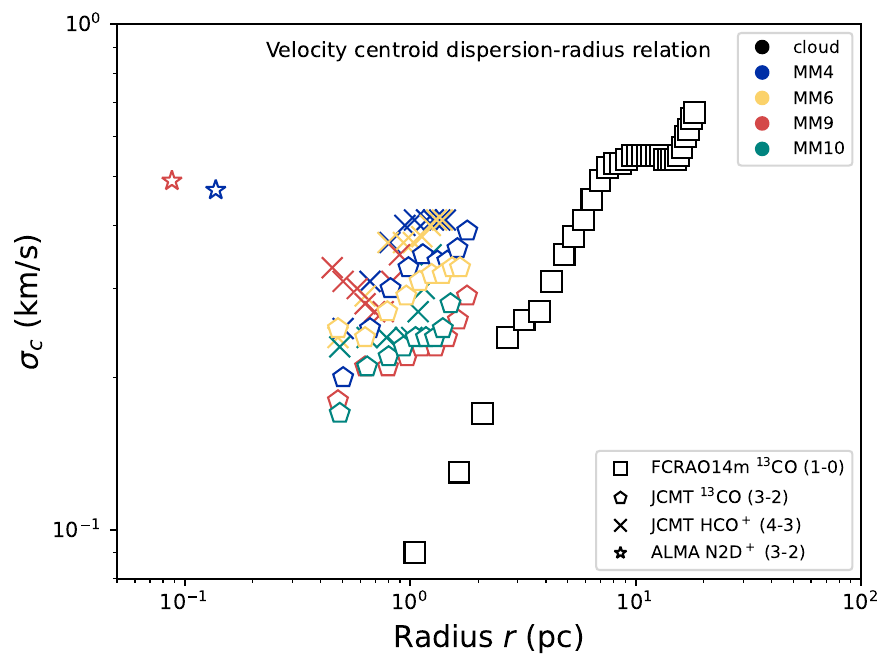}{0.48\textwidth}{(b)}
 }
\caption{(a). Velocity dispersion-radius relation. The blue solid line indicates Larson's first law \citep[$\sigma_{1D,Larson}=0.83$ km s$^{-1}(r/pc)^{0.38}$,][]{1981MNRAS.194..809L}. The original 3D velocity dispersion in Larson's first law is converted to 1D velocity dispersion with the relation $\sigma_{3D,Larson} = \sqrt{3}\sigma_{1D,Larson}$. (b). Velocity centroid dispersion-radius relation. Only data points with $l$ greater than the beam resolution and smaller than the effective radius of the considered region are shown. Different colors represent different regions. Different symbols indicate different instruments and lines. \label{fig:G28_dv_r}}
\end{figure}

Figure \ref{fig:G28_dv_r}(a) shows the velocity dispersion-radius relation. At first glimpse, it is obvious that the $\sigma_v-r$ relations from different instruments and lines are not continuous. The discontinuity of $\sigma_v-r$ relation is similar to what \citet{2023MNRAS.525.2935P} have recently found in a large sample of 27 infrared dark clouds. While \citet{2023MNRAS.525.2935P} interpreted the discontinuous $\sigma_v-r$ relation as clump dynamics decoupling from those of the parental clouds, we suggest that the discontinuity is more likely due to the fact that the physical conditions traced by different line data are different and that the $\sigma_v-r$ relation does not correctly present the true velocity dispersion-size relation of the gas in 3D. 

As seen in Figure \ref{fig:G28_dv_r}(a), the $\sigma_v-r$ relation for each region seems to be relatively flat with small slopes or even negative slopes for the single-dish (FCRAO-14m and JCMT) observations. The flat slopes suggest that our measured velocity dispersion is likely dominated by the LOS integration up to the largest depth of the gas traced by specific lines, which is similar to what was found in the low-mass cloud Polaris Flare \citep{2002A&A...390..307O}. The JCMT velocity dispersions are lower than the FCRAO-14m $^{13}$CO (1-0) velocity dispersions. This might be due to the higher $n_c$ of the $^{13}$CO (3-2) and HCO$^{+}$ (4-3) lines, which trace fewer substructures along the LOS. On the other hand, the single-dish observations filter out the diffuse emission at the extent of the off position. The particular JCMT observations may have set a nearer off position than the FCRAO-14m observations, which could filter out more diffuse emissions and is also plausible to account for the lower velocity dispersions revealed by JCMT. For the JCMT observations, the velocity dispersions of HCO$^{+}$ (4-3) are lower than that traced by $^{13}$CO (3-2), which might be due to the higher $n_c$ and fewer LOS substructures traced by HCO$^{+}$ (4-3). In addition, the ALMA velocity dispersions are much lower than those traced by JCMT and FCRAO-14m, which may mainly be because the emissions from lower-density substructures are filtered out by the interferometric observations. 
%, which deviates from a single continuous power law across different scales as predicted by the Larson's first law

We also study the scaling relation of velocity fields with the dispersion (i.e., standard deviation) of velocity centroids ($\sigma_c$). The velocity centroid is insensitive to thermal broadening and thus primarily reveals non-thermal motions \citep{2022ApJ...935...77L}. Moreover, velocity centroids present averaged statistics along the LOS and are not affected by the broadening effect of LOS depth as for the linewidth. Thus the radius in the $\sigma_c-r$ relation correctly reflects the true scale in the scaling relation. Due to the LOS averaging, for a single-component turbulent structure, the turbulent velocity centroid dispersion is smaller than the turbulent velocity dispersion by a factor of $\sigma_v/\sigma_c=\sqrt{N_{turb}}=\sqrt{L/\lambda_c}$ \citep{1985ApJ...295..479D}, where $N_{turb}$ is the number of independent turbulent cells along the LOS, $L$ is the effective depth, and $\lambda_c$ is the turbulent correlation length. 
Previous observational $V_c$ studies usually implicitly assume that $N_{turb}$ is constant or does not vary too much within an individual cloud \citep[e.g.,][]{2002A&A...390..307O, 2023ApJ...949...30L}. Considering that each line tracer mainly traces a layer of gas with densities between $n_c$ and 100$n_c$, the validity of this assumption could be unclear. 

%The velocity centroid autocorrelation function (VCACF) is given by

Figure \ref{fig:G28_dv_r}(b) shows the velocity centroid dispersion-radius relation. The $\sigma_c-r$ relation in each region is complex and we refrain from interpreting them in detail in this work. The $\sigma_c-r$ relations from different instruments and lines are not continuous as well. The $\sigma_c$ in most clumps traced by JCMT is comparable to $\sigma_c$ in the environmental gas traced by FCRAO-14m, but lower than the $\sigma_c$ of the dense cores traced by ALMA. If $N_{turb}=\sqrt{L/\lambda_c}$ is similar for different line data or $N_{turb}$ is larger for higher-resolution data, the observed $\sigma_c-r$ relation could suggest that the early-stage massive star formation activities have already increased the non-thermal motions in small-scale and high-density regions in IRDC G28.34. However, we cannot rule out the possibility that the velocity centroid dispersion at a smaller $r$ is less averaged due to a smaller $N_{turb}$.

In summary, we find the velocity statistics are not continuous at different scales revealed by different instruments and lines, nor universal in different regions. 
However, due to the unsolved issues on the LOS length of linewidth and the averaging of velocity centroids, we cannot rule out the possibility that the actual multi-scale velocity statistics in G28.34 could still follow a continuous and universal power-law relation. Solving those issues requires the application of more advanced analysis methods \citep[e.g.,][]{2006ApJ...652.1348L}.

\subsection{Magnetic field strength}\label{sec: B}
The Davis-Chandrasekhar-Fermi (DCF) method \citep{1951PhRv...81..890D, 1953ApJ...118..113C} and its modified forms have been widely used to estimate the plane-of-sky (POS) magnetic field strength in molecular clouds. The basic assumptions of the DCF method are: there is an equipartition between the transverse (i.e., perpendicular to the ordered field) turbulent magnetic and kinetic energies; the turbulence is isotropic; and the turbulent-to-ordered or turbulent-to-total magnetic field ratio can be traced with the statistics of magnetic field orientations. Under these assumptions, the POS ordered and total magnetic field strengths are given by
\begin{equation}\label{eq:Bu}
B_0 = \sqrt{\mu_0 \rho }\frac{\sigma_v}{B_{\mathrm{t}}/B_0}
\end{equation}
and 
\begin{equation}\label{eq:Btot}
B = \sqrt{\mu_0 \rho }\frac{\sigma_v}{B_{\mathrm{t}}/B},
\end{equation}
respectively, where $\mu_0$ is the permeability of
vacuum, $\rho = \mu_{\mathrm{H_2}} m_{\mathrm{H}} n_{\mathrm{H_2}}$ is the gas density, $n_{\mathrm{H_2}}$ is the volume density, and $\sigma_{\mathrm{v}}$ is the line-of-sight turbulent velocity dispersion. In small angle approximation (i.e., the ordered magnetic field is prominent), the POS turbulent-to-ordered field ratio ($B_{\mathrm{t}}/B_0$) or turbulent-to-total field ratio ($B_{\mathrm{t}}/B$) are usually estimated with the angular dispersion ($\sigma_\theta$) of POS magnetic field orientations: $B_{\mathrm{t}}/B_0 \sim B_{\mathrm{t}}/B \sim \sigma_\theta$. A recent review of the DCF method can be found in \citet{2022ApJ...925...30L, 2022FrASS...9.3556L}.

\subsubsection{Environmental gas}\label{sec:largeB}
We estimate the magnetic field strength in the environmental gas of IRDC G28.34 with the Planck polarization data and the FCRAO-14m $^{13}$CO (1-0) data. As demonstrated later (see Section \ref{sec:VChG}), the Planck polarization detection toward G28.34 mainly traces the emission from the surrounding gas of this IRDC. Thus it is appropriate to use the Planck polarization map to obtain the magnetic field information of the cloud environment. The radius of the polarization detection region must be $\gtrsim$2 of the beam size to obtain meaningful statistics of the turbulent magnetic field \citep{2021ApJ...919...79L}. Thus we consider a circular region with a radius of $r=$15 pc centered at G28.34. The angular dispersion of the Planck magnetic field is $\sigma_\theta = 3.9\degr$ within the considered region. The almost straight POS magnetic field lines at 15-pc scale and the small angular dispersion imply $B_0 \sim B$. As mentioned in Section \ref{sec:N_envir}, the density estimated from the $^{13}$CO (1-0) data tends to be underestimated while the density estimated from the Planck map tends to be overestimated. Thus we adopt the density extrapolated from the power-law relation ($n=6398$ cm$^{-3} (r/\mathrm{pc})^{-1.41}$) fitted from the COMB data and ALMA data. At $r=15$ pc, we obtain $n_{ \mathrm{H_2} } \sim 144$ cm$^{-3}$. 
Similar to the turbulent kinetic field, the turbulent magnetic field as well as its ratio with respect to the ordered or total field could be underestimated by a factor of $\sqrt{N_{B}}$ if there are $N_{B}$ independent magnetic turbulent cells along the LOS \citep{1990ApJ...362..545Z}. The 3D unaveraged turbulent magnetic and kinetic fields within the considered region are not directly measurable from observations. As the DCF method assumes equipartition between turbulent magnetic and kinetic energies, it is reasonable to further assume that the turbulent magnetic and kinetic fields are averaged by a similar extent (i.e., $N_{B} \sim N_{turb}$) due to the LOS signal integration. Although $N_{turb}$ and $N_{B}$ may not be accurately measured from our observational data, the corrections of the LOS signal integration effect for the turbulent kinetic field (as traced by $\sigma_c$) and turbulent magnetic field (as traced by $\sigma_\theta$) could cancel out with $(\sqrt{N_{turb}}\sigma_c)/(\sqrt{N_{B}}\sigma_\theta) = \sigma_c/\sigma_\theta$ \citep{2016ApJ...821...21C}. Within our studied region, the velocity centroid dispersion is $\sigma_c=0.54$ km s$^{-1}$. Incorporating the correction for the LOS signal integration, the magnetic field strength of the environmental gas within the 15-pc region is given by $B \sim B_0 \sim \sqrt{\mu_0 \rho }\sigma_c/\sigma_\theta \sim 0.074$ mG. Alternatively, using a model with $n_{ \mathrm{H_2} } \sim 100$ cm$^{-3}$ and $l=8$ pc, \citet{2001ApJ...546..980O} numerically investigated the uncertainties in the DCF estimation of clouds and derived a correction factor of 0.5 (hereafter the Ost01 correction factor). As the density of their simulation is comparable to that of the region of our interest, we apply the Ost01 correction factor for an alternative magnetic field strength estimation. Adopting $\sigma_v = 3.93$ km s$^{-1}$ at $r=15$ pc, we obtain $B \sim B_0 \sim 0.5\sqrt{\mu_0 \rho }\sigma_v/\sigma_\theta \sim 0.27$ mG. Note that the critical density of $^{13}$CO (1-0) is an order of magnitude higher than the density of our considered region, so the turbulent velocity dispersion here may be underestimated, which might lead to underestimation of the magnetic field strength as well.
%Here we adopt another assumption that the Planck polarized emission within the 15-pc region is mainly from the surrounding gas of G28.34. This is a reasonable assumption since the JCMT observation recovers most of the Planck polarization emission (Section \ref{sec:obs_planck}). 
%numerical correction by Ostriker 2001. 

%Thus we convolve the $^{13}$CO (1-0) line data to the same angular resolution as the Planck data (5$\arcmin$). With a simple Gaussian fit for the averaged spectra of the optically thin $^{13}$CO (1-0) line, we obtain $\sigma_{v,mol}=4.0$ km s$^{-1}$. The non-thermal velocity dispersion is given by where $m_\mathrm{mol}$ is the mass of the molecule. Then we assume the non-thermal motion is dominated by pure turbulent motions at this scale (i.e., $\sigma_{\mathrm{v}} \sim \sigma_{\mathrm{v,nt}}$) and obtain $\sigma_v\sim4.0$ km s$^{-1}$. With these estimated parameters, the magnetic field strength is calculated as $B \sim B_0 \sim \sqrt{\mu_0 \rho }\sigma_v/\sigma_\theta \sim 0.18$ mG. 

%Vc without convolving is 0.65 km/s. significant non-thermal energy in small-scale? or beam smoothing? how to explain the larger N after convolving? 

At the considered scale, additional uncertainty on the estimated field strength could come from the anisotropy of MHD turbulence when gravity is negligible \citep{2022ApJ...935...77L}. The correction for the turbulence anisotropy requires knowledge of the mean-field inclination and the fraction of turbulence modes. However, both parameters cannot be easily measured with our existing data. Thus we are unable to correct this effect in this work. 

Some recent theoretical and numerical studies suggest that in a non-gravitational and very sub-Alfv\'{e}nic environment, the turbulent kinetic energy is in equipartition with the coupling-term magnetic field energy fluctuation rather than with the turbulent magnetic energy  \citep{2016JPlPh..82f5301F, 2021A&A...647A.186S, 2022MNRAS.515.5267B}, which could bring another uncertainty to the DCF method. However, it has been debated whether their proposed coupling-term energy equipartition is valid for molecular clouds \citep{2022ApJ...935...77L, 2022MNRAS.510.6085L, 2022ApJ...925...30L, 2022FrASS...9.3556L}. Firstly, the physical interpretation for including the coupling-term field in the energy equipartition is unclear. Even if the reference velocity is set as the cloud average velocity, the substructures moving at a different velocity from the cloud average velocity or the ordered non-turbulent motions (e.g., infall or rotation) due to star formation activities could still generate a significant amount of coupling-term velocity field fluctuation within molecular clouds. There is no reason to only consider the coupling-term magnetic field but not consider the coupling-term velocity field. Secondly, all the numerical works supporting the coupling-term energy equipartition have adopted the whole simulation-averaged mean magnetic field, which represents the interstellar medium (ISM) magnetic field at approximately the turbulence injection scale ($\sim$100 pc), in the calculation of the coupling-term magnetic field. While it is probably fine to adopt the global mean field as the local mean field in an extremely sub-Alfv\'{e}nic case \citep[e.g., $\mathcal{M}_{A}=0.01$,][]{2022MNRAS.515.5267B}, the star-forming 10-pc clouds are only trans-to-sub-Alfv\'{e}nic \citep[e.g., $\mathcal{M}_{A}\gtrsim0.6$,][]{2019NatAs...3..776H}. In such cases, the 10-pc local mean field could have significantly deviated from the 100-pc global mean field and thus it would be inappropriate to adopt the global mean field of the ISM in the calculation of the cloud energetics. In summary, more work needs to be done to understand the validity of the coupling-term energy equipartition assumption in molecular clouds. 

\subsubsection{Clump}\label{sec:B_clump}

For the dense clumps revealed by JCMT observations, the magnetic field shows non-linear ordered field structures due to gravitational effects and star formation activities, which overestimates the angular dispersion that should be only attributed to turbulence. In G28.34, all clumps have $\sigma_\theta > 25\degr$ within $r=1$ pc, but using $\sigma_\theta$ to estimate $B$ is only valid when $\sigma_\theta < 25\degr$ in both low and high density regions \citep{2001ApJ...546..980O, 2021ApJ...919...79L}. To account for the ordered field contribution, we use the angular dispersion function (ADF) method \citep[][]{2009ApJ...696..567H, 2009ApJ...706.1504H, 2016ApJ...820...38H}, a modified DCF method, to estimate the magnetic field strength of these clumps. 

The ADF accounting for the ordered field contribution, beam-smoothing effect, and turbulent correlation effect is given by
\begin{equation}
1 - \langle \cos \lbrack \Delta \Phi (l)\rbrack \rangle \simeq \frac{\langle B_\mathrm{t}^2 \rangle}{\langle B^2 \rangle} \times \lbrack 1 - e^{-l^2/2(l_\delta^2+2l_W^2)}\rbrack + a_2' l^2,
\end{equation}
where $\Delta \Phi (l)$ is the angular difference of two position angles separated by $l$, $l_W = l_{\mathrm{beam}}/\sqrt{8 \ln{2}}$ is the standard deviation of the beam size, $a_2' l^2$ is the second-order term of the Taylor expansion for the ordered field, and $l_\delta$ is the turbulent correlation length. Here the POS turbulent-to-total magnetic energy ratio $\langle B_\mathrm{t}^2 \rangle/\langle B^2 \rangle$ does not consider the effect of LOS signal integration. Note that the second-order term $a_2' l^2$ is only valid to represent the ordered field at small $l$ when the higher-order terms are negligible. Also note that polarization observations can only trace the magnetic field orientation with a 180$\degr$ ambiguity, so $\Delta \Phi (l)$ is constrained to be within $[-90, 90]$ degrees \citep{2009ApJ...696..567H, 2009ApJ...706.1504H}. Because the actual magnetic field direction angle in the range of -180$\degr$ to 180$\degr$ is approximated by the position angle in the range of -90$\degr$ to 90$\degr$, the ADF method implicitly assumes that the turbulent field is smaller than the ordered field (i.e., sub-Alfv\'{e}nic). 

\begin{figure}[!tbp]
\gridline{\fig{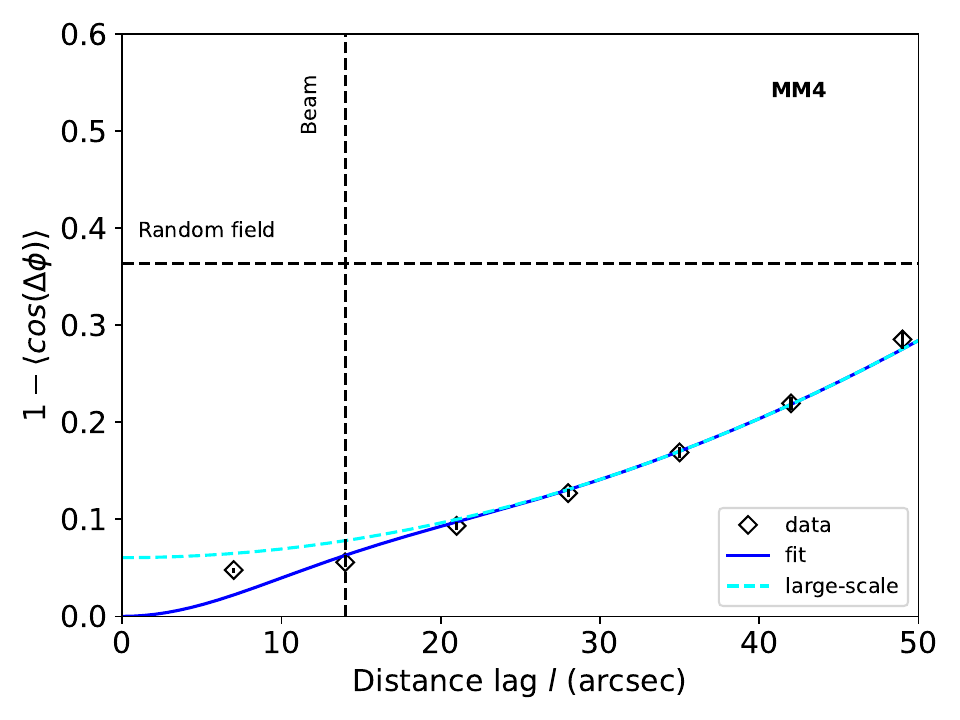}{0.45\textwidth}{}
\fig{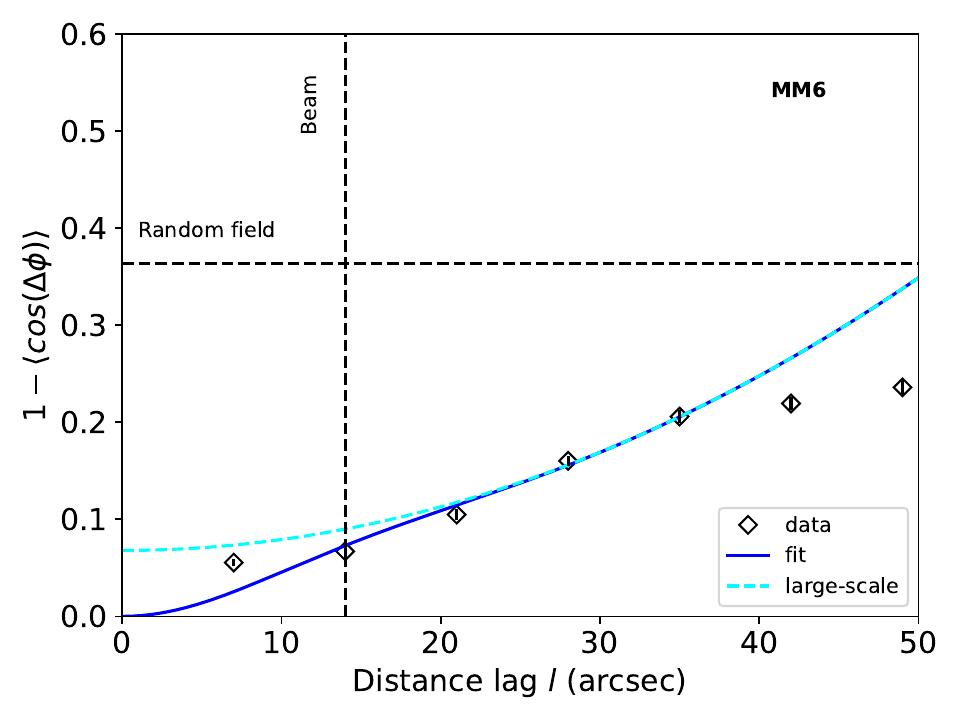}{0.45\textwidth}{}}
\gridline{\fig{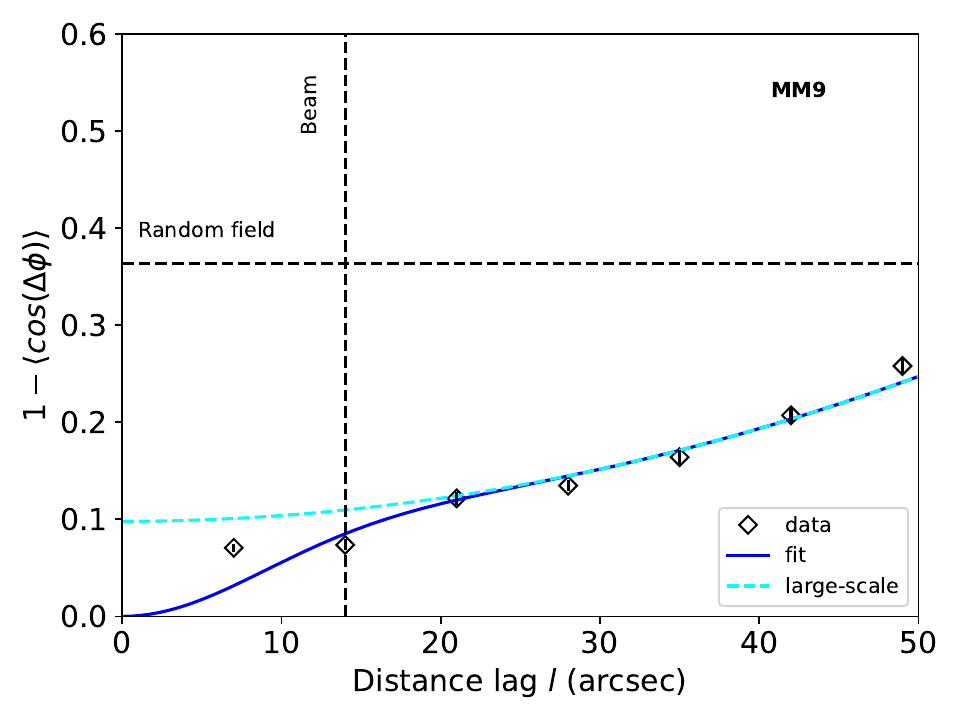}{0.45\textwidth}{}
 \fig{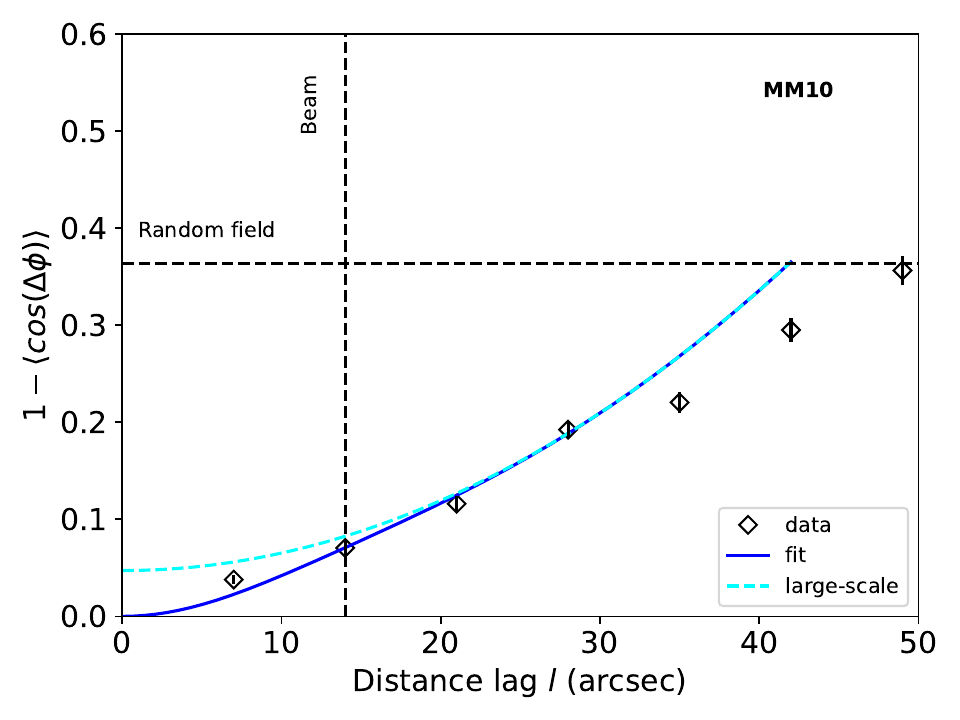}{0.45\textwidth}{}}
\caption{Angular dispersion functions for MM4, MM6, MM9, and MM10. The diamond symbols represent the observed data points. The error bars indicate the statistical uncertainties propagated from the observational uncertainty \citep{2009ApJ...706.1504H}. The blue dashed line indicates the best-fitted results. The cyan dashed line shows the large-scale component ($\langle B_\mathrm{t}^2 \rangle/\langle B^2 \rangle + a_2' l^2$) of the best fit. The horizontal dashed line indicates the ADF value corresponding to a random field \citep[0.36, ][]{2021ApJ...919...79L}. \label{fig:jcmt_adf}}
\end{figure}

Figure \ref{fig:jcmt_adf} shows the ADFs for clumps MM4, MM6, MM9, and MM10. We adopt a binning interval of $l_{\mathrm{bin}} = l_{\mathrm{beam}}/2$ for the distance lag. We fit each ADF via reduced $\chi^2$ minimization. We fit the ADFs over different maximum $l$ ranges and obtain the best fit with the smallest reduced $\chi^2$. The best fitted results are $\langle B_\mathrm{t}^2 \rangle/\langle B^2 \rangle^{0.5} = 0.25, 0.26, 0.31, 0.22$ for MM4, MM6, MM9, and MM10, respectively. 

Due to the effect of LOS signal integration, the statistics of POS polarization position angles could underestimate the turbulent magnetic field. The ADF method proposes that the turbulent magnetic energy is averaged by a factor of 
\begin{equation}
N_{B} = \frac{(l_\delta^2 + 2l_W^2)l_\Delta}{\sqrt{2\pi}l_\delta^3}
\end{equation}
for single-dish data due to the LOS integration \citep{2009ApJ...706.1504H}, where $l_\Delta$ is the LOS cloud effective depth. \citet{2009ApJ...706.1504H} suggested that $l_\Delta$ could be estimated as the width at half of the maximum of the normalized autocorrelation function for the integrated normalized polarized flux. However, the normalized autocorrelation function of the integrated normalized polarized flux is greater than half of the maximum for all the clumps, so $l_\Delta$ in our case cannot be derived in this way. Moreover, the numerical study by \citet{2021ApJ...919...79L} found that the ADF method may not work well for the effect of LOS signal integration. So we refrain from adopting the correction for this effect suggested by \citet{2009ApJ...706.1504H}. 

Alternatively, we adopt the analytical corrections suggested by \citet{2016ApJ...821...21C} (CY16) or numerical corrections by \citet{2021ApJ...919...79L} (Liu21). \citet{2016ApJ...821...21C} suggested using the turbulent velocity centroid dispersion instead of the turbulent velocity dispersion in the DCF equation to account for the LOS averaging effect. Here we derive $\sigma_c$ from the $^{13}$CO (3-2) line data because its critical density is closer to the clump densities. The velocity centroid dispersions are $\sigma_c=$0.34, 0.30, 0.22, and 0.23 km s$^{-1}$ for MM4, MM6, MM9, and MM10, respectively. Adopting  $n = 6.3 \times 10^{3}$, $5.8 \times 10^{3}$, $5.7 \times 10^{3}$, and $6.0 \times 10^{3}$ cm$^{-3}$, we obtain $B \sim \sqrt{\mu_0 \rho }\sigma_c(\langle B_\mathrm{t}^2 \rangle/\langle B^2 \rangle)^{-0.5} \sim$0.081, 0.065, 0.039, and 0.060 mG for MM4, MM6, MM9, and MM10, respectively. Here we assume that the velocity centroid dispersion is dominated by turbulent motions rather than non-turbulent motions. The magnetic field strength could be overestimated if $\sigma_c$ is mostly non-turbulent. On the other hand, \citet{2021ApJ...919...79L} has numerically derived the average correction factors for the clump-to-core scale magnetic field strength estimated with the ADF method using the velocity dispersion. The velocity dispersions are $\sigma_{v}=$2.32, 2.54, 1.91, and 2.70 km s$^{-1}$ for MM4, MM6, MM9, and MM10, respectively. Adopting a numerical correction factor of 0.21 \citep[with a 45\% uncertainty,][]{2021ApJ...919...79L}, we obtain $B \sim 0.21 \sqrt{\mu_0 \rho }\sigma_{v}(\langle B_\mathrm{t}^2 \rangle/\langle B^2 \rangle)^{-0.5} \sim$0.120, 0.120, 0.075, and 0.154 mG for MM4, MM6, MM9, and MM10, respectively. The magnetic field strengths of the clumps estimated with either the CY16 or Liu21 corrections are comparable to the magnetic field strength of the environmental gas, which means the magnetic field does not significantly scale with density in the low-density gas. This behavior is consistent with previous magnetic field strength estimations in other regions \citep{2019FrASS...6...15P, 2022ApJ...925...30L, 2022FrASS...9.3556L}. Using the DCF methods to derive the ordered field strength could have large uncertainties in self-gravitating regions \citep{2021ApJ...919...79L}, so we only derive the total field strength for the clumps. 
% within a factor of 2
% As discussed in Section \ref{sec:vstat}, the line velocity dispersion probed by JCMT traces the LOS integration to the largest depth, so the average spectra correspond to a 3D cylinder rather than a 3D sphere. The magnetic field strengths derived from the Liu21 correction and from the CY16 correction generally agree with each other within a factor of 2.5.  Note that \citet{2021ApJ...919...79L} has substracted rotational motions from the velocity dispersion while we do not. The magnetic field strength could be overestimated if there are significant non-turbulent non-thermal motions in the clumps.

Within self-gravitating clumps and cores, the uncertainty from anisotropic turbulence on the DCF method is negligible \citep{2022FrASS...9.3556L}. Thus we do not consider the correction for this effect. Because there is no evidence supporting coupling-term energy equipartition in self-gravitating regions \citep{2022FrASS...9.3556L}, we keep our assumption of turbulent energy equipartition. Note that if the clumps are super-Alfv\'{e}nic, the turbulent magnetic energy could be smaller than the turbulent kinetic energy, and our derived magnetic field strengths for the clumps could be overestimated. 

\subsubsection{Core}

From our ALMA observations, only one core (MM4-core4) has sufficient polarization detections to derive the magnetic field strength. \citet{2020ApJ...895..142L} has estimated the ordered magnetic field strength of MM4-core4 using the ADF method. The kinetic information adopted in \citet{2020ApJ...895..142L} was from the EVLA NH$_3$ line data \citep{2012ApJ...745L..30W}, which had a different resolution and filtering scale larger than the ALMA data. Here we recalculate the magnetic field strength with the CY16 and Liu21 corrections and with the velocity information from the ALMA N$_2$D$^+$ line data.

In MM4-core4, the velocity dispersion and velocity centroid dispersion of N$_2$D$^+$ are 0.61 and 0.37 km s$^{-1}$, respectively. We adopt the radius ($r = 0.053$ pc), density ($n = 1.1 \times 10^{6}$ cm$^{-3}$), and turbulent-to-total magnetic field strength ratio without correction for LOS integration ($(\langle B_\mathrm{t}^2 \rangle/\langle B^2 \rangle)^{0.5} \sim 1$. i.e., the total field is dominated by the turbulent field) estimated in \citet{2020ApJ...895..142L}. Adopting the CY16 and Liu21 corrections, we obtain $B \sim \sqrt{\mu_0 \rho }\sigma_c(\langle B_\mathrm{t}^2 \rangle/\langle B^2 \rangle)^{-0.5} \sim$0.29 mG and $B \sim 0.21 \sqrt{\mu_0 \rho }\sigma_{v}(\langle B_\mathrm{t}^2 \rangle/\langle B^2 \rangle)^{-0.5} \sim$0.10 mG, respectively. It should be noted that both the CY16 and Liu21 corrections may not well account for the LOS integration effect at $<$0.1 pc and $n > 10^{6}$ cm$^{-3}$ \citep{2021ApJ...919...79L}. Also, note that the turbulent velocity motions may be overestimated due to the existence of non-turbulent non-thermal motions. And if the core is super-Alfv\'{e}nic, the energy equipartition assumption could break down. Those above-mentioned factors could all overestimate the B strength for MM4-core4. On the other hand, the ALMA observations filter out the large-scale emission, but \citet{2021ApJ...919...79L} did not perform the filtering when deriving the velocity dispersion from the simulations. This inconsistency could underestimate the B strength estimated from the Liu21 correction when using the line velocity dispersion from interferometric observations, which could explain the smaller B value from the Liu21 correction than that from the Cy16 correction. Overall, the B value in core MM4-core4 is comparable to or only slightly larger than that in clump MM4, which may suggest that gravity does not significantly compress and amplify the magnetic field strength in the clumps in the early massive star formation stage. 

%\subsubsection{Angular dispersion function}

\subsection{Velocity Gradient Technique}\label{sec:VGT}

\subsubsection{Alfv\'{e}nic Mach number}
Using the angular dispersion of polarization position angles to derive the Alfv\'{e}nic Mach number $\mathcal{M}_{A}$ requires corrections for the LOS signal integration and other effects. It is inappropriate to just equal the angular dispersion and $\mathcal{M}_{A}$ as some previous studies have done. Based on the property of MHD turbulence \citep{1995ApJ...438..763G}, \citet{2018ApJ...865...46L} proposed an alternative method (Velocity Gradient Technique, VGT) to estimate $\mathcal{M}_{A}$ in non-self-gravitating regions with the statistics of velocity gradient orientation ($\theta_{\mathrm{VG}}$) from molecular line observations. The advantage of VGT is that the velocity gradient orientation histogram is independent of the LOS integration \citep{2018ApJ...865...46L}. It was numerically shown that the top-to-bottom ratio ($N_{top}/N_{bot}$) of the velocity gradient orientation histogram has the relation with $\mathcal{M}_{A}$ in a sub-Alfv\'{e}nic region \citep{2021ApJ...912....2H}
\begin{equation}
\mathcal{M}_{A} \simeq 1.6 (N_{top}/N_{bot})^{\frac{1}{-0.6}}.
\end{equation}
The Top ($N_{top}$) and Bottom ($N_{bot}$) values of the histogram can be obtained by fitting a Gaussian profile (i.e., $(N_{top}-N_{bot})\exp(-\alpha(\theta_{\mathrm{VG}} - \theta_{\mathrm{0}})^2) + N_{bot}$, where $\alpha$ and $\theta_{\mathrm{0}}$ are coefficients) for the $V_c$ histogram.

\begin{figure}[!tbp]
\gridline{\fig{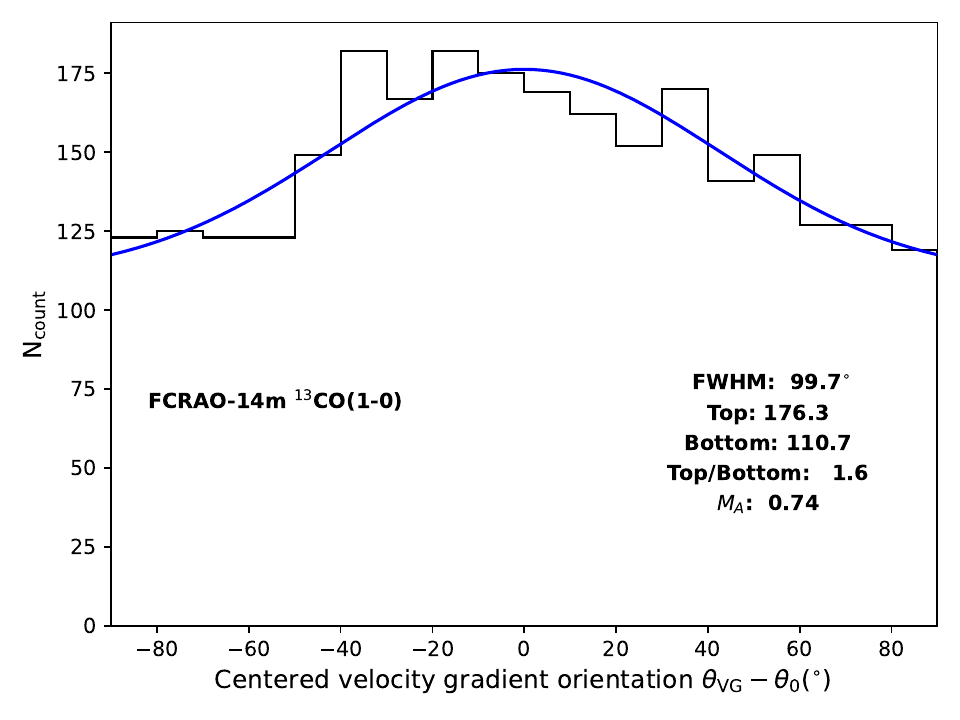}{0.45\textwidth}{}}
\caption{Histogram of the centered velocity gradient orientation for the FCRAO-14m $^{13}$CO (1-0) line data within the $r=15$ pc region. The blue line indicates the best-fitted result. \label{fig:vghist}}
\end{figure}

Figure \ref{fig:vghist} shows the histogram of the centered velocity gradient orientation ($\theta_{\mathrm{VG}} - \theta_{\mathrm{0}}$) for the FCRAO-14m $^{13}$CO (1-0) line data within the $r=15$ pc region centered at G28.34. We fit the histogram with a Gaussian profile and obtain $\mathcal{M}_{A}=0.74$, which is slightly sub-Alfv\'{e}nic.

Within molecular clouds, the velocity gradient could be significantly affected by gravitational effects and star formation activities, which makes the property of velocity gradient statistics overall deviate from what is expected for pure MHD turbulence. Thus we refrain from applying the VGT to smaller scales. 
%Also, the size-to-resolution ratio of the JCMT and ALMA line data is not sufficient for a VGT analysis. 

\subsubsection{Magnetic field orientation} \label{sec:VChG}
The Planck polarization map probes all the LOS structures, which makes it hard to separate the emission from the cloud environment and the foreground/background. Here we independently derive the magnetic field orientation from the FCRAO-14m $^{13}$CO (1-0) data with the VGT to confirm whether the Planck polarization toward G28.34 mainly traces emission from the cloud environment. The basic principle behind VGT is that the magnetized turbulence eddy is anisotropic and elongated along the magnetic field, so the velocity gradient of MHD turbulence should be perpendicular to the magnetic field \citep{1995ApJ...438..763G} in the absence of gravity. By selecting different velocity ranges, this technique can separate LOS velocity components corresponding to different distances, thus avoiding contamination from the foreground/background.

We implement the velocity channel gradients \citep[VChG,][]{2018ApJ...853...96L} on the FCRAO-14m data with approaches similar to \citet{2022MNRAS.511..829H} and \citet{2023MNRAS.524.2379H}: (1) We convolve the line intensity map at each velocity channel with a Sobel kernel to obtain the intensity gradient map of each channel (i.e., channel gradient). Pixels with S/N$<$5 are not considered in this step. (2) In the POS, we bin the Planck map over 3$\times$3 pixels and perform sub-block averaging \citep{2017ApJ...837L..24Y} for the line channel gradients within the area of each binned Planck pixel. The ratio between the block size and the line spatial resolution is $\sim$6.7, which is similar to previous VGT studies \citep[e.g.,][]{2022MNRAS.511..829H} and is sufficient for the statistics required by sub-block averaging. With a circular Gaussian fit for the histogram of channel gradients, we find the averaged gradient orientation $\Psi_{gs}$ for each block. Note that due to sub-block averaging, the resolution of the spectral data is degraded to the block size ($\sim$5$\arcmin$). (3) Integrating channels along the LOS, we calculate pseudo-Stokes parameters $Q_g$ and $U_g$ with 
\begin{equation}
Q_g = \Sigma_i^{N_{\mathrm{ch}}} I_i \cos (2\Psi_{gs,i}),
\end{equation}
\begin{equation}
U_g = \Sigma_i^{N_{\mathrm{ch}}} I_i \sin (2\Psi_{gs,i}),
\end{equation}
where $I_i$ is the block-averaged intensity. Here we integrate the velocity channels within three different velocity ranges: the cloud velocity range (from 71 to 86 km s$^{-1}$), $\pm$5 km s$^{-1}$  (from 74 to 84 km s$^{-1}$), and $\pm$3 km s$^{-1}$ (from 76 to 82 km s$^{-1}$). (4) The orientation of VChG is given by $\theta_{\mathrm{VChG}} = 0.5 \arctan(U_g/Q_g)$. The pseudo-magnetic field orientation is then given by $\theta_{\mathrm{B,VGT}} = \theta_{\mathrm{VChG}} + 90\degr$. 
% within the cloud velocity range (from 71 to 86 km s$^{-1}$).  (between 71 and 86 km s$^{-1}$ in our case)

\begin{figure*}[!htbp]
\gridline{\fig{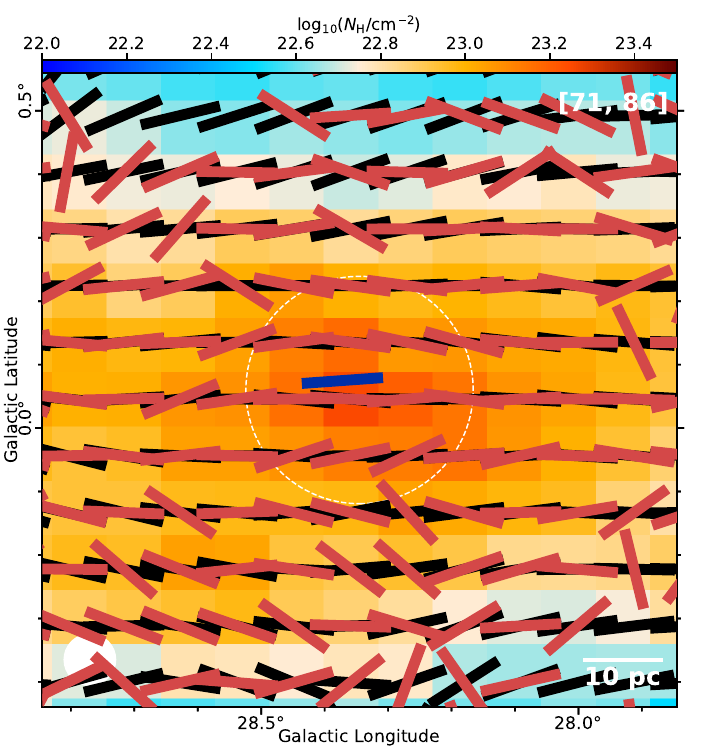}{0.33\textwidth}{}
 \fig{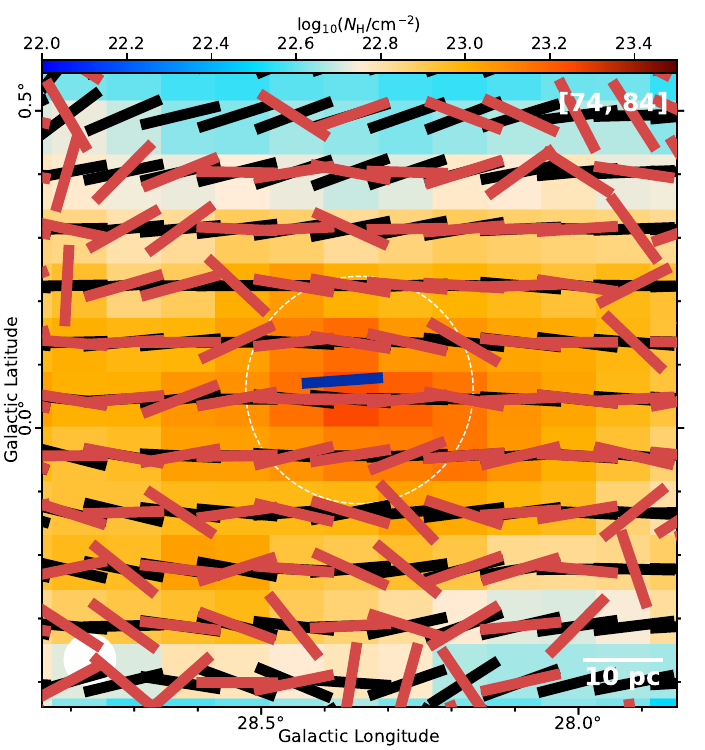}{0.33\textwidth}{}
 \fig{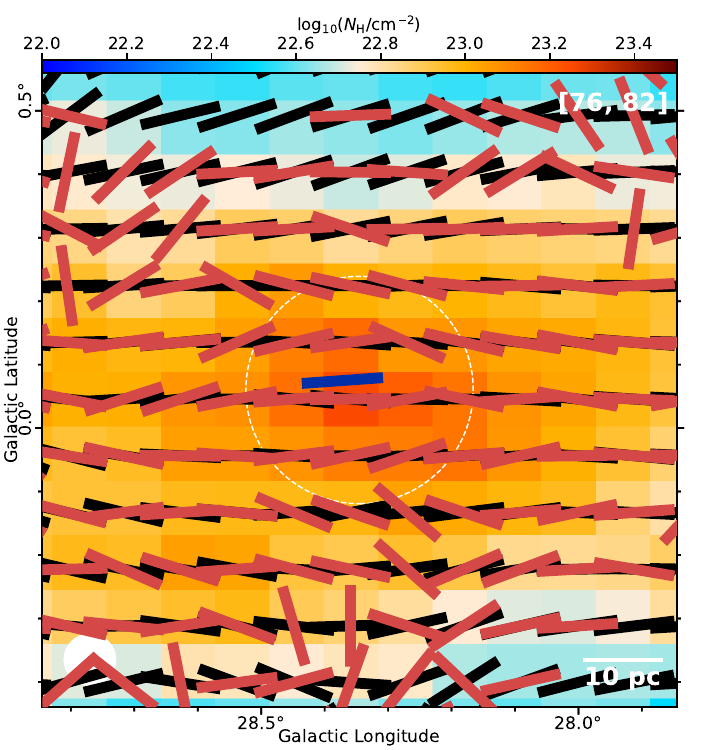}{0.33\textwidth}{}
 }
\caption{Pseudo magnetic field orientations inferred from VChG (red lines) within different velocity ranges (indicated in each panel) overlaid on the column density map. The Magnetic field orientations revealed by the Planck dust polarization are shown in black lines. The blue line indicates the average magnetic field orientation of the convolved JCMT polarization map. Line lengths are arbitrary. The white dashed circle marks the $r=15$ pc region surrounding G28.34 within which we perform the DCF analysis. Within $r=15$ pc, the pseudo magnetic field orientations derived from VGT within different velocity ranges are very similar.  \label{fig:vchg}}
\end{figure*}

Figure \ref{fig:vchg} compares the magnetic field orientations inferred from the VChG with those traced by the Planck dust polarization. The two orientations are well aligned with each other in bright regions near G28.34. As we only select the cloud velocity range in the VGT analysis, the field orientation inferred from VChG should not contain contributions from the foreground/background. Moreover, we convolve the JCMT maps to the same resolution as Planck, then estimate the average polarization position angle within 3$\arcmin$ of the peak of the convolved total intensity map. The magnetic field orientation measured from the convolved JCMT map is only 9$\degr$ different from the Planck measurement and 7$\degr$ different from the VGT measurement at the nearest pixel (see Figure \ref{fig:vchg}). The close alignment between measurements from VChG, JCMT, and Planck tends to suggest that the Planck dust polarization toward G28.34 mainly traces the emission from G28.34 and its close environment, but not the foreground/background. \citet{2017arXiv170303026Y} suggested that the magnetic field orientation and the block-averaged velocity gradient could transit to a perpendicular alignment (i.e., re-rotation) when gravity becomes important in high-density regions. Although there are some slight angular differences between the field orientations revealed by VChG and Planck, overall there is a lack of evidence for perpendicular alignments toward the brightest pixels. This suggests that gravitational motions do not dominate in the large-scale diffuse gas. On the other hand, the field orientations traced by VChG and Planck show larger differences in weak-emission regions, which may suggest that the Planck dust polarization mainly traces the materials outside of the cloud velocity range in those regions or those weak-emission areas are significantly influenced by noise. 

In Section \ref{sec:largeB}, we used the angular dispersion measured from the Planck polarization data to estimate the magnetic field strength of the cloud environment. Alternatively, if we use the angular dispersion from the VGT measurements in the DCF calculations, the estimated magnetic field strength will decrease by a factor of $\sim$3. However, this alternative field strength estimation should be regarded as a lower limit. This is because the angular dispersion from VGT is expected to be affected by ordered velocity fields. i.e., although not dominant, the gravitational motions may have already contributed to the orientations of VChG, which increases the angular dispersion of VGT measurements. 

\subsection{Relative orientation analysis}\label{sec:roa}

Here we investigate the multi-scale physical properties of G28.34 with a synergistic local relative orientation analysis \citep{2023ApJ...945..160L} combining the approaches of the polarization-intensity gradient (Koch-Tang-Ho or KTH) method \citep{2012ApJ...747...79K} and the Histogram of Relative Orientation (HRO) analysis \citep[][]{2013ApJ...774..128S}. Basically, we characterize the relative angle between magnetic fields and other orientations (column density gradient and local gravity) with the alignment measure (AM) parameter \citep{2018ApJ...853...96L}. 
The $AM$ is given by
\begin{equation} \label{eq:am}
AM = \langle \cos (2\phi_{o1}^{o2}) \rangle,
\end{equation}
where $\phi_{\mathrm{o1}}^{\mathrm{o2}} = \vert \theta_{\mathrm{o1}} - \theta_{\mathrm{o2}} \vert$ is the angle between two orientations and is in the range of 0 to 90$\degr$. The uncertainty of $\phi_{\mathrm{o1}}^{\mathrm{o2}}$ is given by $\delta\phi_{\mathrm{o1}}^{\mathrm{o2}} = \sqrt{\delta\theta_{\mathrm{o1}}^2+\delta\theta_{\mathrm{o2}}^2}$.  Data points with $\delta\phi>10\degr$ are excluded from our analysis.
The uncertainty of $AM$ is given by \citep{2023ApJ...945..160L}
\begin{equation} \label{eq:dam}
\delta AM = \sqrt{(\langle (\cos (2\phi_{o1}^{o2}))^2 \rangle - AM^2 + \Sigma_i^{n'} (2\sin (2\phi_i) \delta \phi_i)^2)/n'},
\end{equation}
where $n'$ is the number of data points considered. For the Planck data, we consider area within $r=15$ pc centered at G28.34. For the JCMT data, we consider area with S/N($I$)$>$5 and S/N($PI$)$>$2. For the ALMA data, we consider area with S/N($I$)$>$3 and S/N($PI$)$>$2. We calculate AM in 6, 10, and 15 different column density bins for the Planck, JCMT, and ALMA data, respectively. The typical numbers of pixels are $\sim$20, $\sim$20-80, and $\sim$30-180 per bin for the Planck, JCMT, and ALMA data, respectively. 
% and FCRAO-14m /FCRAO-14m /FCRAO-14m

\subsubsection{Magnetic field versus column density gradient}\label{sec:B_NG}

\begin{figure*}[!htbp]
 \gridline{\fig{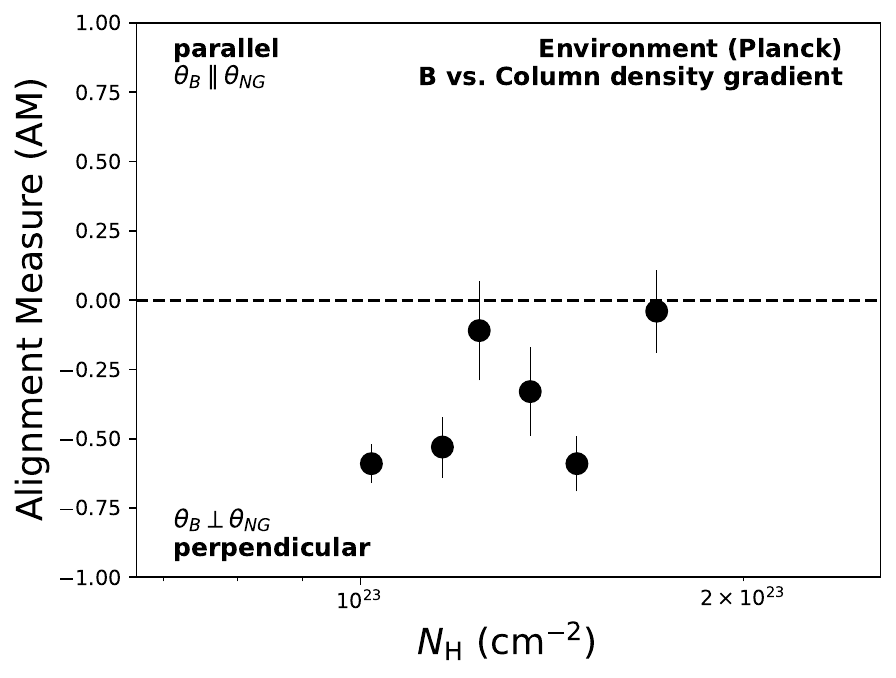}{0.33\textwidth}{}
 \fig{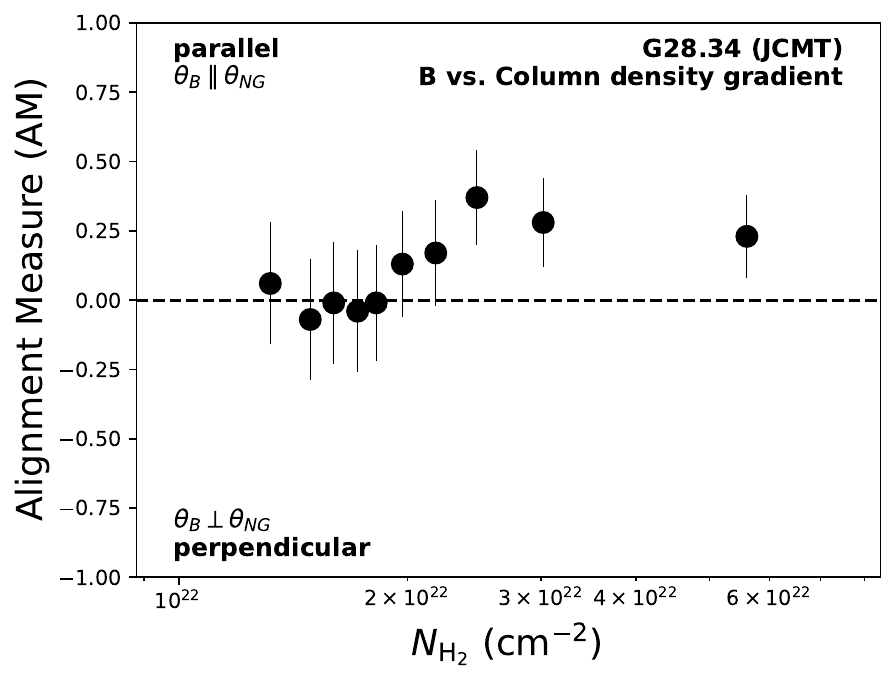}{0.33\textwidth}{}
 \fig{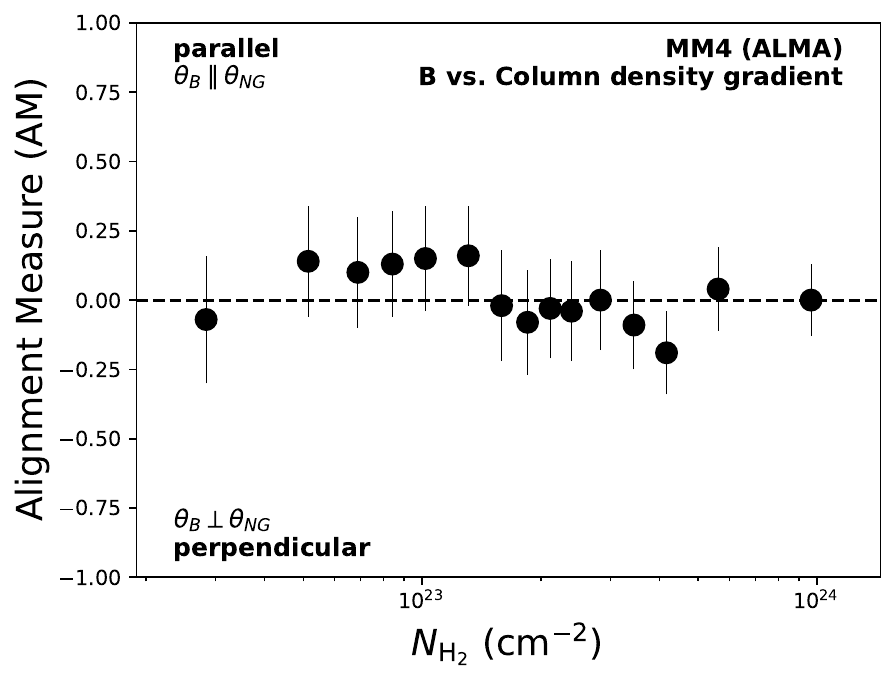}{0.33\textwidth}{}
 }
\caption{Relative orientations (characterized by $AM$. See Equation \ref{eq:am}) between magnetic field ($\theta_{\mathrm{B}}$) and column density gradient ($\theta_{\mathrm{NG}}$) as a function of column density for Planck (left), JCMT (middle), and ALMA (right) observations. Due to the filtering of large-scale emissions for JCMT and ALMA, the absolute column densities from different instruments are not comparable. $AM>0$ and $AM<0$ indicate a preferentially parallel and perpendicular alignment, respectively. \label{fig:G28_B_NG}}
\end{figure*}

Figure \ref{fig:G28_B_NG} shows the AM-N relation for the angle between the magnetic field ($\theta_{\mathrm{B}}$) and column density gradient ($\theta_{\mathrm{NG}}$). The column density gradient is obtained by applying a $3\times3$ Sobel kernel to the column density map. The uncertainty of $\theta_{\mathrm{NG}}$ is derived following \citet{2023ApJ...945..160L}. As $\theta_{\mathrm{NG}}$ is perpendicular to the column density contour, studying the AM-N relation for $\phi_{B}^{NG}$ is equivalent to the HRO analysis \citep{2013ApJ...774..128S}. As seen in Figure \ref{fig:G28_B_NG}, the relative orientation changes from a statistically slightly more perpendicular alignment ($AM_{B}^{NG} \lesssim 0$) in the environment at low column densities as revealed by Planck to a slightly more parallel alignment ($AM_{B}^{NG} \gtrsim 0$) in G28.34 at intermediate column densities as revealed by JCMT. This transition can only be reproduced in trans-to-sub-Alfv\'{e}nic simulations in numerical HRO studies \citep[see a review of the HRO studies in][]{2022FrASS...9.3556L}, which suggests that G28.34 is in a trans-to-sub-Alfv\'{e}nic environment. This is in agreement with the result of our VGT analysis in Section \ref{sec:VGT}. The reasons for the different alignment at different column densities and the transition of alignment are still under debate \citep{2022FrASS...9.3556L}. As discussed below (see Section \ref{sec:B_LG}), the transition to $AM_{B}^{NG} \gtrsim 0$ may be related to the distortion of gravity. At high column densities revealed by ALMA, the two angles tend to transit back to $AM_{B}^{NG}\sim 0$ as $N$ increases, which may be due to the influence of early massive star formation activities (e.g., infall, rotation, outflows, accretion, and et al.). Overall, the alignment between $\theta_{\mathrm{B}}$ and $\theta_{\mathrm{NG}}$ in IRDC G28.34 at different column densities is very similar to what we have found in the evolved massive star formation region NGC 6334 \citep{2023ApJ...945..160L}. 

\subsubsection{Magnetic field versus local gravity}\label{sec:B_LG}

\begin{figure*}[!htbp]
 \gridline{\fig{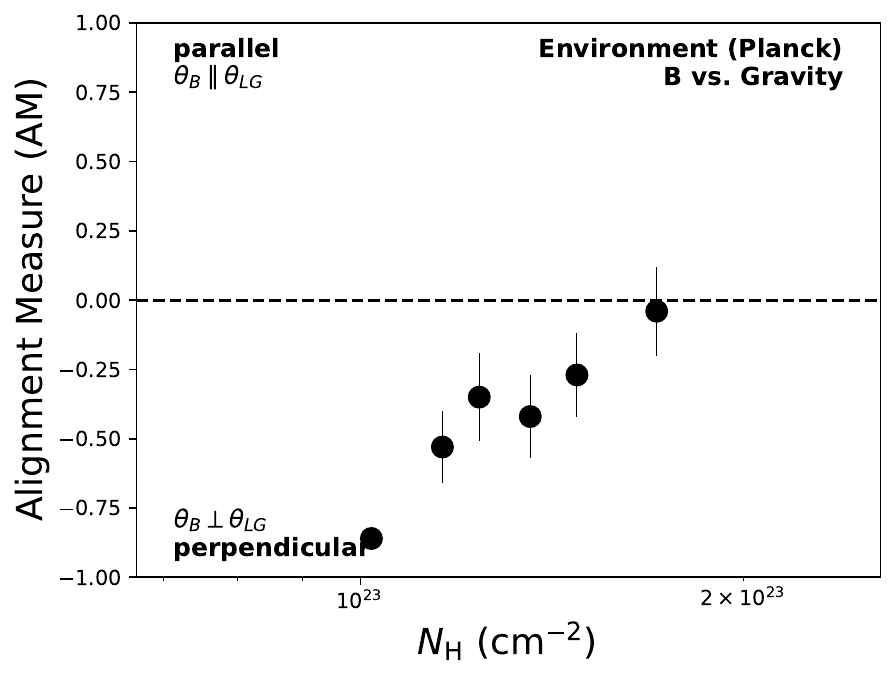}{0.33\textwidth}{}
 \fig{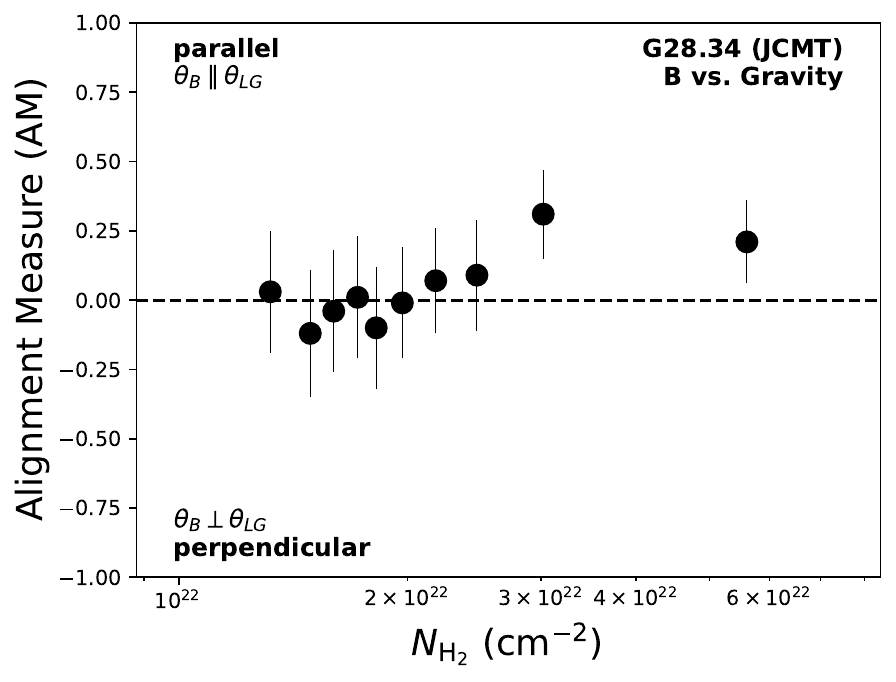}{0.33\textwidth}{}
 \fig{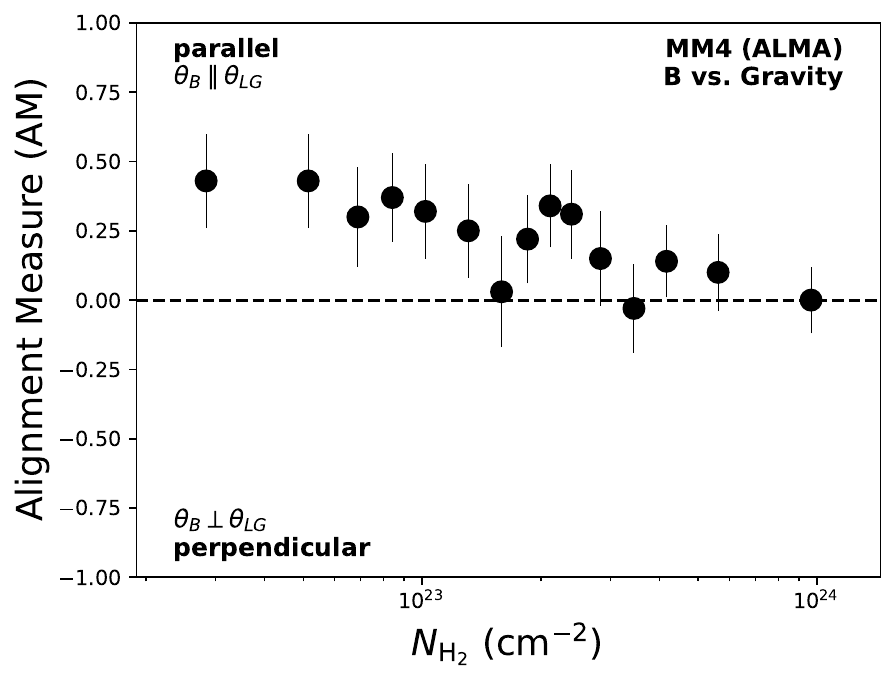}{0.33\textwidth}{}
 }
\caption{Relative orientations (characterized by $AM$) between magnetic field ($\theta_{\mathrm{B}}$) and local gravity ($\theta_{\mathrm{LG}}$) as a function of column density for Planck (left), JCMT (middle), and ALMA (right) observations. \label{fig:G28_B_LG}}
\end{figure*}

Figure \ref{fig:G28_B_LG} shows the AM-N relation for the angle between the magnetic field ($\theta_{\mathrm{B}}$) and local gravity ($\theta_{\mathrm{LG}}$). Considering gas mass at S/N($I$)$>$3, the 2D local gravity direction ($\theta_{\mathrm{LG}}$) at position $\boldsymbol{r_j}$ is calculated with the standard formula of gravitation:
\begin{equation}
\boldsymbol{g_{j}(r)} = G \sum\limits_{i=1}^{n} \frac{m_{i}m_{j}}{\vert \boldsymbol{r_j} - \boldsymbol{r_i} \vert ^2} \boldsymbol{e_{ij}},
\end{equation}
where $\boldsymbol{e_{ij}}$ is the direction between position $\boldsymbol{r_i}$ and $\boldsymbol{r_j}$, $G$ is the gravitational constant, $m_{j}$ is the mass at position $\boldsymbol{r_j}$, and $\theta_{\mathrm{LG}}$ is the direction of $\boldsymbol{g_{j}(r)}$.
Overall, the AM-N trend for $\phi_{B}^{LG}$ is similar to the trend for $\phi_{B}^{NG}$, where $AM_{B}^{LG}$ transits from a statistically more perpendicular alignment to a statistically more parallel alignment, then transit back to $AM_{B}^{LG}\sim0$ as $N$ increases. As $\phi_{B}^{LG}$ directly traces the correlation between magnetic fields and gravity, the observed AM-N trend for $\phi_{B}^{LG}$ suggests that the magnetic field is resisting gravitational distortion at lower density, but is dragged by gravity as density increases within the cloud, and might be affected by early massive star formation activities near the central young stellar objects at even higher densities \citep{2023ApJ...945..160L}. 
%\citep[e.g.,][]{2012ApJ...747...79K, 2018ApJ...855...39K, 2020ApJ...895..142L, 2022ApJ...940...89K, 2022ApJ...931..115W, 2023ApJ...945..160L}

\subsubsection{Normalised mass-to-flux ratio}\label{sec:lambda_KTH}

Based on ideal MHD equations, \citet{2012ApJ...747...80K} suggested that the normalized mass-to-flux ratio can be estimated with
\begin{equation}\label{eq:lambdaKTH}
    \lambda_{\mathrm{KTH}} = \langle \Sigma_B^{-1/2} \rangle \pi^{-1/2},
\end{equation}
where $\Sigma_B$ is the local ratio between the magnetic field force ($F_{B}$) and the gravitational force ($F_{G}$). $\Sigma_B$ can be estimated with
\begin{equation}\label{eq:sigmaB}
    \Sigma_B = \frac{\sin \phi_{LG}^{NG}}{\sin (90\degr - \phi_{B}^{NG})} = \frac{F_B}{\vert F_G\vert}
\end{equation}
if the hydrostatic gas pressure is negligible.  

\begin{figure}[!htbp]
 %\gridline{\fig{./A_N6334_lambda.pdf}{0.48\textwidth}{}
 %}
  \gridline{\fig{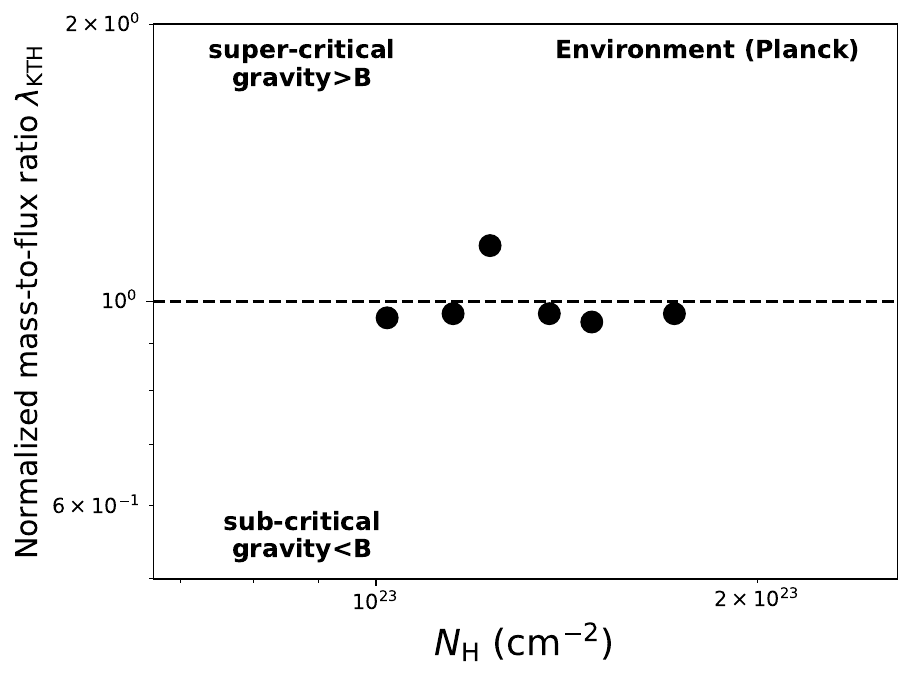}{0.33\textwidth}{}
 \fig{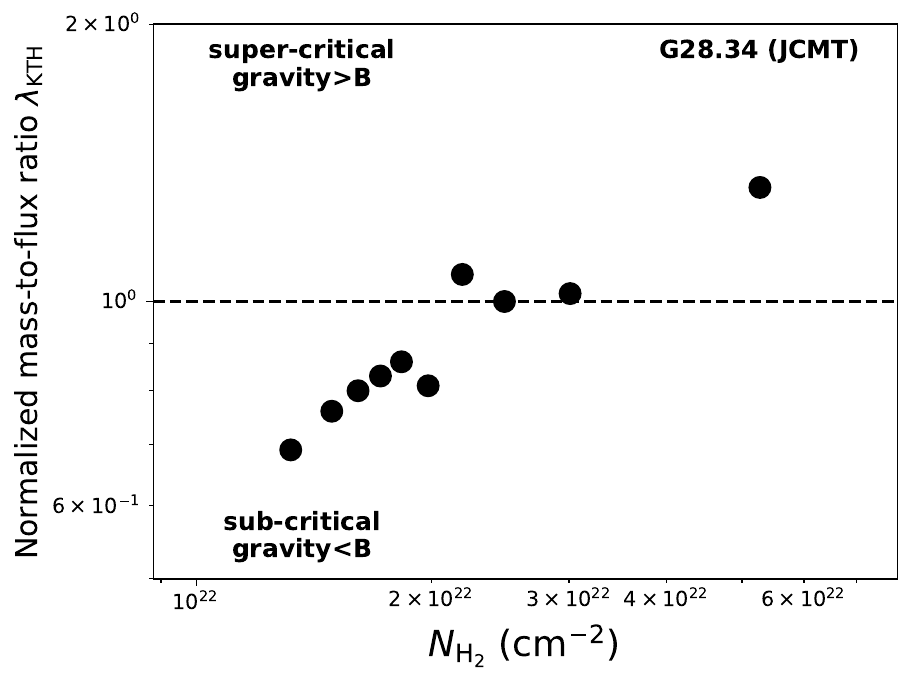}{0.33\textwidth}{}
 \fig{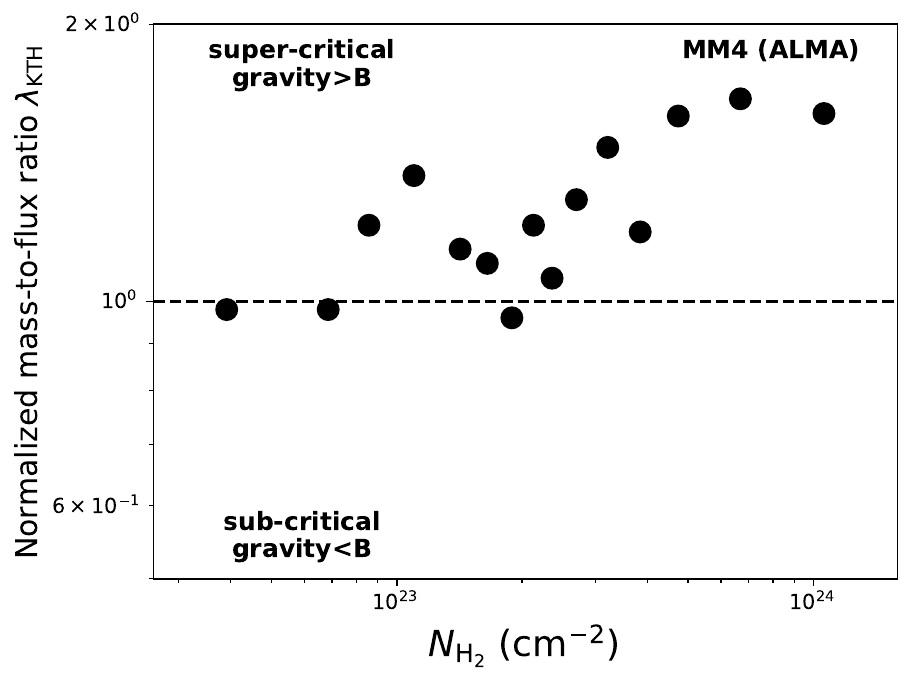}{0.33\textwidth}{}
 }
\caption{Normalised mass-to-flux ratio derived from the KTH method as a function of column density for Planck (left), JCMT (middle), and ALMA (right) observations. \label{fig:G28_lambda_kth}}
\end{figure}

Figure \ref{fig:G28_lambda_kth} shows the $\lambda_{\mathrm{KTH}}$ derived from the KTH method. For the Planck data, there is $\lambda_{\mathrm{KTH}} \sim 1$, which suggests a magnetically trans-critical state in the environment. For the JCMT data, the $\lambda_{\mathrm{KTH}}$ increases with $N$ and transits from $\lambda_{\mathrm{KTH}}<1$ to $\lambda_{\mathrm{KTH}}>1$. The discrepancy between the Planck and JCMT data might be because the JCMT data filters out the large-scale emission and underestimates the gravitational force at lower column densities. For the ALMA data, overall there is $\lambda_{\mathrm{KTH}} \gtrsim 1$. However, the magnetic field may be affected by star formation feedback from central young stellar objects, which could make the KTH method not applicable to the ALMA data. Note that the systematic uncertainty of the $\lambda_{\mathrm{KTH}}$ value estimated with the KTH method is unclear due to the lack of numerical tests. 

\section{Discussion} \label{sec:discussion}

\subsection{Equilibrium state}
Different star formation theories \citep[e.g.,][]{1997MNRAS.285..201B, 2002Natur.416...59M} have distinct predictions over the energy balance between gravity, magnetic fields, and turbulence in different scales and evolutionary stages of massive star formation. Observationally determining the energy budget of massive star formation regions is key to distinguishing between those theories. Here we use the virial theorem to characterize the multi-scale equilibrium state of G28.34. Neglecting the surface term, tension term, and ordered velocity motion, the virial theorem is written as
\begin{equation}
\frac{1}{2}\frac{d^2 I}{dt^2} = E_\mathrm{G} + 2E_{\mathrm{th}} + 2E_{\mathrm{turb}} + E_\mathrm{B},
\end{equation}
where $I$ is the moment of inertia, $E_\mathrm{G}$ is the gravitational energy, $E_{\mathrm{th}}$ is the thermal energy, $E_{\mathrm{turb}}$ is the turbulent kinetic energy, and $E_\mathrm{B}$ is the magnetic energy. $\frac{1}{2}\frac{d^2 I}{dt^2}\gtrsim0$ indicates a supported state in Quasi-Equilibrium, while $\frac{1}{2}\frac{d^2 I}{dt^2}<0$ indicates a dynamical collapsing state. For a spherical structure with density profile $n\propto r^{-i}$, the gravitational energy is given by $E_{\mathrm{G}} = -GM^2/(k_iR)$, where $k_i = (5-2i)/(3-i)$, $R$ is the radius, and $G$ is the gravitational constant. The thermal energy is given by $E_{\mathrm{th}} = 1.5 n k_{B} T V$, where $V=4\pi R^3/3$ is the Volume. The turbulent kinetic energy is given by $E_{\mathrm{turb}} = 1.5 M \sigma_{turb}^2$, where $\sigma_{\mathrm{turb}}$ is the 1D turbulent velocity dispersion. The magnetic energy is given by $E_{\mathrm{B}} = B^2 V/(2\mu_0)$.

%compare with Liu 2022a,b IRDCs

\subsubsection{Gravity vs Thermal force}
The relative importance between gravity and the thermal term in the virial theorem can be characterized with the ratio $2E_{\mathrm{th}}/\vert E_\mathrm{G} \vert$. Using the fitted mass profile $M\propto r^{1.59}$ and density profile $n\propto r^{-1.41}$  (see Section \ref{sec:N}) and assuming a constant temperature ($T=15$ K), we have derived this thermal-to-gravitational ratio as a function of radius. 

\begin{figure}[!htbp]
 \gridline{\fig{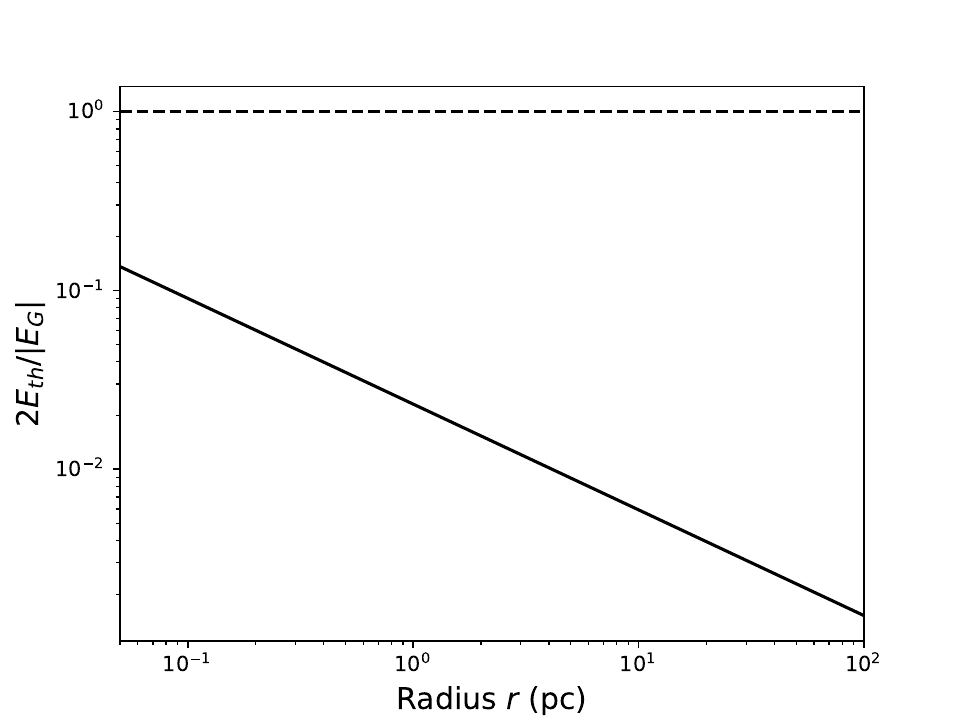}{0.48\textwidth}{(a)}
 \fig{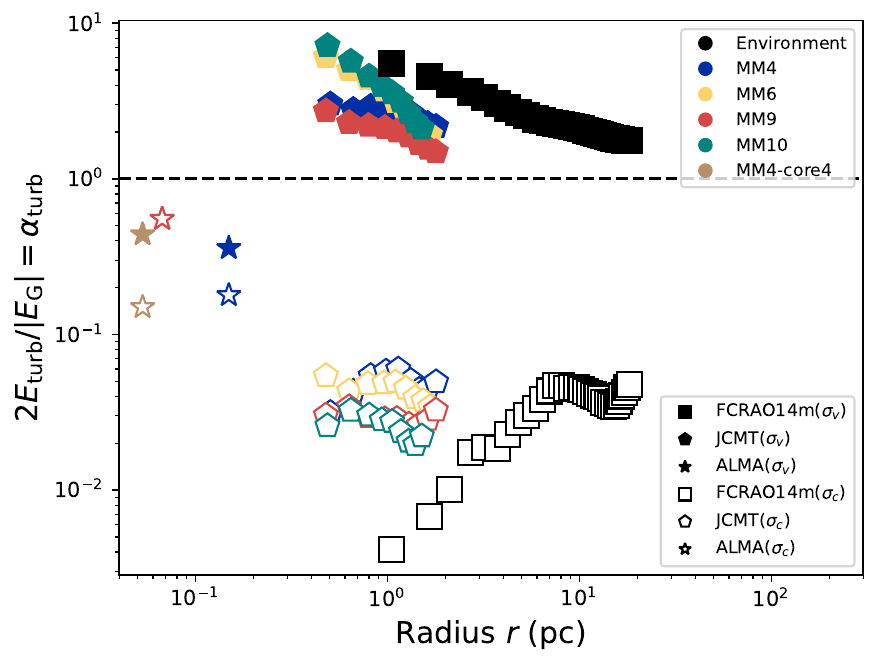}{0.44\textwidth}{(b)}
 }
  \gridline{\fig{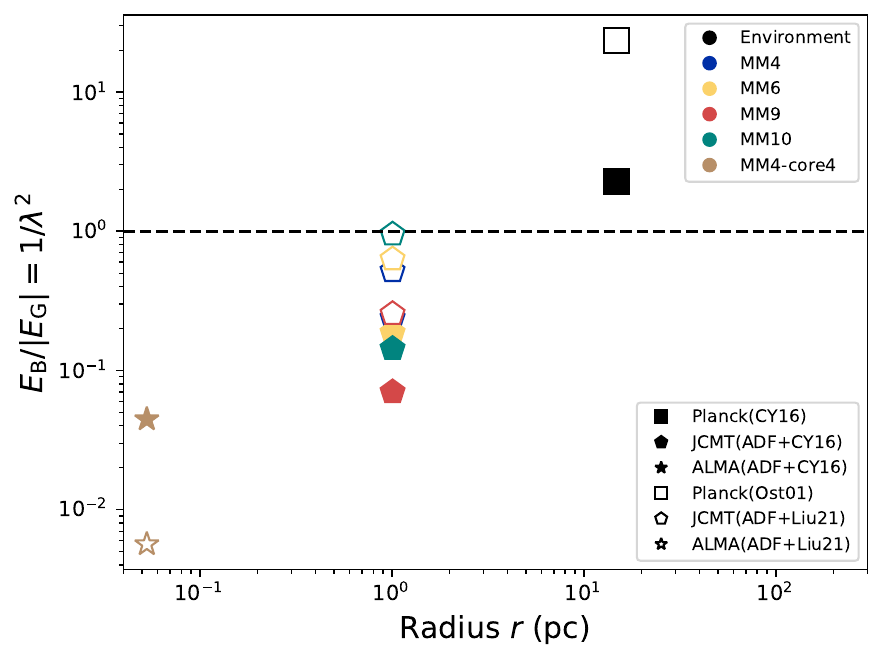}{0.44\textwidth}{(c)}
  }
\caption{(a). Thermal-to-gravitational ratio $2E_{\mathrm{th}}/\vert E_\mathrm{G} \vert$ as a function of radius. The horizontal dashed line indicates a balance between the thermal term $2E_{\mathrm{th}}$ and gravitational term $E_\mathrm{G}$ in the virial equation. (b). Turbulent-to-gravitational ratio $2E_{\mathrm{turb}}/\vert E_\mathrm{G} \vert = \alpha_{\mathrm{turb}}$ (i.e., turbulent virial parameter) as a function of radius. Different colors represent different regions. Different symbols indicate different instruments. Close symbols indicate $\alpha_{\mathrm{turb}}$ estimated with LOS velocity dispersion $\sigma_v$, while open symbols indicate $\alpha_{\mathrm{turb}}$ estimated with velocity centroid dispersion $\sigma_c$. The horizontal dashed line indicates a balance between the turbulent term $2E_{\mathrm{turb}}$ and gravitational term $E_\mathrm{G}$ in the virial equation. (c). Magnetic-to-gravitational ratio $E_{\mathrm{B}}/\vert E_\mathrm{G} \vert = 1/\lambda^2$ as a function of radius. Different colors represent different regions. Different symbols indicate different instruments and modified DCF methods. The horizontal dashed line indicates a balance between magnetic energy and gravitational energy.  \label{fig:G28_E_EG_r}}
\end{figure}

Figure \ref{fig:G28_E_EG_r}(a) shows the derived thermal-to-gravitational ratio as a function of radius $r$. The thermal force is negligible compared to the gravitational force throughout the scales of our interest which is critical for star formation. This is reasonable because the thermal energy is usually less dominant compared to other forces in massive star formation \citep{2014prpl.conf..149T}. There is a trend that $2E_{\mathrm{th}}/\vert E_\mathrm{G} \vert$ increases as $r$ decreases. At a very small scale near the central Young Stellar Objects, the thermal force might surpass the gravitational force and establish an equilibrium state solely supported by thermal pressure. Because the power-law indexes for the mass profile and density profile may change at small scales, higher-resolution observations are required to study whether a small-scale equilibrium between thermal force and gravity could be achieved, and if so, at what scale this equilibrium might occur.

\subsubsection{Gravity vs Turbulence}
The relative importance between gravity and turbulence is usually characterized by the turbulent virial parameter $\alpha_{\mathrm{turb}} = 2E_{\mathrm{turb}}/\vert E_{\mathrm{G}} \vert = M_{\mathrm{turb}}/M$ \citep{1992ApJ...395..140B}. For a spherical structure with density profile $n\propto r^{-i}$, the turbulent virial mass $M_{\mathrm{turb}}$ is given by 
\begin{equation}
M_{\mathrm{turb}} = \frac{3k_i R \sigma_{\mathrm{turb}}^2}{G}.
\end{equation}
$\alpha_{\mathrm{turb}}<1$ suggests the turbulence cannot solely resist gravitational collapse (i.e., sub-virial), and vice versa. 

We have calculated the turbulent virial parameter $\alpha_{\mathrm{turb}}$ for the multi-scale structures of G28.34. For the density profile, we adopt $i=1.41$ (see Section \ref{sec:N}). For the clumps and cores, we adopt the mass estimated in Section \ref{sec:N}. For the environmental gas, we adopt the mass extrapolated from the fitted mass-radius relation $M\propto r^{1.59}$ (see Section \ref{sec:N}). As discussed in Section \ref{sec:line}, the LOS velocity dispersion $\sigma_v$ tends to overestimate the actual 1D velocity dispersion at scale $r$ because the LOS substructures are superposed and the LOS depth is larger than $r$, while the velocity centroid dispersion $\sigma_c$ tends to underestimate the actual 1D velocity dispersion at scale $r$ due to the LOS averaging. As it is hard to derive the pure turbulent velocity dispersion corresponding to a specific scale ($r$) in molecular clouds, we adopt the LOS velocity dispersion $\sigma_v$ as the upper limit of the 1D turbulent velocity dispersion and the velocity centroid dispersion $\sigma_c$ as the lower limit, with the assumption that the non-turbulent part of $\sigma_c$ is much weaker than the turbulent part. For the JCMT line data, we only adopt $\sigma_v$ and $\sigma_c$ estimated from $^{13}$CO (3-2) data as its critical density is closer to the clump densities. 

Figure \ref{fig:G28_E_EG_r}(b) shows the estimated turbulent virial parameter $\alpha_{\mathrm{turb}}$ as a function of radius $r$. It is clear that the cores in MM4 and MM9 are sub-virial. For the clumps, the cloud, and the environmental gas, the $\alpha_{\mathrm{turb}}$ estimated with $\sigma_v$ is super-virial, while the $\alpha_{\mathrm{turb}}$ estimated with $\sigma_c$ is sub-virial. Due to the uncertainties in the turbulent velocity dispersion, we are unable to determine whether these large-scale structures are turbulence-supported or not. 
%turb and non-thermal

\subsubsection{Gravity vs Magnetic fields}
The relative significance of gravity and magnetic fields can be assessed using their energy ratio $E_{\mathrm{B}}/\vert E_{\mathrm{G}} \vert$. Alternatively, this comparison is more usually expressed by the normalized mass-to-flux ratio $\lambda$ \citep{2004ApJ...600..279C}. For a spherical structure with density profile $n\propto r^{-i}$, $\lambda$ is given by \citep{2022ApJ...925...30L}
\begin{equation} \label{eq:lambliu22}
\lambda = \mu_{\mathrm{H_2}} m_{\mathrm{H}} \sqrt{1.5\mu_0\pi G/k_i} \frac{N_{\mathrm{H_2}}}{B_{3d}}. 
\end{equation}
$\lambda > 1$ indicates magnetic fields cannot solely resist gravitational collapse (i.e., magnetically super-critical), and vice versa. For a spherical structure, the relation between $\lambda$ and the magnetic-to-gravitational energy ratio is $E_{\mathrm{B}}/\vert E_{\mathrm{G}} \vert = 1/\lambda^2$.
%Alternatively, the magnetic virial parameter $\alpha_{B} \sim 1/\lambda$ can also be used to compare gravity and magnetic fields. 

With the magnetic field strengths estimated in Section \ref{sec: B}, we have calculated the energy ratio $E_{\mathrm{B}}/\vert E_{\mathrm{G}}\vert$ as well as the normalized mass-to-flux ratio $\lambda$ for the environmental gas within 15-pc, the clumps, and MM4-core4. In the calculation, the POS total magnetic field $B$ is converted to the 3D total magnetic field $B_{3d}$ with the relation $B_{3d} \sim B \times 1.25$ \citep{2022FrASS...9.3556L}. Similarly, we adopt $i=1.41$ for the density profile and adopt the extrapolated mass and density for the environmental gas. Table \ref{tab:sumphy} summarizes the calculated $\lambda$ values.

\begin{deluxetable}{ccccccc}[t!]
\tablecaption{Summary of physical parameters \label{tab:sumphy}}
\tablecolumns{7}
\tablewidth{0pt}
\tablehead{
\colhead{Region} &
\colhead{$n$} &
\colhead{$N$} & 
\colhead{$r$} &
\colhead{$M$} & 
\colhead{$B$  \tablenotemark{a}} & 
\colhead{$\lambda$  \tablenotemark{a}}  \\
\colhead{} &  \colhead{(cm$^{-3}$)} & \colhead{(cm$^{-2}$)} &  \colhead{(pc)}  & \colhead{($M_{\odot}$)} & \colhead{(mG)}  & \colhead{}
}
\startdata
MM4-core4 &   1.1 $\times$ 10$^{6}$ &   1.5 $\times$ 10$^{23}$ & 0.053 &       43.0  &  0.29, 0.10  &  3.32, 9.32 \\ \hline
  MM4 &   6.3 $\times$ 10$^{3}$ &   2.6 $\times$ 10$^{22}$ &     1 &     1875.8  &  0.08, 0.12  &  2.09, 1.41 \\ 
  MM6 &   5.8 $\times$ 10$^{3}$ &   2.4 $\times$ 10$^{22}$ &     1 &     1711.6  &  0.07, 0.12  &  2.36, 1.28 \\ 
  MM9 &   5.7 $\times$ 10$^{3}$ &   2.4 $\times$ 10$^{22}$ &     1 &     1681.9  &  0.04, 0.08  &  3.85, 2.02 \\ 
 MM10 &   6.0 $\times$ 10$^{3}$ &   2.5 $\times$ 10$^{22}$ &     1 &     1789.2  &  0.06, 0.15  &  2.69, 1.04 \\ \hline
Environment &   1.4 $\times$ 10$^{2}$ &   8.6 $\times$ 10$^{21}$ &    15 &   133159.6  &  0.07, 0.27  &  0.76, 0.21 \\ 
\enddata
\tablenotetext{a}{The former value adopted the CY16 correction and the latter value adopted the Liu21 or Ost01 correction.} 
%\tablenotetext{b}{Extrapolated from the mass-radius and density-radius relations at smaller scales.}

\end{deluxetable}

\begin{figure}[!htbp]
  \gridline{\fig{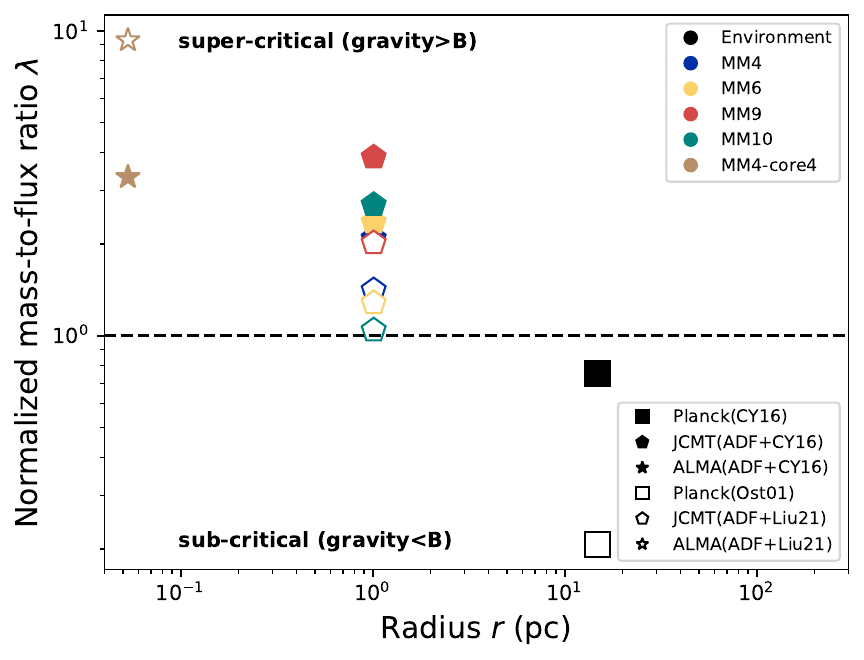}{0.45\textwidth}{(a)}
 \fig{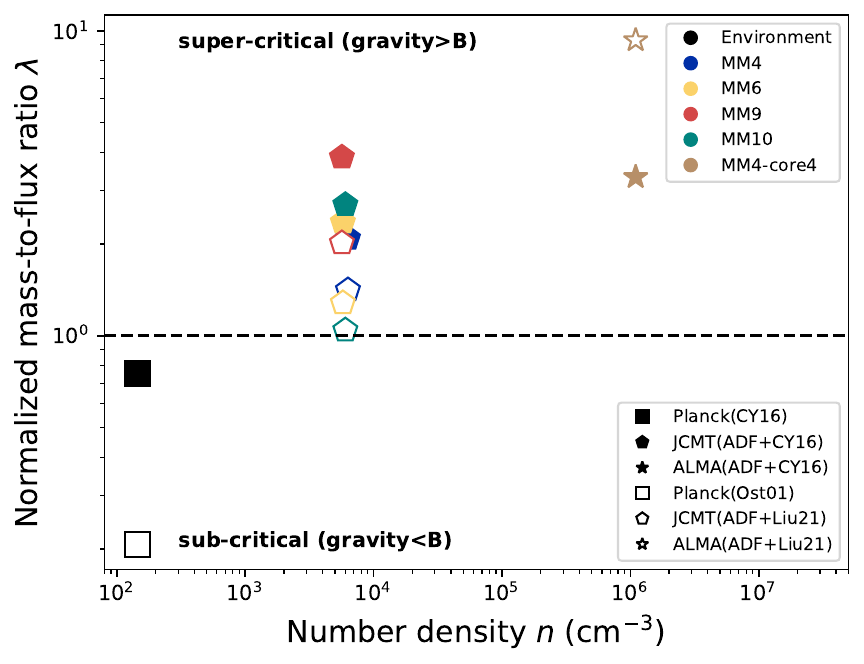}{0.45\textwidth}{(b)}
 }
  \gridline{\fig{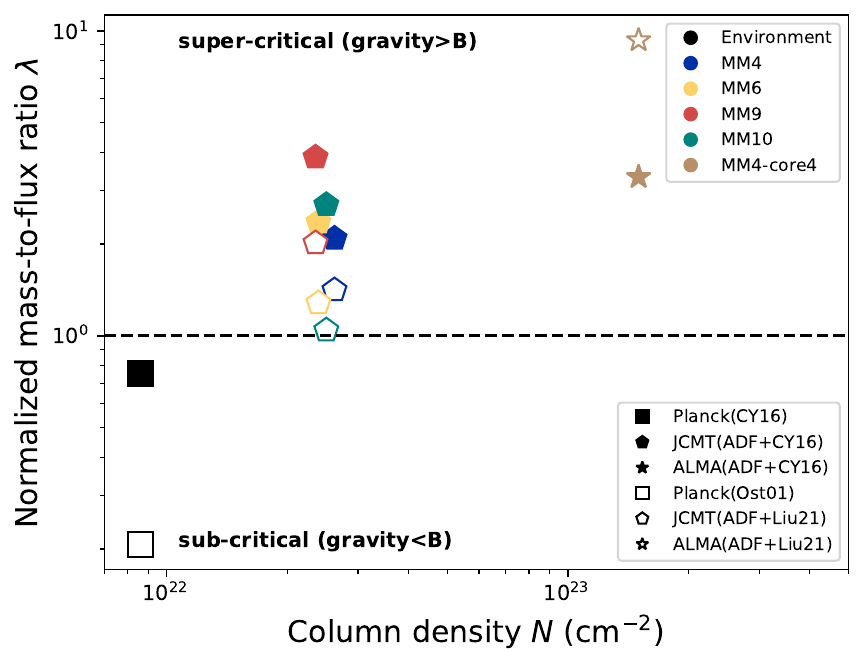}{0.45\textwidth}{(c)}
  \fig{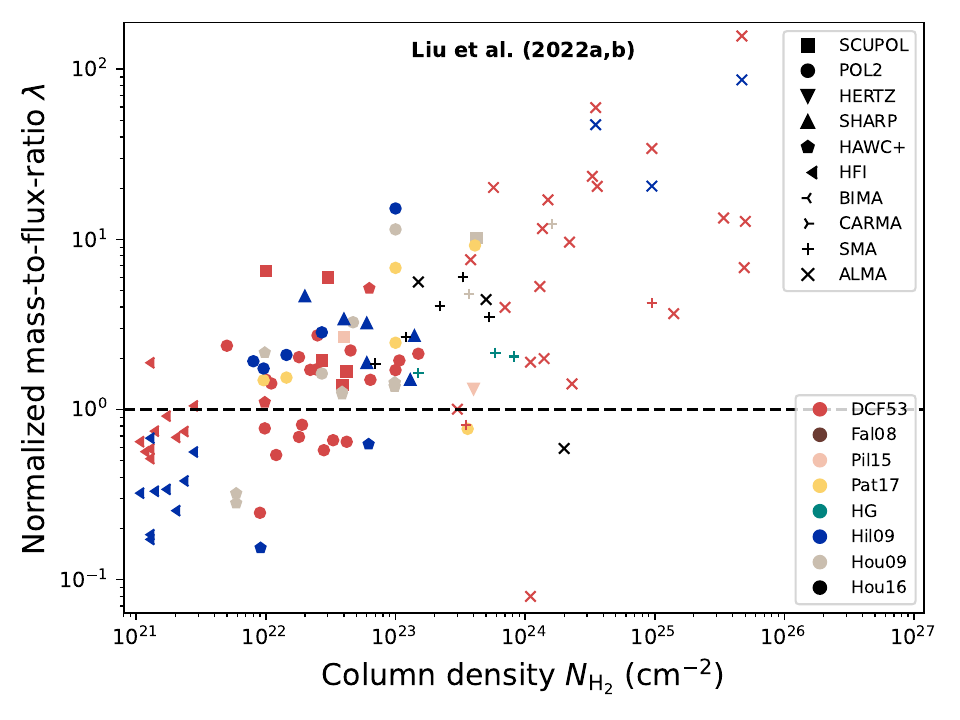}{0.465\textwidth}{(d)} 
 }
\caption{(a)-(c). Normalised mass-to-flux ratio derived from the DCF method as functions of radius, number density, and column density for Planck, JCMT, and ALMA observations toward G28.34. Different colors represent different regions. Different symbols indicate different instruments and modified methods. The horizontal dashed line indicates an energy balance between magnetic energy and gravitational energy (i.e., $\lambda=1$). (d). Normalised mass-to-flux ratio derived from the compilation of DCF estimations of different star formation regions in the literature before June 2021 \citep[reproduced with permission. See reviews in][and see references therein]{2022ApJ...925...30L, 2022FrASS...9.3556L}. Different colors represent different DCF variants and different symbols indicate different observational instruments \citep[see Figure 3 of][]{2022FrASS...9.3556L}. 
 \label{fig:G28_lambda_dcf}}
\end{figure}

Figure \ref{fig:G28_E_EG_r}(c) shows the energy ratio $E_{\mathrm{B}}/\vert E_{\mathrm{G}}\vert$ as a function of radius. As most previous studies have used $\lambda$ to compare magnetic fields with gravity, we also show the estimated $\lambda$ values as functions of $r$, $n$, and $N$ in Figures \ref{fig:G28_lambda_dcf}(a)-(c). We see that $\lambda$ increases with increasing density and decreases with increasing radius. This trend is consistent with that derived from the KTH method (see Figure \ref{fig:G28_lambda_kth}). The environmental gas tends to be magnetically sub-critical, which means the magnetic field is strong enough to resist gravitational collapse in the large-scale diffuse gas. In the clumps/cores (as revealed by JCMT and ALMA), the state transits to super-critical, allowing gravitational collapse to happen. The transition of the magnetic critical state occurs at the cloud-to-clump scale. At the core scale, the magnetic support is notably weaker, even weaker than the thermal support (see Figures \ref{fig:G28_E_EG_r}(a) and (c)). Note that although the magnetic field strengths estimated with the CY16 and Liu21/Ost01 corrections are different, the critical states derived with the two different corrections are consistent. Although it is hard to assess the uncertainty of the $\lambda$ values estimated from the modified DCF methods due to the many assumptions adopted, it is reasonable to think that the systematic uncertainties from those assumptions may shift $\lambda$ toward larger or smaller values, but does not significantly change the general trend of $\lambda$. The trend of increasing $\lambda$ with density in G28.34 is in agreement with the $\lambda-N$ trend from the previous DCF estimations in the literature \citep[reviewed in][]{2022ApJ...925...30L, 2022FrASS...9.3556L}, which is shown in Figure \ref{fig:G28_lambda_dcf}(d) for comparison. Note that the sample of the DCF compilation statistics in \citet{2022ApJ...925...30L, 2022FrASS...9.3556L} are mostly from different regions, while our analysis presents the multi-scale magnetic critical state in an IRDC for the first time. The similarity between the multi-scale magnetic critical state in IRDC G28.34 and in other star formation regions may suggest that early massive star formation regions follow a similar evolution route as in other star formation regions of different evolutionary stages and masses. The general trend of $\lambda$ appears to be consistent with magnetic field-controlled star formation theories \citep{2006ApJ...646.1043M}, where magnetically sub-critical clouds gradually form super-critical substructures that collapse. The increase of $\lambda$ at higher densities may be due to the dissipation of magnetic flux \citep[e.g., ambipolar diffusion or magnetic reconnection,][]{1999ASIC..540..305M, 1999ApJ...517..700L}, or mass accumulation along magnetic field lines. 

%agree with Section \ref{sec:B_LG}

\subsubsection{Gravity vs Sum of competing forces}

The total equilibrium state of star formation regions can be determined by comparing $\vert E_{\mathrm{G}}\vert$ with the competing terms ($2E_{\mathrm{th}} + 2E_{\mathrm{turb}} + E_\mathrm{B}$) in the virial equation. Similar to the turbulent virial parameter, we could define a total virial parameter as the ratio of energies: 
\begin{equation}\label{eq:virialtotalE}
\alpha_{\mathrm{total,E}} = \frac{2E_{\mathrm{th}} + 2E_{\mathrm{turb}} + E_\mathrm{B}}{\vert E_{\mathrm{G}}\vert}.
\end{equation}
Alternatively, the total virial parameter can be defined as the ratio between the total virial mass and mass:
\begin{equation}\label{eq:virialtotal}
\alpha_{\mathrm{total,M}} = \frac{M_{\mathrm{total}}}{M}.
\end{equation}
Note that the values of $\alpha_{\mathrm{total,E}}$ and $\alpha_{\mathrm{total,M}}$ are not equal. The total virial mass is given by \citep{2020ApJ...895..142L}
\begin{equation}
M_{\mathrm{total}} = \sqrt{M^2_{\mathrm{B}} + (\frac{M_{\mathrm{turb}}+M_{\mathrm{th}}}{2})^2} + \frac{M_{\mathrm{turb}}+M_{\mathrm{th}}}{2},
\end{equation}
where the thermal virial mass is estimated with
\begin{equation}
M_{\mathrm{th}} = \frac{3k_iR }{G}\frac{k_{B} T}{\mu_{\mathrm{H_2}} m_{\mathrm{H}}},
\end{equation}
and the magnetic virial mass\footnote{Some previous studies wrote the magnetic virial mass of a uniform spherical structure as $M'_{\mathrm{B}} = (5V_{\mathrm{A}}^2R)/(6G)$, where $V_{\mathrm{A}}$ is the 3D Alfv\'{e}nic velocity. As demonstrated in the Appendix of \citet{2020ApJ...895..142L}, $M'_{\mathrm{B}}$ underestimates $M_{\mathrm{B}}$ when $M>M_{\mathrm{B}}$, and vice versa. Thus, $M'_{\mathrm{B}}$ is wrong and should not be used in the virial calculation.} is estimated with
\begin{equation}
M_{\mathrm{B}} = \frac{\pi R^2 B}{\sqrt{1.5\mu_0\pi G/k_i}}.
\end{equation}

The environmental gas should have $\alpha_{\mathrm{total}}>1$ since the magnetic field itself is stronger than gravity. i.e., gravity is unimportant in the large-scale diffuse gas. At the intermediate scale, although it is clear that magnetic fields are dominated by gravity in the clumps (i.e., $\lambda > 1$), it is hard to assess their total equilibrium states (considering both the turbulent and magnetic supports). This is because there are some uncertainties in the derivation of the pure turbulent velocity dispersion at specific scales for the single-dish data (see Section \ref{sec:line}). Approximately adopting an average value between the LOS velocity dispersion and the velocity centroid dispersion as the turbulent velocity dispersion, we obtain that some clumps have $\alpha_{\mathrm{total}}<1$ while some clumps have $\alpha_{\mathrm{total}}>1$, with the average state being near quasi-equilibrium (i.e., $\overline{\alpha_{\mathrm{total}}}$ not far from 1). Alternatively, assuming that the magnetic support is comparable to the turbulent support at the clump scale, we obtain similar results that the average state of clumps is near quasi-equilibrium. At the core scale, even when considering the upper limit of supporting forces, the total virial parameter for core MM4-core4 is only $\alpha_{\mathrm{total,E}}\sim 0.5$. Thus, it is safe to say that gravity dominates over the combination of competing forces for MM4-core4. 

In summary, we suggest that the G28.34 cloud is located in a globally-supported environment and its clumps are likely in an approximate quasi-equilibrium state, but the cores therein are undergoing dynamic collapse. 
%Considering that G28.34 is in a trans-to-sub-Alfv\'{e}nic environment and may transit into a trans-to-super-Alfv\'{e}nic state within the cloud (see Section \ref{sec:roa}), it is reasonable to assume that the magnetic support is comparable to the turbulent support at the clump scale. Adopting this assumption, the clumps tend to have $\alpha_{\mathrm{turb+B}}$ around 1 (i.e., near quasi-equilibrium).

\subsubsection{Implications on massive star formation}
It has been long debated whether molecular clouds and their substructures are in equilibrium or not. The Planck dust polarization survey \citep{2016AA...586A.138P} has found that the Gould Belt Clouds, including one massive star formation region (Orion), are in magnetically sub-critical (i.e., $\lambda<1$) states with DCF estimations. Later, \citet{2023ApJ...945..160L} analyzed the Planck dust polarization data in another massive star formation region NGC 6334 and found that this region is also in a magnetically sub-critical state at large scale. Combined with our finding of magnetically sub-critical state in the environment of IRDC G28.34, we suggest that massive star-forming clouds may be globally supported by magnetic fields both in the early and late evolutionary stages. The quasi-equilibrium state of clouds is essential to explain the low star formation rate observed and to allow ambipolar diffusion to happen \citep{2007ARA&A..45..565M}, but contradicts the idea that clouds are short-lived and are undergoing global dynamical collapse \citep[i.e., the global hierarchical collapse model,][]{2019MNRAS.490.3061V}. On the other hand, the sub-critical state does not necessarily mean the region will expand or disperse. Sub-critical clouds could still create overdense regions via other mechanisms such as large-scale turbulent inertial flows \citep[i.e., the inertial-flow model,][]{2020ApJ...900...82P} or local infall through magnetic channels \citep{2018ApJ...855...39K, 2022ApJ...940...89K}, instead of via symmetric gravitational collapse. 

Within the G28.34 cloud, while the clumps may not be far from equilibrium, the cores are dominated by gravity. This is consistent with the findings of previous studies that gravity is more important in higher-density regions, and that cores in both early and evolved massive star formation regions tend to be averagely dominated by gravity even considering both the magnetic and turbulent supports \citep[see reviews in][]{2022ApJ...925...30L, 2022FrASS...9.3556L}. The dynamic state of cores is inconsistent with the turbulent core accretion model \citep{2002Natur.416...59M}, which predicts $\alpha_{\mathrm{total}}\sim1$ across different scales. The near-equilibrium state of clumps tends to be inconsistent with the competitive accretion model \citep{1997MNRAS.285..201B}, which requires $\alpha_{\mathrm{total}}<1$ for cloud substructures \citep{2005Natur.438..332K}. Thus, both the two major massive star formation models may need some modifications to be consistent with the observational results. 

%There are two major massive star formation models. The turbulent core accretion model and the competitive accretion model.

%\subsection{Multi-scale magnetic fields and velocity fields}

\subsection{Magnetic fields or turbulence?}
Different star formation theories have distinct explanations for the controlling factor of the formation and evolution of molecular clouds and cloud substructures: some emphasize the role of magnetic fields \citep{2006ApJ...646.1043M}, while others highlight the role of turbulence \citep{2004RvMP...76..125M}. Qualitative and quantitative comparisons between magnetic fields and turbulence are needed to determine their relative importance in star formation. 

There is no way to observationally compare the turbulent kinetic energy and the turbulent magnetic energy yet \citep{2022FrASS...9.3556L}. People usually assume an equipartition between turbulent kinetic and magnetic energies, which implicitly assumes that the total magnetic energy is larger than the turbulent kinetic energy. On the other hand, the relative importance between the ordered magnetic field and the turbulence can be directly derived from the dust polarization maps, without the need for molecular line observations. The relation between the 3D Alfv\'{e}nic Mach number and the POS angular dispersion is $\mathcal{M}_{A} \sim (f_t/f_u/Q_c/f_o) \sigma_\theta$ \citep{2022FrASS...9.3556L}, where $f_u$ and $f_t$ are factors for the 3D-to-POS conversion of $B_0$ and $B_{\mathrm{t}}$, $f_o$ is a correction factor for the ordered field contribution, and $Q_c$ is a correction factor to account for other effects (e.g., the LOS signal integration, the difference between the orientation and direction, et al.). However, the actual value of $f_t$, $f_u$, and $f_o$ is usually unclear in individual regions. Statistical values exist for the correction factors \citep[e.g.,][]{2004ApJ...600..279C, 2021ApJ...919...79L}, but those statistical values (especially for $f_u$ and $f_o$) may only be appropriate for statistical studies \citep[e.g.,][]{2022ApJ...925...30L, 2023ASPC..534..193P}. Thus, we refrain from deriving $\mathcal{M}_{A}$ from polarization angular dispersions in G28.34. 

The relative orientation analysis (Section \ref{sec:roa}) offers an alternative way to qualitatively compare magnetic fields and turbulence. With an approach equivalent to the HRO analysis, we find that the magnetic field and column density gradient transits from statistically more perpendicular to more parallel as column density increases for the Planck and JCMT observations (Section \ref{sec:B_NG}), which is a sign of trans-to-sub-Alfv\'{e}nic turbulence at large scales. This transition of alignment is consistent with previous studies in low-mass star formation regions and in evolved massive star formation regions \citep{2016AA...586A.138P, 2023ApJ...945..160L}. 
%On the other hand, we find that the magnetic field and velocity gradient transits from statistically more perpendicular to randomly aligned as column density increases for the Planck and JCMT observations (Section \ref{sec:VG_B}). The decrease in turbulence anisotropy may indicate that the state transits to super-Alfv\'{e}nic within the cloud, although we cannot rule out the possibility that this loss of anisotropy is due to gravity. At higher densities revealed by ALMA, we find that the magnetic field and velocity gradient transit from statistically more parallel to more perpendicular as column density increases, which may be due to a complex interaction between star formation activities and magnetic fields, but the exact reason for the alignment is still unclear and needs future investigations. This trend of alignment within G28.34 is different from what was found in the evolved massive star formation region NGC 6334 \citep{2023ApJ...945..160L}, where the magnetic field and velocity gradient stay perpendicular within the cloud. The difference between G28.34 and NGC 6334 may be due to their different local physical conditions and star formation history. However, future studies with more samples will be needed to draw more general conclusions.

The statistical tool VGT offers another way to compare magnetic fields and turbulence at large scales. Implementing the VGT, we find $\mathcal{M}_{A}=0.74$ within $r=15$ pc, which implies that G28.34 is located in a slightly sub-Alfv\'{e}nic environment. The sub-Alfv\'{e}nic state is consistent with the results from previous VGT studies in nearby low-mass star formation regions \citep{2019NatAs...3..776H}.

In summary, we conclude that both low-mass and high-mass star formation happens in a trans-to-sub-Alfv\'{e}nic environment, which means magnetic fields play a more important role than turbulence in controlling star formation at large scales. Since gravity is not important in large-scale diffuse gas, cloud formation and evolution at the cloud scale should be mainly controlled by the property of trans-to-sub-Alfv\'{e}nic MHD turbulence. Within the cloud, the situation is more complicated at small scales. More reliable analysis methods are required to compare magnetic fields and turbulence in high-density and small-scale regions with significant gravity. 
%This is also evidenced by the lack of clear continuous power-law relation for the multi-scale velocity statistics in G28.34 (Section \ref{sec:line}), which contradicts the expectation of the turbulence-controlled star formation model \citep{2004RvMP...76..125M}. 
%The cloud formation is mostly controlled by the interplay between magnetic fields and turbulence.  It is reasonable to think that some cloud substructures are sub-Alfv\'{e}nic at small scales, while some others are in super-Alfv\'{e}nic states. 

%\subsection{Magnetic fields and outflow}

\section{Summary} \label{sec:summary}
With JCMT and ALMA dust polarization observations as well as Planck dust polarization data, we present a study of the multi-scale magnetic fields in IRDC G28.34. We also have studied the multi-scale velocity fields in G28.34 with molecular line data from FCRAO-14m, JCMT, and ALMA. The findings are:
\begin{enumerate}
    \item Within our JCMT detection region, 5 clumps (MM1, MM2, MM11, MM16, and MM17) have average magnetic fields aligned within 30$\degr$ of the cloud-scale magnetic field, while 5 clumps (MM4, MM6, MM9, MM10, and MM14) have averaged magnetic fields misaligned at 60$\degr$-90$\degr$ with respect to the cloud-scale magnetic field. The bimodal distribution suggests that the clump-scale magnetic field is organized with respect to the cloud-scale magnetic field. In MM4, the core-to-condensation scale magnetic field is preferentially aligned with the clump-scale magnetic field and the clump-scale magnetic field is perpendicular to the chain of fragments. This may suggest that the magnetic field plays a crucial role in the clump collapse and fragmentation process. 
    \item With a simple power-law fit for the mass-radius and density-radius relations of G28.34, we obtain $M\propto r^{1.59}$ and $n\propto r^{-1.41}$ between $\sim$0.07 pc and $\sim$7 pc. 
    %\item We have studied the multi-scale velocity fields in G28.34 with velocity dispersion-radius ($\sigma_v-r$) relation, velocity centroid dispersion-radius ($\sigma_c-r$) relation, and velocity centroid structure function (VDF). The velocity statistics are not continuous at different scales revealed by different instruments and lines, nor universal in different regions. This does not support the idea that the non-thermal motions in star formation regions are mainly due to the cascade of interstellar supersonic turbulence and that the star formation within molecular clouds is mainly regulated by large-scale interstellar turbulence. We suggest that the LOS velocity dispersion within a cloud measured from single-dish data may systematically overestimate the true velocity dispersion at corresponding scales.
    \item We have studied the multi-scale relative orientations between magnetic fields, column density gradients, and local gravity in G28.34. As column densities increase, the magnetic field and column density gradient transit from preferentially more perpendicular to more parallel, then transit back to a random alignment. The alignment between the magnetic field and local gravity shows a similar varying trend with column density. The results of the relative orientation analysis suggest that G28.34 is located in a trans-to-sub-Alfv\'{e}nic environment, the magnetic field is resisting gravitational collapse in the large-scale diffuse gas, the magnetic field is distorted by gravity within the cloud, and the magnetic field is affected by star formation activities in high-density regions.
    %On the other hand, as column density increases, the magnetic field and velocity gradient transit from statistically more perpendicular to a random alignment for the Planck/FCRAO-14m and JCMT data, but their alignment transits from statistically more parallel to more perpendicular for the ALMA data. 
    \item We have measured the magnetic field strengths in the environmental gas, infrared dark clumps, and core MM4-core4 of G28.34 with modified DCF analysis. In this early massive star formation region, the magnetic field strengths do not significantly increase as density increases. With the estimated magnetic field strength, we find that the normalized mass-to-flux ratio $\lambda$ increases with density but decreases with radius, and transits from magnetically subcritical ($\lambda<1$) in the environmental gas to supercritical ($\lambda<1$) at clump/core scales. This is in agreement with the prediction of magnetic field-controlled star formation theories. With an alternative analysis using the KTH method, we find a similar increasing trend of $\lambda$ with density, except for the Planck data where $\lambda\sim1$. 
    % We emphasize that it could be inappropriate to assume equipartition between turbulent kinetic energy and coupling-term magnetic field fluctuation within molecular clouds in the magnetic field strength calculation.
    \item Combining the thermal, turbulent, and magnetic supports, we find that the environmental gas is super-virial ($\alpha_{\mathrm{total}}>1$, supported) and MM4-core4 is sub-virial ($\alpha_{\mathrm{total}}<1$, gravity-dominant). The infrared dark clumps may be averagely in a near-equilibrium state  ($\alpha_{\mathrm{total}}\sim1$). The transition from $\alpha_{\mathrm{total}}>1$ to $\alpha_{\mathrm{total}}<1$ is inconsistent with either the competitive accretion model or the turbulent core accretion model, suggesting that the two major massive star formation models may need some modifications. 
    \item With a VGT analysis, we find $\mathcal{M}_{A}=0.74$ within $r=15$ pc centered at G28.34. The slightly sub-Alfv\'{e}nic state is consistent with the relative orientation analysis and implies that magnetic fields regulate cloud formation and evolution at large scales. More reliable analysis methods are required to compare magnetic fields and turbulence in high-density regions where gravity is dominant. 
    %regulated by sub-Alfvenic turbulence
\end{enumerate}
%In summary, suggests that magnetic fields dominate star formation in large-scale diffuse gas. while gravity dominates star formation in small-scale dense gas.

%\begin{acknowledgments}
\section*{}
We thank the anonymous referee for the constructive and in-depth comments that have improved the clarity of this work, especially on the discussions of the large-scale LOS contamination, the turbulence anisotropy, and the small-scale thermal effects. J.L. thanks Prof. Hua-Bai Li for helpful comments on the corresponding scales of linewidth and Ms. Yuchen Xing for helpful discussions on the mass-size relation. J.L. also thanks Prof. Patricio Sanhueza, Dr. Paulo Cortes, and the active JCMT BISTRO team members for helpful general discussions on magnetic field properties. 
%We thank the anonymous referee for the detailed comments. High-mass protostellar Object (HMPO)  and Q.Z. thank Mr. Yue Hu for helpful discussions on the velocity anisotropy 

J.L. is partially supported by a Grant-in-Aid for Scientific Research (KAKENHI Number JP23H01221) of JSPS and was supported by the EAO Fellowship Program under the umbrella of the East Asia Core Observatories Association. K.Q. is supported by National Key R\&D Program of China grant No. 2022YFA1603100. K.Q. acknowledges the support from National Natural Science Foundation of China (NSFC) through grant Nos. U1731237, 11590781, and 11629302. H.B.L. is supported by the National Science and Technology Council (NSTC) of Taiwan (Grant Nos. 111-2112-M-110-022-MY3). Z.Y.L. is supported in part by NSF AST-2307199 and NASA 80NSSC20K0533. J.M.G. acknowledges the support from the program Unidad de Excelencia María de Maeztu CEX2020-001058-M and the grant PID2020-117710GB-I00 (MCI-AEI-FEDER, UE). TGSP gratefully acknowledges support by the National Science Foundation under grant No. AST-2009842 and AST-2108989.

This paper makes use of the following ALMA data: ADS/JAO.ALMA\#2016.1.00248.S and ADS/JAO.ALMA\#2017.1.00793.S. ALMA is a partnership of the ESO (representing its member states), NSF (USA) and NINS (Japan), together with NRC (Canada), MOST and ASIAA (Taiwan), and KASI (Republic of Korea), in cooperation with the Republic of Chile. The Joint ALMA Observatory is operated by ESO, AUI/NRAO, and NAOJ. The National Radio Astronomy Observatory is a facility of the National Science Foundation operated under cooperative agreement by Associated Universities, Inc.
The JCMT is operated by the EAO on behalf of NAOJ; ASIAA; KASI; CAMS as well as the National Key R\&D Program of China (No. 2017YFA0402700). Additional funding support is provided by the STFC and participating universities in the UK and Canada. Additional funds for the construction of SCUBA-2 were provided by the Canada Foundation for Innovation.
This work is based on observations obtained with Planck (http://www.esa.int/Planck), an ESA science mission with instruments and contributions directly funded by ESA Member States, NASA, and Canada. 
%\end{acknowledgments}

\vspace{5mm}
\facilities{Planck, JCMT, ALMA}


\software{Astropy \citep{2013A&A...558A..33A,2018AJ....156..123A},  
Matplotlib \citep{2007CSE.....9...90H}.
          }

\appendix
\section{Polarization percentage}\label{sec:PM}
\begin{figure}[!tbp]
\gridline{\fig{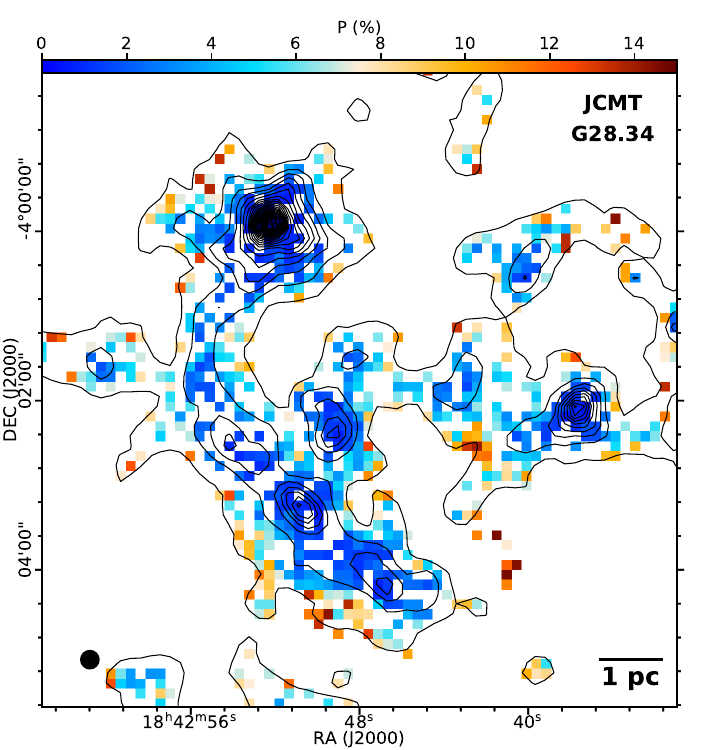}{0.45\textwidth}{}}
\caption{Polarization percentage (colorscales) of our POL-2 data (S/N($PI$)$>$2). Black solid contours indicate the JCMT Stokes-I intensities. Contour starts at 50 mJy beam$^{-1}$ and continues at 150 mJy beam$^{-1}$.  \label{fig:PMpol2}}
\end{figure}

\begin{figure}[!tbp]
\gridline{\fig{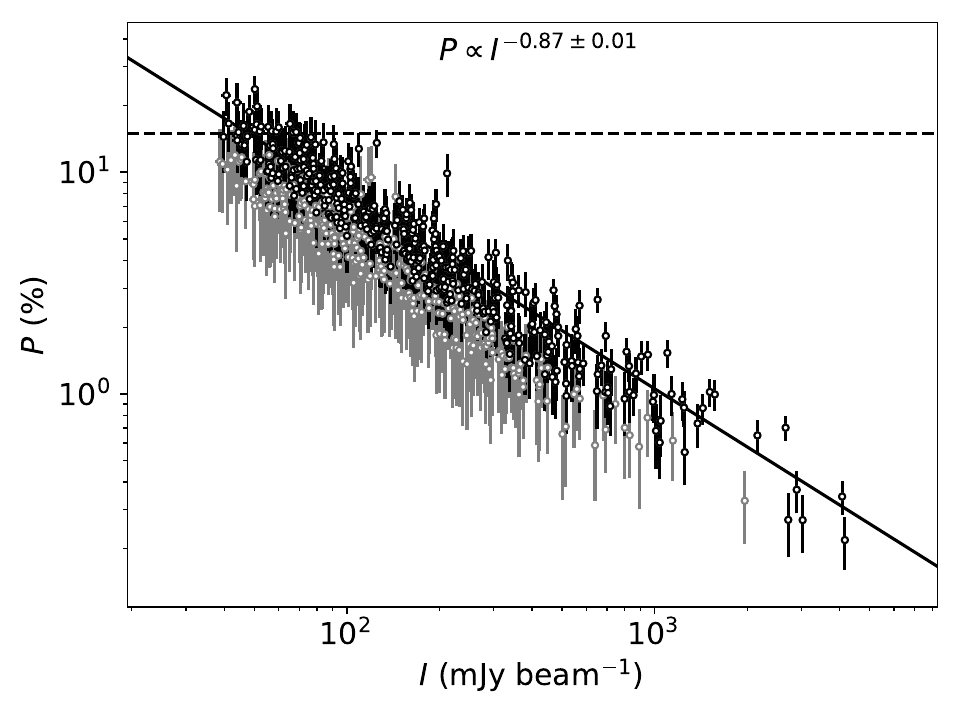}{0.45\textwidth}{}}
\caption{Relation between polarization percentage and total intensity for our POL-2 data (S/N($I$)$>$3) within the central r=3$\arcmin$ region. Grey and black colors correspond to 2$<$S/N($PI$)$<$3 and S/N($PI$)$>$3, respectively. The horizontal dashed line indicates $P=15$\%. The solid line indicates the result from a simple power-law fit for the $P-I$ relation for data points with S/N($PI$)$>$3. \label{fig:P_I_pol2}}
\end{figure}

Figure \ref{fig:PMpol2} shows the polarization percentage map of our POL-2 data. Figure \ref{fig:P_I_pol2} shows the $P-I$ relation. The polarization percentage clearly decreases with increasing intensity. A small portion of data points have $P>15\%$, but previous theoretical models suggested that the polarization percentage of submm observations should not exceed 15\% \citep{2009ApJ...696....1D}. We conservatively excluded data points with $P>15\%$ in our analysis. Note that the large-scale extended emissions are filtered out during the JCMT data reduction processes. The $I$ emission may be more extended than the polarized emission in G28.34. Thus, the JCMT data reduction may have removed more $I$ emissions than $Q$ and $U$, which could lead to a systematical overestimation of $P$. This presents a plausible explanation for the high $P$ values observed in weak-emission regions. 

\section{Molecular line}\label{sec:appline}

\begin{figure*}[!htbp]
 \gridline{
 \fig{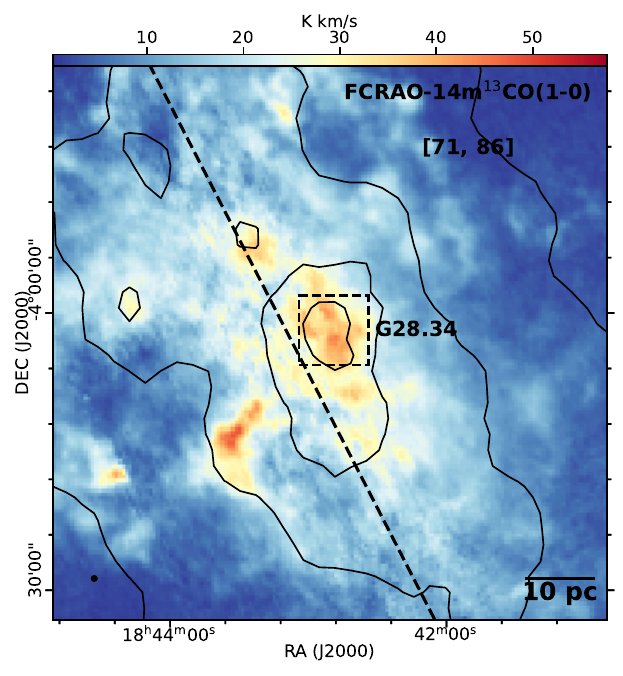}{0.33\textwidth}{(a)}
 \fig{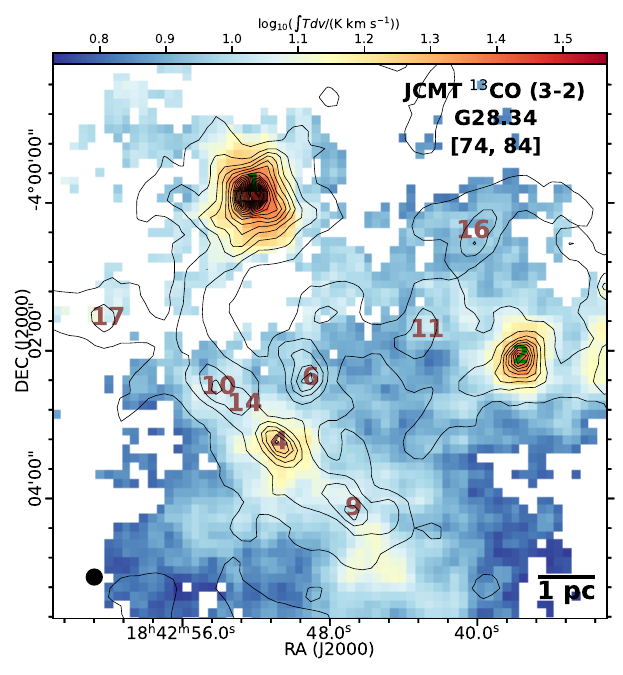}{0.33\textwidth}{(b)}
 \fig{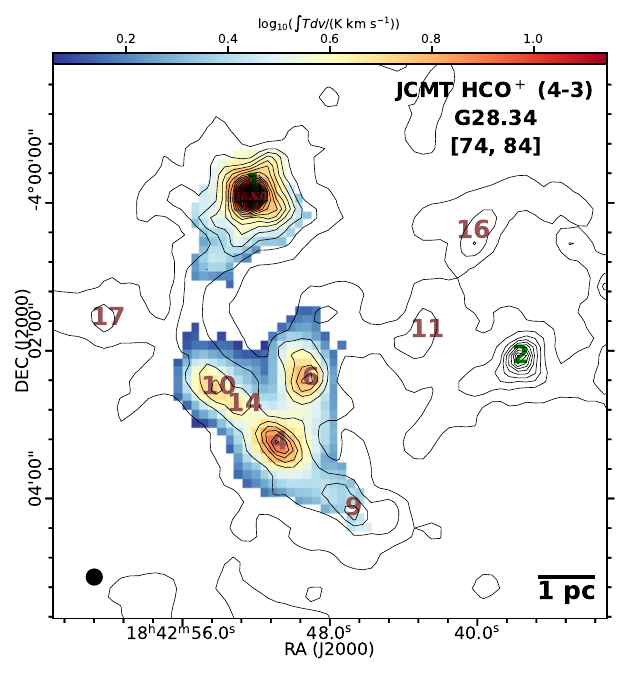}{0.33\textwidth}{(c)}
 }
\caption{(a) Integrated intensity of FCRAO-14m $^{13}$CO (1-0) in the surrounding of G28.34. Black contours correspond to the Planck $\tau_{353}$ map, starting from 0.0005 and continuing with an interval of 0.0005. A dashed line indicates the galactic plane ($b=0\degr$). The dashed rectangle indicates the JCMT map area in (b) and (c). (b)-(c) Integrated intensity of JCMT $^{13}$CO (3-2) and HCO$^{+}$ (4-3) toward the IRDC G28.34. Black contours correspond to the JCMT 0.85 mm dust continuum map. Contour starts at 50 mJy beam$^{-1}$ and continues at 150 mJy beam$^{-1}$. \label{fig:G28_large_line_m0}}
\end{figure*}

\begin{figure*}[!htbp]
 \gridline{
 \fig{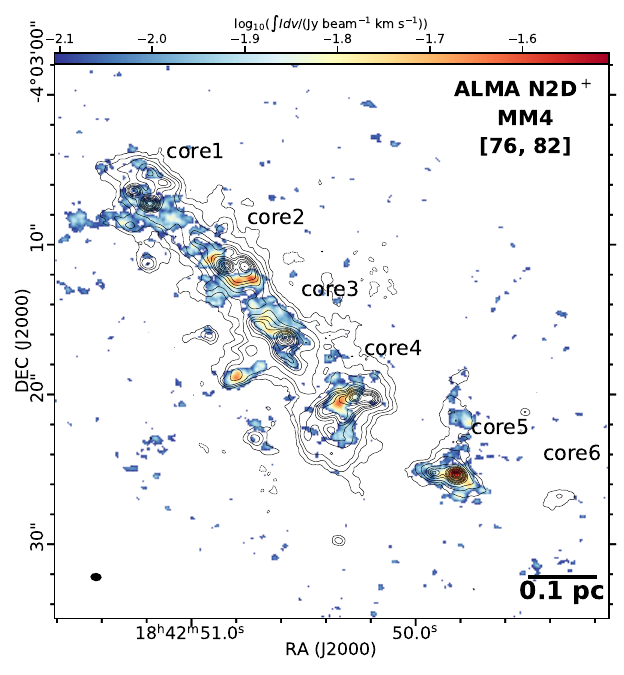}{0.4\textwidth}{(a)}
 \fig{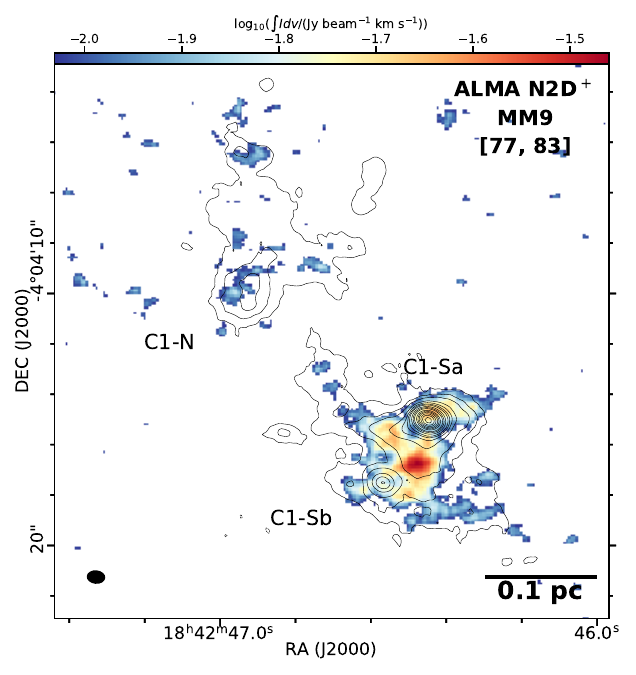}{0.4\textwidth}{(b)}
 }

\caption{Integrated intensity of ALMA N$_2$D$^{+}$ (3-2) toward MM4 and MM9. Contours correspond to the ALMA 1.3 mm dust continuum map. Contour levels are ($\pm$3, 6, 10, 20, 30, 40, 50, 70, 90, 110, 150, 180, 210, 250, 290, 340, 390, 450) $\times \sigma_{I}$. \label{fig:G28_alma_line_m0}}
\end{figure*}

Figures \ref{fig:G28_large_line_m0} and \ref{fig:G28_alma_line_m0} show the integrated intensity maps of the FCRAO-14m $^{13}$CO (1-0), JCMT $^{13}$CO (3-2) and HCO$^{+}$ (4-3), and  ALMA N$_2$D$^{+}$ (3-2) data. The integrated velocity ranges are indicated in each panel. 

\begin{figure}[!htbp]
 \gridline{
 \fig{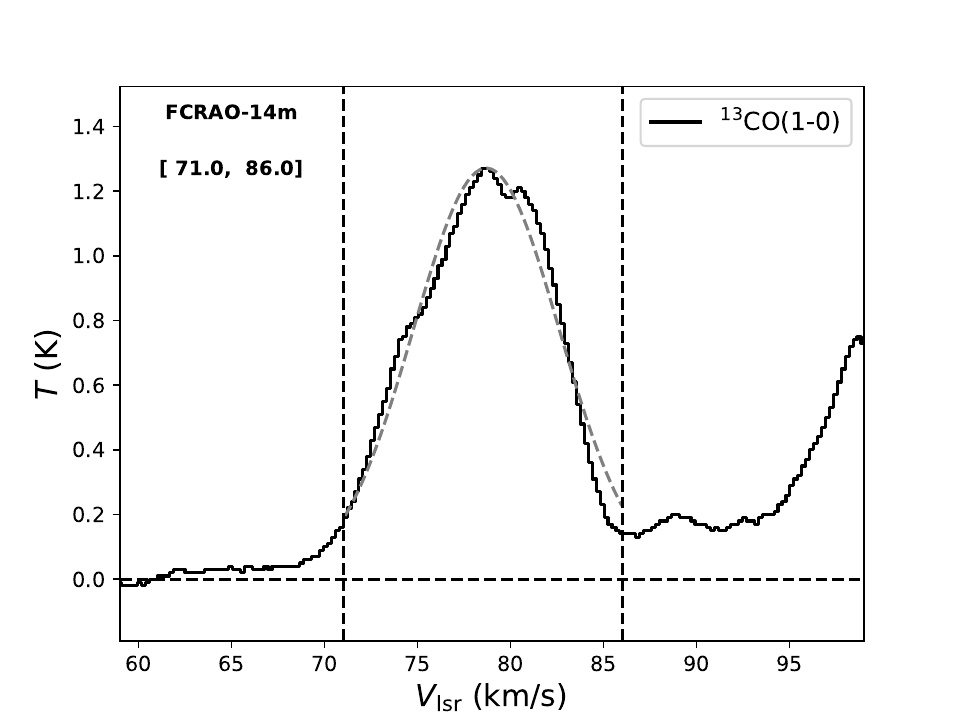}{0.33\textwidth}{} }
\caption{Example of the average FCRAO-14m $^{13}$CO (1-0) spectra within the central 15-pc (radius) area. The dashed single-Gaussian profile indicates the best-fitting result. The vertical dashed lines indicate the velocity ranges used for the analyses.  \label{fig:G28_large_line_spec} }
\end{figure}

\begin{figure}[!htbp]
 \gridline{
 \fig{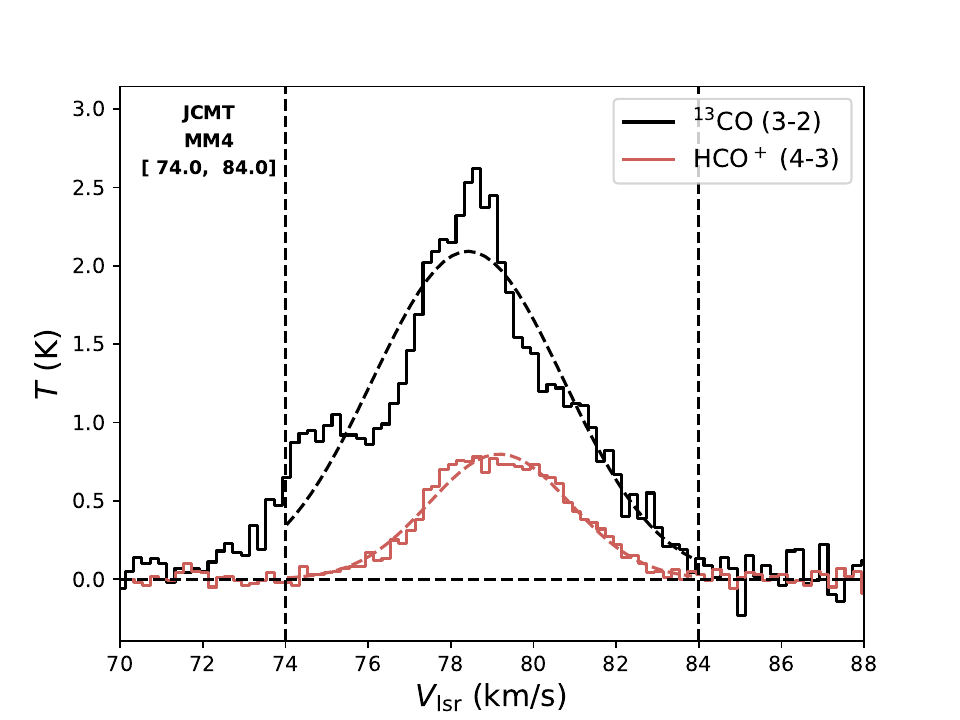}{0.33\textwidth}{}
 \fig{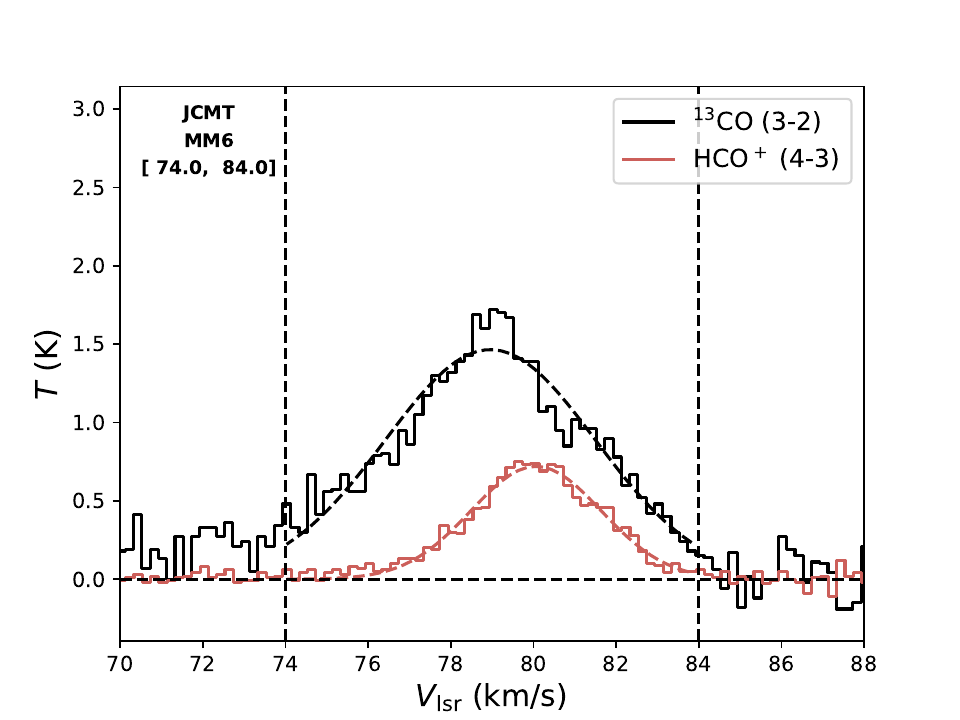}{0.33\textwidth}{}
 }
\gridline{
 \fig{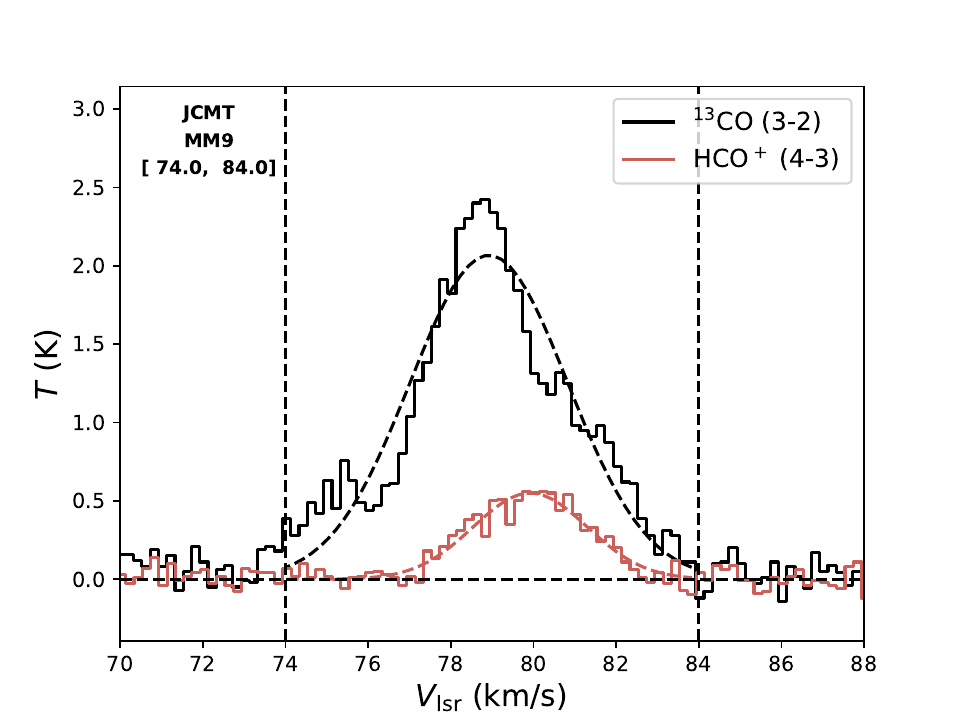}{0.33\textwidth}{}
 \fig{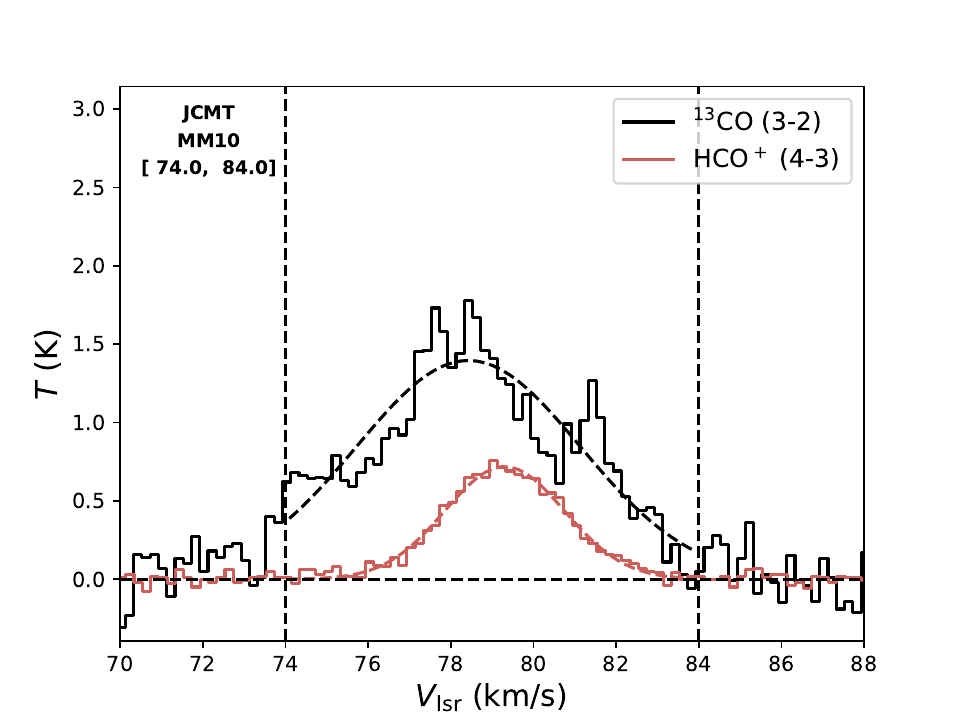}{0.33\textwidth}{}
}
\caption{Example of the average JCMT $^{13}$CO (3-2) (black) and HCO$^{+}$ (4-3) (red) spectra of each clump within the central 1-pc (radius) area. The dashed Gaussian profiles indicate the best-fitting results for $^{13}$CO (3-2). The vertical dashed lines indicate the velocity ranges used for the analyses. \label{fig:G28_jcmt_line_spec}}
\end{figure}

\begin{figure}[!htbp]
 \gridline{
 \fig{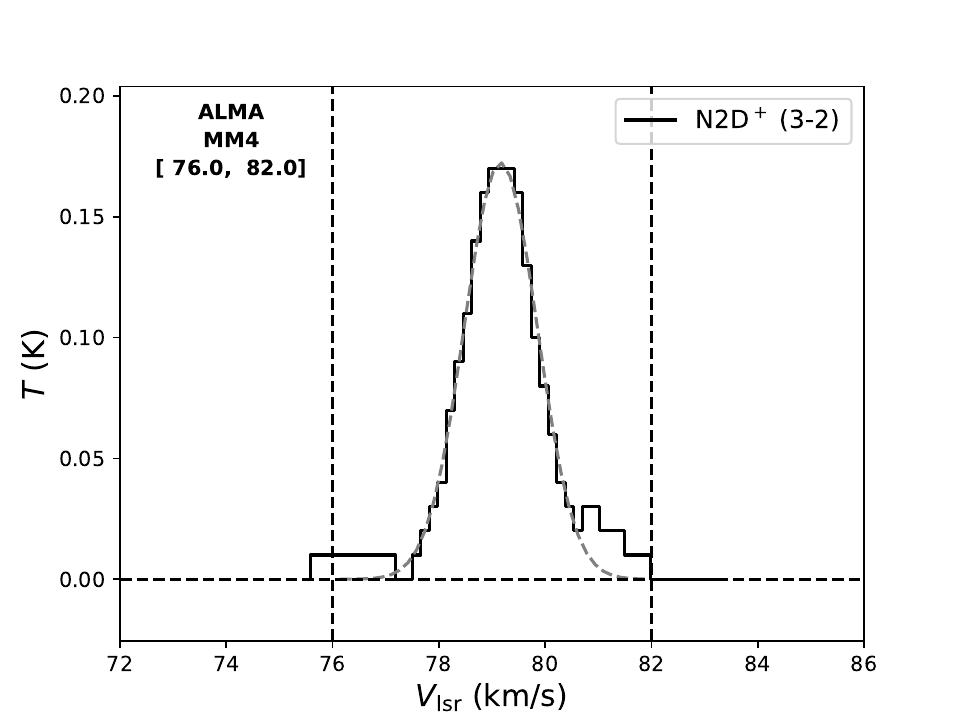}{0.33\textwidth}{}
 \fig{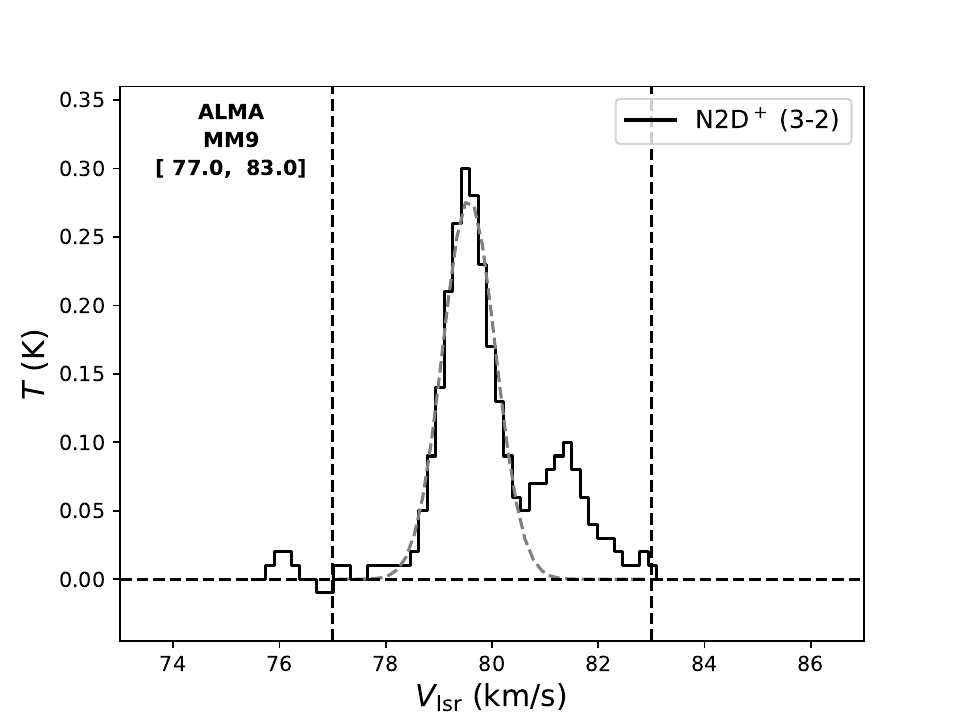}{0.33\textwidth}{}
 }
\caption{Average ALMA N$_2$D$^{+}$ (3-2) spectra of MM4 and MM9 within the whole emission area in the ALMA field. The dashed Gaussian profiles indicate the best-fitting results. The vertical dashed lines indicate the velocity ranges used for the analyses. Our computing ability for ALMA data reduction is limited, so we only imaged the line within velocity ranges from 75 to 83 km s$^{-1}$.  \label{fig:G28_alma_line_spec}}
\end{figure}

Figures \ref{fig:G28_large_line_spec}, \ref{fig:G28_jcmt_line_spec}, and \ref{fig:G28_alma_line_spec} show examples of the average spectra for different line data. Many lines show signs of multiple velocity components, which provides evidence for the superposition of substructures along the LOS. We do not try to identify or separate the multiple velocity components because the combination of those components could appear as a single Gaussian line profile at a coarser spatial resolution, and there are always multiple velocity components seen with higher resolution observations inside a ``single'' velocity component. We perform a single-peak Gaussian fit for each line and obtain the best-fitting result. The fitted velocity dispersion should be the upper limit of the pure turbulent velocity dispersion. 

%\section{Comparing gravity with gas pressure}

%Here we compare the gravitational term 
%$\vert \rho \nabla\Phi \vert$ and gas pressure term $\vert \nabla P \vert$ in the hydrodynamic equation, where $P = n k_{B} T$ is the gas pressure, $\Phi$ is the gravitational potential. For a spherical structure with density profile $n = n_0 (r_0/r)^{i}$, the gravitational potential is 
%\begin{equation}
%\Phi = -\frac{4\pi G \rho r^{2} }{(3-i)(2-i)}.
%\end{equation}
%Inserting the density profile into the gas pressure term, we have 
%\begin{equation}
%\vert \nabla P \vert = (2-i) n k_{B} T r^{-1}
%\end{equation}
%for $0<i<2$ and constant temperature. Inserting the density profile into the gravitational term, we have 
%\begin{equation}
%\vert \rho \nabla \Phi \vert = \frac{4-i}{(3-i)(2-i)} 4 \pi G \rho^2 r
%\end{equation}
%for $0<i<2$ and constant temperature. 


%\nocite{*}

\bibliography{G28mul}{}
\bibliographystyle{aasjournal}

\end{CJK*}
\end{document}